\documentclass[natbib,smallextended]{svjour3}       
\smartqed  
\usepackage{tabularx} 
\usepackage{amsmath}  
\usepackage{amssymb}
\usepackage{graphicx} 
\usepackage{subcaption}
\usepackage[normalem]{ulem}
\usepackage{dirtytalk}
\usepackage{aas_macros}
\usepackage{float}
\usepackage[colorlinks=true,citecolor=blue,urlcolor=blue,breaklinks]{hyperref}
\usepackage[utf8]{inputenc}
\usepackage[english]{babel}

\usepackage{siunitx}
\DeclareSIUnit \h {\ensuremath{\mathit{h}}}
\DeclareSIUnit \pc {pc}

\journalname{The Astronomy and Astrophysics Review}

\begin{document}

\title{A buyer's guide to the Hubble Constant}

\author{Paul Shah \and Pablo Lemos \and Ofer Lahav*}

\institute{P. Shah \and O. Lahav \at 
  Department of Physics and Astronomy, University College London, Gower Street, London, WC1E 6BT, UK \\
\email{paul.shah.19@ucl.ac.uk, o.lahav@ucl.ac.uk}
\and 
P. Lemos \at 
Department of Physics  and Astronomy,
University of Sussex
Brighton, BN1 9QH, UK 
\and 
Department of Physics and Astronomy, University College London, Gower Street, London, WC1E 6BT, UK \\
\email{P.Lemos@sussex.ac.uk} \\ \\
Paul Shah: Orcid 0000-0002-8000-6642
Pablo Lemos: Orcid 0000-0002-4728-8473
Ofer Lahav: Orcid 0000-0002-1134-9035
}

\date{Received: date / Accepted: date}


\maketitle

\begin{abstract}
    Since the expansion of the universe was first established by Edwin Hubble and Georges Lema\^{i}tre about a century ago, the Hubble constant $H_0$ which measures its rate has been of great interest to astronomers. Besides being interesting in its own right, few properties of the universe can be deduced without it. In the last decade a significant gap has emerged between different methods of measuring it, some anchored in the nearby universe, others at cosmological distances. The SH0ES team has found $H_0 = 73.2 \pm 1.3 \; \;\SI{}{\kilo\meter\per\second\per\mega\pc}$ 
    locally, whereas the value found for the early universe by the Planck Collaboration is $H_0 = 67.4 \pm 0.5 \; \;\SI{}{\kilo\meter\per\second\per\mega\pc}$ from measurements of the cosmic microwave background. Is this gap a sign that the well-established $\mathrm{\Lambda CDM}$ cosmological model is somehow incomplete? Or are there unknown systematics? And more practically, how should humble astronomers pick between competing claims if they need to assume a value for a certain purpose? 
    In this article, we review results and what changes to the cosmological model could be needed to accommodate them all. For astronomers in a hurry, we provide a buyer's guide to the results, and make recommendations.
\end{abstract}

\setcounter{tocdepth}{3} 
\tableofcontents

\section{Introduction}
In 1917, Einstein was the first to combine the assumptions of homogeneity and isotropy with his new theory of general relativity, and produce a solution for the universe as a whole \citep{Einstein1917}. 
Einstein imposed his belief in a static universe, and famously introduced the cosmological constant $\Lambda$, to make his equations compatible with this assumption. \citet{Friedmann1922} derived a solution for an expanding (or contracting) universe, but his work remained largely unknown until after his death. Establishing expansion as an observational fact was very challenging with the technology of the day. George Lema\^{i}tre published the first estimate of the expansion rate in \citet{Lemaitre1927}. Two years later, Edwin Hubble\footnote{The historical timeline and attribution of credit for the discovery of the expansion of the universe is examined in \citet{Way2013}.} combined his observations of stellar magnitudes using the Mount Wilson telescope with Shapley's, Humanson's and Slipher's redshifts $z$ to a similar result \citep{Hubble1929}. 
Hubble's constant, as it later became known, is then the constant of proportionality between recession speed $v$ and distance $d$:
\begin{equation}
\label{eq:hubblelaw}
    v = H_0 \, d \; .
\end{equation}
\begin{figure}[h!]
    \centering
    \includegraphics[width=\textwidth]{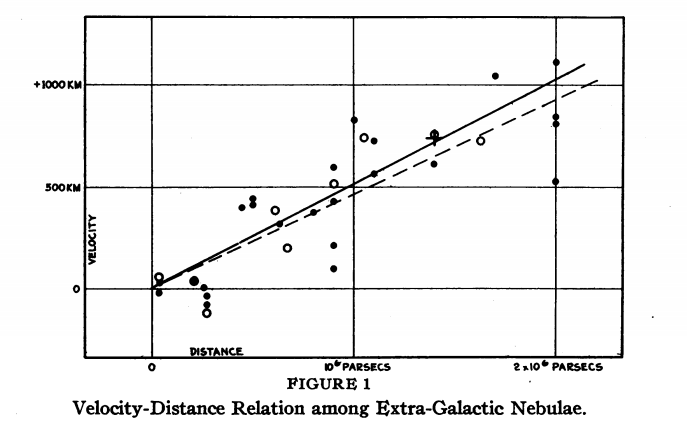}
    \label{fig:hubblediagram}
    \caption{Hubble's original diagram from \citet{Hubble1929}. Despite the typo on the labelling of the y axis, which should read $\SI{}{\kilo\meter\per\second}$ and not $\SI{}{\kilo\meter}$, it is still easy to read off $H_0 \simeq 500\; \;\SI{}{\kilo\meter\per\second\per\mega\pc}$. Hubble and Humanson were using the largest telescope in the world at the time. M31 is recognisable as the lowest black dot in the bottom left; Humanson had determined it is velocity as $220 \;\SI{}{\kilo\meter\per\second}$ towards the Milky Way (the modern value is $110 \;\SI{}{\kilo\meter\per\second}$).}
\end{figure}
Surprisingly perhaps, it was not until 1958 that the first recognisably modern value $H_0 \simeq 75 \;\SI{}{\kilo\meter\per\second\per\mega\pc}$ was published \citep{Sandage1958}. Sandage made several corrections to Hubble's earlier results. Firstly, he noted the population of Cepheid variable stars was not as homogeneous as first thought. This added both scatter and bias to distance estimates, compounded by the low numbers of Cepheids observed. Secondly, and more seriously, Hubble had mistaken (far brighter) HII regions as bright stars, and therefore his estimate of the distances to galaxies was too low. 

\begin{figure}[h!]
    \centering
    \includegraphics[width=0.58\textwidth]{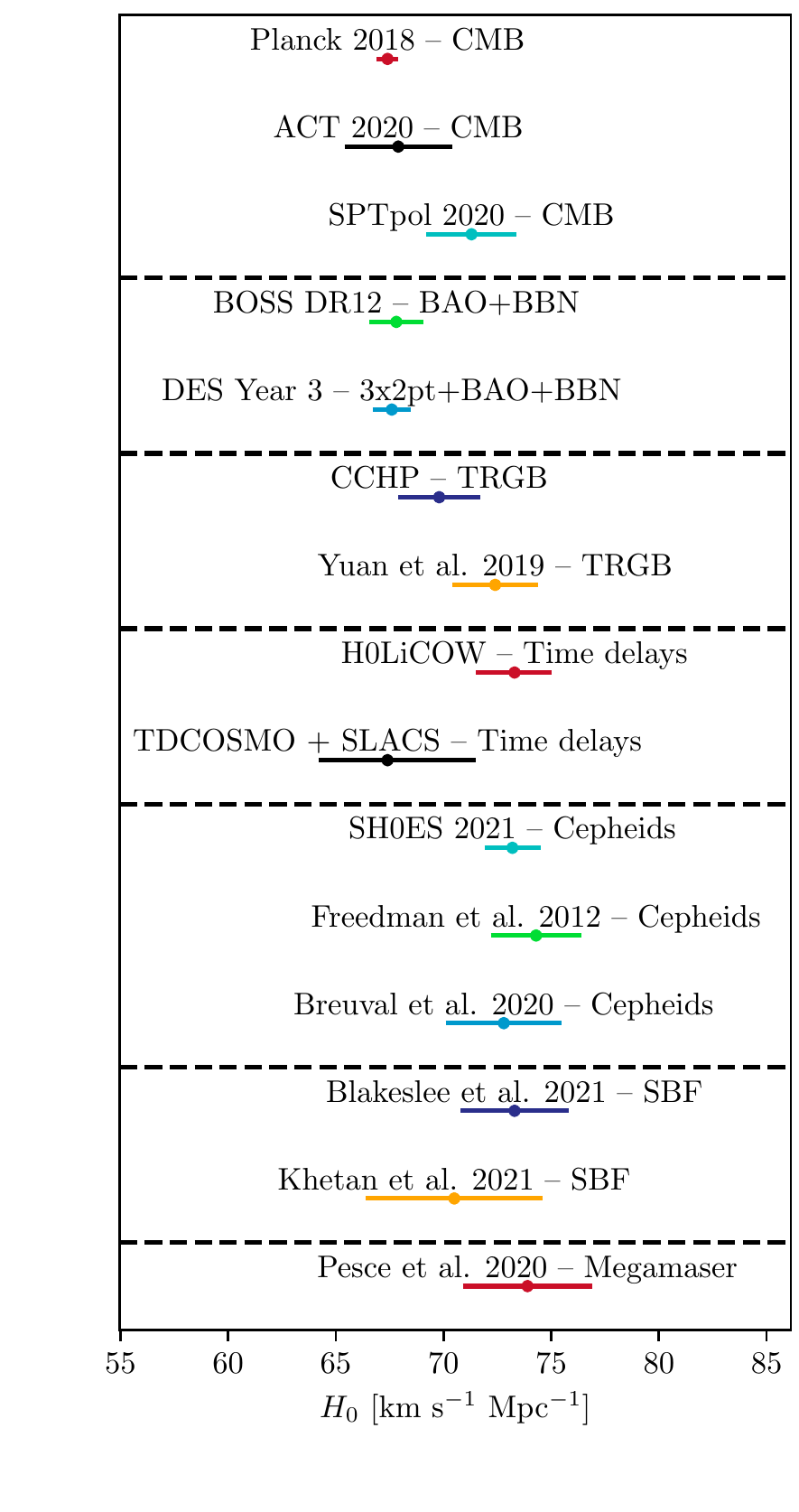}
    \caption{Summary of recent $H_0$ measurements. We have intentionally selected a limited number of results in order to show those that are as independent from each other as possible, in the sense that they use different photometric data, distance calibrators and so on. More comprehensive versions of this plot can be found for example in \citet{DiValentino2021}. Comments:  1) CCHP and Yuan et al. share a common distance to LMC as a calibrator. 2) TDCOSMO is a re-analysis of almost the same data as H0LiCOW, but with changes to the galactic potentials. 3) BOSS and DES share a prior constraint on baryon densities from BBN. 4) The results of Blakeslee et al. use new SBF observations, whereas Khetan et al. use archival SBF distances to calibrate SN Ia. The code use to generate this figure is publicly available at \url{https://github.com/Pablo-Lemos/plot1d}}
    
    \label{fig:H0_tension}
\end{figure}

Fast forward to today, and these historical developments seem to echo some present day debates. $H_0$ is measured in a number of ways, which produce inconsistent values. In particular, in 2018 the Planck collaboration used the Cosmic Microwave Background (CMB) temperature and polarization anisotropies and the $\mathrm{\Lambda CDM}$ cosmological model, to find $H_0 = 67.4 \pm 0.5 \; \;\SI{}{\kilo\meter\per\second\per\mega\pc}$ \citep{PlanckCollaboration2018} whereas in 2021 the SH0ES collaboration \citep{Riess2021} used Cepheids and supernovae to find $H_0 = 73.2 \pm 1.3 \;\SI{}{\kilo\meter\per\second\per\mega\pc}$ (here, and for the rest of the review we quote 68\% confidence limits). We show the main modern results in Fig.~\ref{fig:H0_tension}.

Why should this disagreement matter? Firstly, it may be a sign that the standard cosmological model is incomplete and new physics is required. All $H_0$ results place some reliance on a background cosmology (for example to obtain peculiar velocity adjustments from a model), but the sensitivity is large when comparing results projected over large distances. Secondly, other cosmological parameters such as matter densities, curvature and large scale structure are often degenerate with $H_0$ in observational data; knowing $H_0$ more accurately helps resolve their values too. Thirdly, knowing $H_0$ accurately improves astrophysics, as distances in the universe are $\propto H_0^{-1}$. Its dimensionless cousin $h = H_0 / 100 \;\SI{}{\kilo\meter\per\second\per\mega\pc}$ is ubiquitous in formulae.

We organise our review as follows. The first section defines $H_0$ and distances; it may be skipped by a reader familiar with cosmological models. In Sect.~\ref{sec:two}, we show how $H_0$ is calculated from observational data, and the type of problems that might generally arise. We also briefly discuss the use of Bayesian methods as a tool to discriminate between competing observations and models. In Sect.~\ref{sec:three}, we discuss ways in which $H_0$ has been recently estimated. There is a rich literature on the subject, and it is difficult to cover all papers. Our approach is to cite for each topic a seminal paper, and the recent most significant developments. In Sect.~\ref{sec:four}, we discuss the possibility that measurements are correct, and it is our understanding of cosmology that is wrong. In Sect.~\ref{sec:five} we conclude, and in the spirit of our guide for consumers, we provide our buyer's advice and recommendations. We aim to be impartial, and the views expressed here are solely our own. Busy readers could review Sects.~\ref{sec:two}, \ref{sec:four} and \ref{sec:five}, and dip into Sect.~\ref{sec:three} if more detail is needed.

For the remainder of this review, we adopt $c=1$ throughout the text.

\subsection{Linking $H_0$ to $H(z)$}
Hubble's \say{constant} (Eq.~\ref{eq:hubblelaw}) is not fixed when we observe beyond our local cosmological neighbourhood; which is to say it is not fixed in time. Therefore, we write $H_0 \equiv H(t_0)$, where $t_0$ is the present day and in general $H = H(t)$. The expansion has slowed down in the past, but the universe is now accelerating (since $z \sim 0.6$) and has been dark energy dominated since $z \sim 0.3$. The change in $H(t)$ may be written using the phenomenological \textit{deceleration parameter} $q(t)$ as
\begin{equation}
    \frac{dH}{dt} = -H^2 (t) (1+q(t)) \; .
\end{equation}
In our local universe $q(t)$ is approximately constant, and hence some authors adopt $q_0 \equiv q(t_0)$ as a constant (see for example \citealt{Riess2016} or \citet{Freedman2019}). As redshift is a monotonically decreasing function of time, we can write $H = H(z)$ and approximate  
\begin{equation}
    \label{eq:nearbyH}
    H(z) \simeq  H_0 [1+ (1+q_0)z] \;.
\end{equation}
As $q_0 \simeq -0.55$, light travelling to us from more than $100 \;\SI{}{\mega\pc}$ away will have been emitted when the Hubble constant was more than 1\% different to its current value. The purpose of phrasing cosmography in this way is to avoid explicit assumptions on the matter or energy content of the universe, and it is of course possible to use more general parametrisations to expand $H(z)$. While this approximation is reasonable for $z \ll 0.1$, if we wish to go further, or link phenomenological parameters to physical quantities, we need a cosmological model. 

Cosmological models usually assume the universe is homogenous and isotropic on large scales, and has a space-time metric. Under those assumptions, the Friedmann--Robertson--Walker (FRW) metric is
\begin{equation}
    ds^2 = dt^2 - a^2(t) (\frac{dr^2}{1-k r^2} + r^2 (d\theta^2 + \sin{\theta}^{2} d\phi^2))\;,
\end{equation}
where $a(t)$ is a scale factor defining how physical distances evolve with cosmological time $t$, and $(r, \theta, \phi)$ are comoving polar coordinates centered on ourselves. $k$ is a curvature parameter, which here has units of inverse area as we wish to set $a(t_0) = 1$ (an alternative convention is to set $k = 0, \pm 1$ but it is not in general possible to do both). Results from Planck \citep{PlanckCollaboration2018} for the CMB in isolation show a preference for mildly closed universe where $k>0$, and allowing $k$ to vary from zero lowers the CMB derived $H_0$ to $63.6^{+2.1}_{-2.3} \; \;\SI{}{\kilo\meter\per\second\per\mega\pc}$. However, most other observational evidence points to a flat universe (for example galaxy survey data or gravitational lensing of the CMB -- see for example \citealt{Efstathiou2020a}). Our discussion would not be materially affected by including spatial curvature, and for this review we will assume a flat universe where $k = 0$. We return to the point in Sect.~\ref{sec:four}. 

The Hubble parameter is then defined as
\begin{equation}
    H(t) = \frac{\dot{a}}{a} \;,
\end{equation}
where $\dot{a} \equiv da/dt$. The scale factor and redshift are related by 
\begin{equation}
\label{eq:redshift}
    a(t) = \frac{1}{1+z_{\rm cos}(t)} \;.
\end{equation}
By $z_{\rm cos}$, we mean the redshift that would be seen if both the observer and emitter were stationary in comoving coordinates, which we take to be the frame in which the CMB has no dipole. Peculiar velocity is then the velocity with respect to this frame. We know the solar peculiar velocity relative to the CMB from our observed dipole, but to estimate $z_{\rm cos}$ from the observed redshift $z_{\rm obs}$ we also need the peculiar velocity of the emitter. For example, a 10\% error in a peculiar velocity of $300 \;\SI{}{\kilo\meter\per\second}$ of a galaxy lying at $50 \;\SI{}{\mega\pc}$ away would result in a 1\% error in $H_0$ were we to calculate it solely from that galaxy. For that reason, astronomers seek large numbers of objects distributed across the sky, deep into the \say{Hubble flow}, meaning their peculiar velocities are small compared to Hubble expansion and are assumed to average out. From here, we write $z = z_{\rm cos}$ unless indicated otherwise.

The $\mathrm{\Lambda CDM}$ cosmological model is defined as
\begin{equation}
    \label{eq:Friedmann2}
    \left( \frac{H}{H_0} \right)^{2} = \Omega_{k,0} (1+z)^2 + \Omega_{m,0} (1+z)^3 + \Omega_{r,0} (1+z)^4 + \Omega_{\Lambda,0} \;.
\end{equation}
This can be generalised with an equation of state parameter $p = w \rho$ for dark energy where $p$ is pressure, and the above corresponds to the cosmological constant $w = -1$. The fictitious curvature density is $\Omega_{k} = 1 - \Omega_m - \Omega_r - \Omega_{\Lambda}$, and we have assumed spatial flatness $\Omega_{k} = 0$.

The present day density fractions for matter, radiation and dark energy $\Omega_{i,0} \equiv \Omega_i (z=0)$ are defined in terms of the physical densities $\rho_i$ as
\begin{equation}
    \begin{split}
    \Omega_i(z) &= \frac{\rho_i (z)}{\rho_{\rm crit}(z)}\;, \\
    \rho_{\rm crit}(z) &= \frac{3 H^2(z)}{8 \pi G} \;.\\
    \end{split}
\end{equation}
It is straightforward to expand $H(a(z))$ as a Taylor series in $z$ and obtain Eq.~(\ref{eq:nearbyH}) to first order where
\begin{equation} 
q_0 = \left. -\frac{a \ddot{a}}{\dot{a}^2} \right|_{t=t_0} \;\;.
\end{equation}

So, in a narrow sense, a cosmological model is a function to derive $H(z)$ from $H_0$ and $\Omega_i$ or vice versa (provided nothing converts energy from one type to another). This is how $H_0$ is calculated from $H(z)$ at $z \sim 1100$ for the CMB. Hence, one way to reconcile the Hubble tension is to change the function $H(z, \Omega_i)$ by adding new energy components or novel interactions, and we return to this later.

\subsection{Distances from angles and fluxes}
Luminosity distance is defined to recover the standard inverse square law ratio between the bolometric luminosity $L$ and flux $F$ that would hold in a flat, static universe:
\begin{equation}
\label{Eq:LumDistDef}
    d_L^2 \equiv \frac{L}{4\pi F} \;.
\end{equation}
In a homogeneous and isotropic universe we find 
\begin{equation}
    \label{Eq:LumDist}
    d_L = \frac{(1+z_{\rm obs})}{H_0} \int_{0}^{z} \frac{dz^{'}}{E(z^{'})} \;,
\end{equation}
where it is conventional to make the dependence on $H_0$ explicit by setting $H(z) \equiv H_0 E(z)$. Substituting in Eq.~(\ref{eq:Friedmann2}), for flat $\mathrm{\Lambda CDM}$ we have
\begin{equation}
\label{eq:ez}
    E(z) = \sqrt{\Omega_{\Lambda,0} + \Omega_{m,0} (1+z)^3 + \Omega_{r,0} (1+z)^4} \;.
\end{equation}
Angular diameter distance is the ratio between the physical size $l$ of a distant object, and the small angle $\delta\theta$ it subtends on the sky:
\begin{equation}
    \label{Eq:angulardist}
    d_A = \frac{l}{\delta\theta} \;. 
\end{equation}
The Etherington relation \citep{Etherington1933} 
\begin{equation}
\label{eq:Etherington}
d_L = (1+z_{\rm obs})^2 d_A
\end{equation}
is a useful way to convert between the two\footnote{The relation uses $z_{\rm obs}$ rather than $z_{\rm cos}$ as it is due to the total time dilation and redshift between the source and observer. See \citet{Davis2019} for an informative review of redshifts in cosmology. We also recommend \cite{Hogg1999} for a thorough discussion of distance definitions.}. 

To link $\mathrm{\Lambda CDM}$ to our local universe, we expand Eq.~\eqref{eq:Friedmann2} to second order in $z$ :
\begin{equation}
\begin{split}
\label{Eq:LumDist2}
    H(z ) &= H_0 [1+(1+q_0) z + (j_0 - q_0^2)\frac{z^2}{2}] \;, \\
    d_L &= \frac{z}{H_0}[1+\frac{1}{2}(1-q_0)z - \frac{1}{6}(1 - q_0 - 3q_{0}^{2} + j_0)z^2] \;,
\end{split}
\end{equation}
where the \textit{jerk parameter} $j_0$  is defined as
\begin{equation} 
j_0 =  \left. -\frac{a \dddot{a}}{\dot{a}^3} \right|_{t=t_0} \;.
\end{equation}
Eqns.~\ref{Eq:LumDist2} are now a reasonable approximation to Eq.~\eqref{eq:Friedmann2} and Eq.~\eqref{Eq:LumDist} out to $z \sim 0.6$.

Setting $\Omega_{r,0} \simeq 0$, we then obtain $q_0 = \frac{1}{2}(\Omega_{m,0} - 2 \Omega_{\Lambda,0})$ and $j_0 = \Omega_{m,0} +\Omega_{\Lambda,0} \simeq 1$. 

Hubble's law $v= H\, d$ is implicit in Eq.~\eqref{Eq:LumDist}. Expanding the integral as a Taylor series in $z$, we see that $v = cz$ as used by Hubble and Lema\^{i}tre (and re-introducing $c$ here for clarity) is only valid as a low-$z$ approximation. 

\subsection{The age of the universe}

The age of the universe $t_0$  is inversely proportional to $H_0$, with dependence on other cosmological parameters. 
From the definition of $H(t) = {\dot a}/a$ we can write
\begin{equation} 
t_0 = \int_0^1 \frac{da}{a H(a) }\;.
\end{equation}
In the special case of a flat, radiation-free universe, where we set $\Omega_{m,0} + \Omega_{\Lambda,0} =1$,
it can be written analytically as
\begin{equation} 
t_0 = \frac{2}{3} \frac{1}{H_0} \frac{1}{\sqrt{1-\Omega_{m,0}}} \ln [\frac{1+\sqrt{1-\Omega_{m,0}}}{\sqrt{\Omega_{m,0}}}] \;\;.
\end{equation} 
For a flat universe with $\Omega_{m,0}= 0.3$ this gives $H_0 t_0 \approx 0.96$. Historically (in the 1980s), it was believed  that the universe was Einstein--de Sitter ($\Omega_{m,0}= 1.0$ and 
$\Omega_{\Lambda,0} = 0.0$), which yields the product $H_0 t_0 = \frac{2}{3}$. Then, a low value of 
$H_0 \approx  50 \; \;\SI{}{\kilo\meter\per\second\per\mega\pc}$
was required in order to ensure that the universe is older than the oldest globular clusters. The existence of $\Lambda> 0 $ makes the universe naturally older. The Planck estimate is $t_0 = 13.797 \pm 0.023\mathrm{\ Gyr}$  within $\mathrm{\Lambda CDM}$  (Table 2 for Planck alone, 68\% CL; \citet{PlanckCollaboration2018}). This value for the age of the universe is comfortably larger than the age of any known astronomical object. 

\section{Inference of $H_0$}
\label{sec:two}

\subsection{Standard candles and the nearby distance ladder}
A \textit{standard candle} is any population of stars or events which 
\begin{itemize}
    \item can be reliably identified
    \item have the same characteristics wherever they are seen
    \item have an established luminosity law specifying the absolute magnitude in terms of observable quantities.
\end{itemize}
Although the luminosity law is determined empirically by calibration (except in the case of gravitational waves -- see Sect. 2.3), there is an advantage if there is also a solid understanding of the underlying physics of the standard candle as in that case the calibration can be cross-checked against a theoretically-derived one.

The nearby distance ladder starts with a choice of standard candle, and a calibration of the absolute magnitude $M$, for example using parallax distances and apparent magnitudes $m$. $m$ is defined by 
\begin{equation}
    m_{X} = -2.5 \log_{10} \left( \frac{F_{X}}{F_{X, 0}} \right) \;\;,
\end{equation}
where $F_{X}$ is the energy flux per unit area per second across the wavelength range of the band $X$, and $F_{X,0}$ is the fixed reference flux for the magnitude system being used (for example the Vega system). $M$ is the apparent magnitude the object would have if it were at a distance of $10$ parsecs. Distance is conveniently quoted as the \textit{distance modulus}
\begin{equation}
    \mu = m-M \; ,
\end{equation}
and then the luminosity distance \eqref{Eq:LumDistDef} becomes 
\begin{equation}
\label{eq:dlmodulus}
d_L = 10^{0.2(\mu+5)} \; \mbox{parsecs} \;,
\end{equation}
which can then be substituted into Eq.~\eqref{Eq:LumDist}, \eqref{Eq:LumDist2} or similar relations to obtain $H_0$. 

\begin{figure}[ht!]
    \centering
    \includegraphics[width=\textwidth]{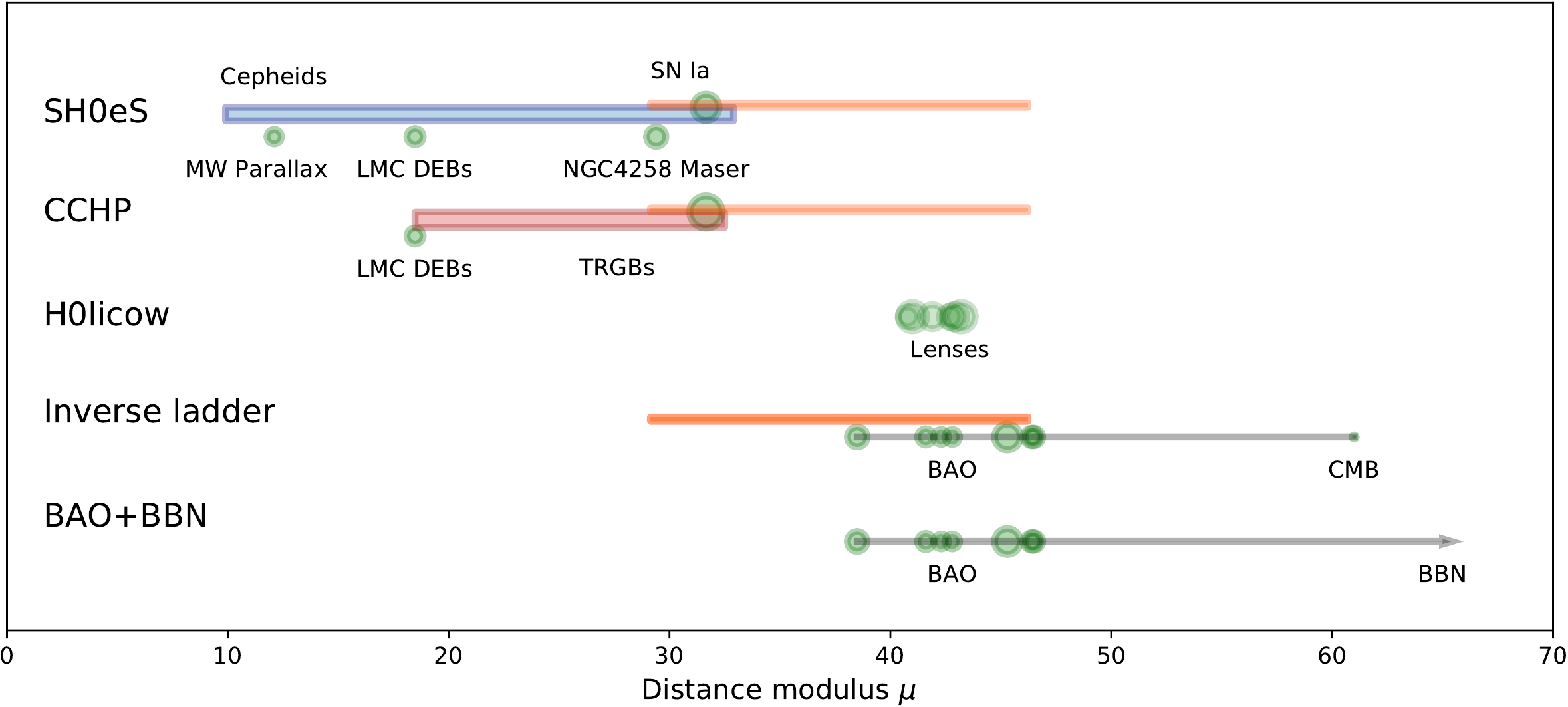}
    \caption{Our schematic illustration of the construction of distance ladders. Green circles represent the calibration of the distance ladder (either the base or overlap of each rung), and the bars are the rungs of the distance ladder. The size of dots or thickness of bars are in proportion to their contribution to the error budget of $H_0$; in the case of the CMB and BBN we have opted for a thin line to show the dependence on $\mathrm{\Lambda CDM}$. }
    \label{fig:Cepheid and Sn1a Ladder}
\end{figure}

A given standard candle seen over a range of distances is termed a \say{rung}, and in turn used as calibrators for the next rung. For example, the SH0ES team \citep{Riess2019} calibrate Classical Cepheids (CC) using parallaxes, a maser distance to NGC4258, and detached eclipsing binaries (DEBs) in the LMC as their first rung. Their next rung is Type Ia supernovae (SN Ia), calibrated using the 19 galaxies in which both Cepheids and SN Ia have been observed. We illustrate their construction in Fig.~\ref{fig:Cepheid and Sn1a Ladder}.

 The quality of the standard candle depends on a number of considerations. Are there enough with good distances to accurately calibrate the absolute magnitude? Can we clearly identify them at large distances? Can they be observed out to a sufficient distance to reach the next rung? Are the objects observed at large distances of the same type as local ones used as calibrators? How to correct for extinction, reddening, metallicity effects and crowded starfields? Which band has the most reliable magnitudes? If data has been combined from different telescopes, have the right adjustments been made to convert photometry? How is magnitude to be defined for variable stars? Each rung depends on the previous one, and errors will propagate up the ladder.

A more subtle issue is that the conversion of observational data to $H_0$ is non-linear. As the expectation $\mathbb{E}[d^n] \neq \mathbb{E}[d]^n$ in general, scatter in observational data will introduce systemic bias. Bias can also be introduced by sample selection unless care is taken as $\mathbb{E}[d_1 < d < d_2] \neq \mathbb{E}[d]$: we average our selected sample but wish to know the expectation of the unselected one. Selection may be overt (for example by cutting outliers) or due to our telescope seeing only up to $m < m_0$ (known as Malmquist bias \citep{Malmquist1922}). Both of these can be (and usually are) corrected for, but require some assumptions and a careful analysis of the data and reduction pipeline. 

\subsection{Standard rulers and the inverse distance ladder}
A \textit{standard ruler} is a feature on the sky of a known physical size $l$, which enables us to calculate the angular diameter distance $d_A$ defined in Eq.~(\ref{Eq:angulardist}) from their angular size. Parallax is an obvious example, and also the size of orbits of masers and detached eclipsing binaries can be determined from their positions, light curves and spectroscopy. In the early universe, acoustic pressure waves in the primordial charged particle and radiation plasma set a physical size called the \textit{sound horizon} $r_s$, which is how far they travel from the initial seeds\footnote{These are thought to be density perturbations originating from quantum fluctuations in an inflationary era.} that generate them. The universe was not that dense at that time, so Thomson scattering by charged particles was necessary to propagate the waves, and hence they are frozen-in by recombination. The sound horizon is then imprinted on the CMB as peaks in the power spectrum of temperature fluctuations, and in the later spatial distribution of galaxies (known as \textit{baryon acoustic oscillations}, or BAO for short)\footnote{The physical size of the sound horizons defined by the CMB and BAO are different by $\sim 4\%$. Loosely speaking, the CMB sound horizon is defined by the redshift $z_{\star}$ where the \say{photons stop caring about the baryons} (otherwise known as the surface of last scattering), and for BAO by the redshift $z_{\rm{drag}}$ where \say{the baryons stop caring about the photons}. The difference between the two is due to their number densities.}. The sound horizon may be calculated in a cosmological model, and depends both on the expansion rate $H(z)$ (the waves are carried along by expanding spacetime) and the matter-energy content of the early universe (determining the sound speed).

The inverse distance ladder, as its name suggests, works in the opposite direction to the nearby distance ladder but on the same principle. It can use the sound horizon for a starting $d_A$, a background cosmology, and the Etherington relation (\ref{eq:Etherington}) to calibrate the luminosity distances $d_L$ of standard candles. For example, \citet{Lemos2019} calibrate BAO at $z \simeq 1$ and SN Ia at $z<1$ using the CMB sound horizon. They replace the standard $\mathrm{\Lambda CDM}$ formula for $H(z)$ with a parametric form. $H(z)$ is extrapolated to today to find $H_0 = 68.42 \pm 0.88 \;\SI{}{\kilo\meter\per\second\per\mega\pc}$. Thus, it is shown that discrepancy between the CMB value of \citet{PlanckCollaboration2018} and \citet{Riess2019} is not caused by assuming the late-universe is $\mathrm{\Lambda CDM}$.

We can express the difference between early and late universe $H_0$ in terms of ruler size or luminosity differences. Planck implies $r_s = 147.27 \pm 0.31$ Mpc in $\mathrm{\Lambda CDM}$; it would need to be $\simeq 10$ Mpc lower \citep{Knox2019} to bring consistency with \citet{Riess2019}. Alternatively, Eqs.~(\ref{eq:dlmodulus}) and (\ref{Eq:LumDist}) show that Cepheids or SN Ia would need to be $\simeq 0.2$ mag brighter than thought to bring consistency with Planck.
 
In summary, we see the two ways of constructing ladders, propagated toward each other by the \say{guard rails} of SN Ia, do not meet! Hence, the $H_0$ tension is 
sometimes characterised as \say{early} versus \say{late}, from which follows the question \say{Is $\mathrm{\Lambda CDM}$ right?}. This may be premature: in fact, few late-universe results in isolation are fully inconsistent with early universe ones, as we discuss later. 

\subsection{No-ladder $H_0$}
We have seen distance ladders require calibration, whether they are nearby or inverse. However, there are some self-calibrating observations from which $H_0$ may be calculated directly.

One example is the CMB. The detailed shape of the power spectrum determines other parameters in $\mathrm{\Lambda CDM}$, and the value of $H_0$ derived from it is best understood as just one part of the simultaneous inference of \textit{all} cosmological parameters. A further example is maser emission systems, which occur in the nucleus of certain galaxies and are bright enough to be seen at cosmological distances. The emission spots appear to follow Keplerian orbits, and so with some disk modelling, the size of the orbit and hence the angular diameter distance can be deduced directly.

Gravitational waves assume general relativity. The masses of merging compact objects and luminosity can be obtained from the shape of the waveform. That is to say, there is no need for an empirical calibration of their intrinsic luminosities, and instead the observational challenge is to determine the redshift of the source. Often referred to as \textit{standard sirens} rather than standard candles, a gravitational wave event whose source galaxy has been identified (by locating the optical counterpart) is referred to as a \say{bright siren}, otherwise it is called a \say{dark siren}. Most gravitational wave events are dark, but progress can be made statistically with them given sufficient numbers. 

For gravitational lenses, general relativity links the time delay caused by the lensing of the background source (a combination of both the longer path and time dilation) and the mass distribution of the lensing galaxies. The absolute time delay is not known, but if a rapidly varying source like a quasar can be seen in multiple images, the relative time delay between images allows the angular diameter distance of the lens to be calculated. In this case, the challenge is to obtain enough constraints on the mass distribution of both the lensing galaxies and the general concentration of matter along the line of sight, using for example the velocity dispersions and surface brightness of the lensing galaxies, and imaging data. 

\subsection{What could cause the tension?}
By using the word \say{tension}, cosmologists mean the discrepancy in measurements of $H_0$ is at a level which is large compared to the reported errors. This means that if the values and errors are correct, this is very unlikely to be the result of chance.
 
Measuring the Hubble constant has always pushed the limits of the available telescopes of the day, and as a consequence observers have to be very careful to avoid bias in their derivations of $H_0$, and have accounted for all errors. Hubble believed he had one population of Cepheids, whereas we know today he had two, and had also confused nebulae with bright stars. 
Alternatively, many researchers interpret the tension in the spirit of the precession of the perihelion of Mercury: something is wrong with the (Newtonian) model and a new one is needed (general relativity). We can categorise explanations as follows: 
\begin{itemize}
\item 
\textbf{Observational bias}. An observational bias is an error in mean photometry that would be expected to increase with magnitude or distance. To give some examples, consider first crowding. For distant stars, resolving them from their neighbours becomes harder, therefore their photometry will be progressively more blended with other stars the more distant they are. Blending increases the apparent magnitude, and changes the colour (see for example the discussion in Sect.~4.2 in \citealt{Javanmardi2021}). For very faint stars, the response of the detector may be non-linear \citep{Riess2019}, and needs to be corrected. Another issue is combining observations between ground and space telescopes, as in general fainter stars will be observed from space, but nearby ones more cheaply from the ground. Aside from atmospheric extinction, each instrument has different passbands, detector response and resolution, meaning the magnitudes of the same star observed in each telescope will be different. Photometry must be transformed to a common system (see for example Eqs. 10--12 in \citealt{Riess2016}), and if not done (or done incorrectly) some bias will likely have been introduced. Any parameters derived from photometry - such as photometric redshifts - would inherit the same propensity to bias. 
\item
\textbf{Astrophysical bias.} An astrophysical bias occurs when the properties of the object being studied are not fully resolved, and those properties differ with distance. For example, consider Cepheids in the LMC and SN Ia hosts. The LMC is close so Cepheids with a full range of periods can be seen, whereas for distant galaxies only the brighter Cepheids with longer periods are seen. Additionally, the LMC is metal-poor compared to a typical spiral galaxy, the Cepheids there may be expected to be relatively metal-poor compared to those in SN Ia hosts. Hence, any curvature in the Leavitt law, or mis-calibrated metallicity dependency could bias distances (see for example \citealt{Freedman2010}). A second example is the step-like link between SN Ia magnitudes and properties of the host galaxy (\citet{Smith2020} and references therein). This could indicate there are two distinct populations of SN Ia. If then the SN Ia in the 19 galaxies where both Cepheids and SN Ia were of mostly one type, whereas the rest a blend of both types, $H_0$ would be biased by the calibration of SN Ia. 
\item
\textbf{Statistical bias.} The main causes of statistical bias will be selection effects and scatter as we discussed in the introduction. Statistical biases can be corrected by running random simulated observations through the same selection and analysis pipeline, but the simulations will themselves need some physical parameter choices, perhaps determined from previous surveys, or fixed in advance to \say{reasonable} levels. In Bayesian analysis, residual dependence on the choice of prior is a feature of sparse observations. Statistical bias correction is subtle and difficult, as we see later in the sections on parallax and SN Ia.
\item
\textbf{Physics of $\mathrm{\Lambda CDM}$.} Before invoking new physics, could the explanation be found within $\mathrm{\Lambda CDM}$? Cosmological formulae such as Eq.~(\ref{eq:Friedmann2}) are derived from a homogeneous, isotropic universe. Could corrections allowing for inhomogeneities  be large enough to explain the tension? A specific example is the \say{Hubble Bubble} or \textit{local void} proposal (for a recent example, see \citealt{Shanks2019}), in which we are by chance located in an under-dense region, and our local $H_0$ is different from the \say{universal} one. Additionally, inhomogeneities mean we must correct redshifts for peculiar velocities, and further the propagation of light through overdense or underdense regions might bias our analysis \citep{Kaiser2016}.
\item
\textbf{New Physics.} If the expansion history of the universe were different to $\mathrm{\Lambda CDM}$, $H_0$ inferred from the CMB or BAO might be brought into alignment with the local value. Alternatively, performing our analysis of a non-$\mathrm{\Lambda CDM}$ universe using $\mathrm{\Lambda CDM}$ formulae may have confused us. For example, an extra particle species would increase the pre-recombination expansion speed, so reduce the size of the sound horizon: sound waves have less time to propagate before they are frozen in. To keep the same observed angular size of the CMB temperature fluctuations, the value of $H_0$ calculated from the CMB will increase (see Eqs.~\ref{Eq:LumDist}, \ref{Eq:angulardist}, \ref{eq:Etherington}). However, as we see later $\mathrm{\Lambda CDM}$ makes many other successful predictions, such as the CMB spectrum itself or primordial element abundances, and is not lightly tampered with. 
\end{itemize}

\subsection{Is the tension significant?}
The tension is often quoted as the number of standard deviations \say{$m\sigma$}. In particle physics, the meaning is clear: there are millions of collisions and probability is frequentist. There is no need to work in a Bayesian framework, with some prior assumption of a parameter to update with new data. The law of large numbers drives distributions to Gaussian normal shape, and we can translate $m\sigma$ to a probability of occurrence by chance. At the $5\sigma$ level, it is unanimously agreed new physics has been detected. None of the above applies in cosmology! 

Nevertheless, that the tension is significant should not be in doubt: see Fig.~\ref{fig:H0_tension}. But we are looking for more from our data analysis, and there are three ways in which Bayesian statistics are helpful, which we now briefly survey.

Bayes' theorem states that the posterior probability distribution is 
\begin{equation}
    P (\theta \mid D, \mathcal{M}) = \frac{P (D \mid \theta , \mathcal{M}) P(\theta \mid \mathcal{M})}{P(D \mid \mathcal{M})} \;,
\end{equation}
where $D$ is the data and $\theta$ are the parameters of the model $\mathcal{M}$ of interest. For example $\mathcal{M}$ might represent $\mathrm{\Lambda CDM}$ with its associated parameters including $H_0$ and $\mathcal{M}_{1}$ some extension of it with additional parameters. $P(\theta \mid \mathcal{M})$ is any prior belief in the parameters of the model and $P (D \mid \theta , \mathcal{M})$ is called the \textit{likelihood}, typically specified by the team analysing the data. The denominator
\begin{equation}
    P(D \mid \mathcal{M}) = \int_{\Theta} P (D \mid \theta, \mathcal{M}) P(\theta \mid \mathcal{M}) d\theta \;,
\end{equation}
is called the \textit{evidence}. An extended cosmological model will have a smaller evidence if there exist large values of the parameter space with low likelihood, even if it agrees better with the data. Bayesian evidence then naturally embodies Ockham's razor: a simpler model will have a larger evidence, unless the extended model has a significantly better fit to the data. 

\begin{figure}[ht!]
    \centering
    \includegraphics[width=\textwidth]{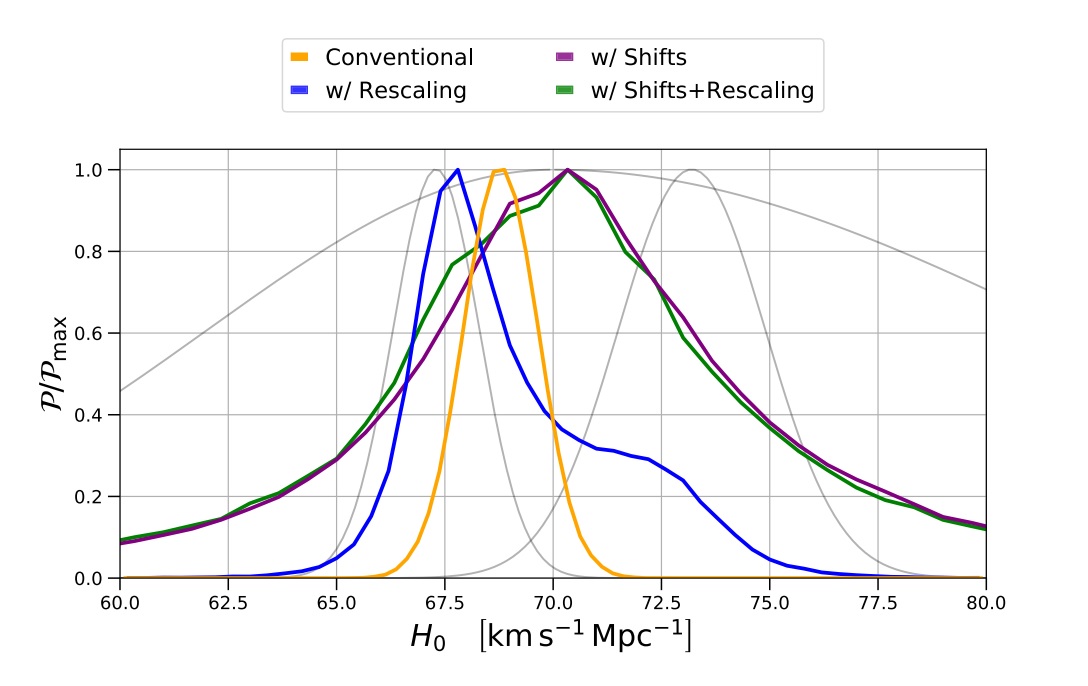}
    \caption{An illustration of BACCUS applied to Planck, H0LiCOW and SH0ES data, whose posteriors are shown as thin grey lines. Conventional (orange) is the standard Bayesian combination assuming equal weighting, w/Rescaling (blue) is equivalent to hyper-parameters, w/Shifts (brown) adds an unknown systemic error offset to each dataset, and w/Shifts+Rescaling (green) combines hyper-parameters and systemic offsets. Figure from \citet{Bernal2018}.}
    \label{fig:H0Baccus}
\end{figure}

Secondly, Bayesian statistics can also help in re-analysis, in the hope that the data itself may reveal issues. A Bayesian hierarchical analysis was used by \citet{Feeney2018} to test relaxing distributional and outlier assumptions embedded in the $\chi^2$ fit used in SH0ES analyses. \citet{Cardona2017} have re-analysed SH0ES data using \say{hyperparameters}, which are weightings of datasets proposed as a measure of credibility by \citet{Lahav2000}. An agnostic prior for the weights is set, and the hyperparameters are marginalised over. Both results are consistent with SH0ES. \citet{Bernal2018} extend the hyperparameter method by adding a free parameter shift in the mean of each dataset to account for unknown systematics, which they dub \say{BACCUS}. Using this to combine Planck, SH0ES and other datasets, produces a compromise. As shown in Fig.~\ref{fig:H0Baccus}, the posterior middles the two with much larger error bars, which perhaps is unsurprising given the agnosticism of the method. Such types of analysis need to be taken with a grain of salt : data that is in tension should not be combined, and BACCUS is not a substitute for a critical analysis of why the tension has happened. However, if one \textit{demands} a method to merge disparate results in a Bayesian framework, BACCUS is a way to achieve that. 

Thirdly, we may want to know how valid a combination of data sets is. If two posteriors for the same parameter barely overlap, a Bayesian analysis will seriously mislead with error bars that are too small, as shown by the yellow line in Fig.~\ref{fig:H0Baccus}. We seek a statistic that is symmetric, (reasonably) independent of prior assumptions and models, and straightforward to calculate and interpret. The R statistic
\begin{equation}
R = \frac{P(D_A, D_B)}{P(D_A) P(D_B)} = \frac{P(D_A \mid D_B)}{P(D_A)} = \frac{P(D_B \mid D_A)}{P(D_B)} \;
\end{equation}
compares the evidence of dataset $D_A$ in light of knowing $D_B$ to that of $D_A$ alone, but is dependent on the prior and so isn't usually comparable between different papers. \citet{Handley2019} define a new statistic called \say{suspiciousness} as
\begin{equation}
    \log S = \log R - \log I \;,
\end{equation}
where $I$ is the information ratio $\log I = \mathcal{D}_{A} +\mathcal{D}_{B} - \mathcal{D}_{AB}$ which quantifies the information gain between prior and posterior. $\mathcal{D_A}$ is defined as
\begin{equation}
    \mathcal{D}_A = \int_{\Theta} P(\theta \mid D_A) \log \frac{P(\theta \mid D_A)}{P(\theta)} d\theta\;,
\end{equation}
and $\mathcal{D}_B$ and $\mathcal{D}_{AB}$ are defined similarly by replacing $A \rightarrow B, AB$ respectively. This is independent of the prior and (being an integral) the choice of parameters. We can interpret $\log S \ll 0$ as the two data sets being in tension: loosely speaking, the evidence of combining them does not exceed the information of considering them separately. Therefore, suspiciousness fits the criteria of simplicity and interpretation we outlined above. 

Finally, even in light of the tension, we cannot dodge the question posed in our introduction: \say{All this debate is interesting, but which value for $H_0$ should I use, and is it valid to use that in $\mathrm{\Lambda CDM}$?}. We suspend judgement until after we have surveyed the data and potential new models. 

\section{Measuring $H_0$}
\label{sec:three}

\subsection{Parallax}

Parallax is both the oldest astrometric technique, and the easiest to understand. Hold out your thumb at arm's length, relative to some fixed point on the wall, and alternately close one eye and then the other. Relative to the fixed wall, the apparent position of your thumb will change, and this is how our depth perception works: the smaller the change in position, the longer your arm must be.  The same principle works with stellar distance, where now our \say{binoculars} correspond to the Earth's position on opposite sides of its orbit. The change in a fixed star's position $2\varpi = \theta_1 - \theta_2$ arcseconds, due to the change in the Earth's position by 2 a.u. over 6 months leads to the distance $d = 1/\varpi$ parsecs\footnote{Many authors use $\pi$ to denote parallax. We aim to avoid any potential confusion by using $\varpi$.}. Our measurement of position must be very precise. Although the nearest star, Proxima Centauri, has a parallex of approximately 0.77 arcseconds, modern measurements target a remarkable 10 $\mu$as, which is the size of a thumbnail on the Moon as seen from Earth. 
 
Modern parallax measurements began with the satellite Hipparcos (the name alludes to the ancient Greek astronomer Hipparchus, who measured the distance to the Moon). Launched by the European Space Agency in 1989, it measured the parallaxes of 100,000 stars at an accuracy of up 0.5 milliarcsecond (mas), the fixed background frame now being extragalactic sources such as quasars. Although undoubtedly impressive, at 1,000 light years an error of 0.5 mas would still be a distance error of 15\%. A further drawback is Cepheids, an important part of the distance ladder we discuss shortly, are relatively rare stars and only a handful are located in our neighbourhood of the Milky Way. 
 
Moving on to the present day, our two best current sources of parallax are Gaia and the Fine Guidance Sensor/Wide Field Cameras (FGS/WFC3) aboard the Hubble Space Telescope (HST). Gaia was launched in 2013 and the mission goal is to measure over a billion stars (including 9,000 Cepheids and half a million reference quasars), both in our galaxy and satellites like the LMC. The mission-expectation precision is 7$\mu$as at $m=12$, rising to 26$\mu$as at $m=20$ \citep{Gilmore2018}. Gaia does this by slowly scanning the sky with two telescopes set at relative angles of $\ang{106.5}$, to make a one-dimensional measurement of the time and position of each star that slowly drifts across the CCD. Up to 70 measurements will be made for each star, which allows the additional calculation of proper motions, and even small changes in position caused by the gravitational tug of planets orbiting the star. The HST operates on similar principles in \say{spatial scanning} mode; although it cannot survey like Gaia, when focused on nearby Cepheids its errors appear competitive \citep{Riess2018a}.

Gaia's high precision is dependent on a very stable mechanical structure of the spacecraft. A variation in the angle between the two fields (which might be caused by thermal expansion) could cause spatial variations in apparent parallax, or if synchronous with the scan period even a fixed systematic offset. Indeed, such a variation has been inferred from the interim Data Release 2 (DR2) \citep{GaiaCollaboration2018}. The average of the quasar parallaxes in it is -29$\mu$as (negative parallaxes can happen when position measurement error is larger than the parallax), and there were indications this \say{zero point} may vary with stellar colour, luminosity and position on the sky \citep{Arenou2018}. \citet{Riess2018b} compared HST Cepheid parallaxes to Gaia, simultaneously solving for the Gaia zero point and Cepheid calibration. They found a difference between them $46\pm15\mu$as, with Gaia parallaxes again appearing too low. As the typical parallax of a Milky Way Cepheid is $400\mu$as, this is very material to distance estimates. \citet{Breuval2020} creatively replaced DR2 Cepheid parallaxes with those of resolved bound binary companions (where available), which being dimmer are closer to the ideal magnitude range for Gaia. 

During the preparation of this review, Gaia Early Data Release 3 was made available, and has already by used by \citet{Riess2021} to revisit Gaia parallaxes. Gaia EDR3 is not intended to be the final word, but indeed Cepheid zero points seem now to be reduced below $10\mu$as. A calibration of Cepheids using 75 Gaia parallaxes only gives $H_0 = 73.0 \pm 1.4 \;\SI{}{\kilo\meter\per\second\per\mega\pc}$, which is slightly lower, but consistent with their previous result based on HST parallaxes. 

Before moving on to discuss alternative calibrations, we will first digress into a discussion of parallax bias. The potential for bias occurs in any astrophysical observation, and is often the subject of lengthy analysis in $H_0$ papers. As it is most easily understood in the context of parallax, it is helpful to discuss it here.

Parallax bias is usually referred to as Lutz--Kelker--Hanson (LKH) bias \citep{Lutz1973, Hanson1978}. This is a summation of three quite different effects: non-linearity of the desired variable (distance) with respect to the observed variable (parallax), population bias (have we observed the object we intended to, or did we confuse it with something else?), and selection bias (our surveys are normally magnitude limited, so we will only \say{select} objects for which $m<m_0$).

To explain non-linear bias, imagine we have a \textit{symmetric} error in our parallax measurement, such as might be caused by an instrumental point spread function. To be concrete, suppose the likelihood of measuring $150\mu$as is the same as measuring $50\mu$as when $100\mu$as is the true value. If we average the distance, we will obtain $\bar{d} = 6666.67 + 20000 = 13,333$ which is biased with respect to the true distance of $10,000$ parsecs. In mathematical terms, because $d = 1/\varpi$, then $\mathbb{E}[d] \neq 1/\mathbb{E}[\varpi]$.

Population bias arises in parallax when we consider a broad survey of stars at a given distance $r$ from our position. Assuming a roughly constant spatial density of similar stars, there are more stars in the shell $(r, r+\Delta r)$ than there are in the shell $(r - \Delta r, r)$ for some finite $\Delta r$. Hence, there are more (further) stars whose parallaxes may be over-estimated to place them at $r$ than (closer) stars whose parallaxes may be over-estimated. Taken to extremes, there are huge numbers of stars with effective parallaxes of zero, waiting for their small but finite chance to \say{crowd in} to a given measured parallax. This will bias observed parallaxes too low. Note that if we were certain of our identification of the star (as we would be for a Milky Way Cepheid close to us), we need not consider population bias: it would stand out from the crowd.
\begin{figure}[h!]
    \centering
    \includegraphics[width=\textwidth]{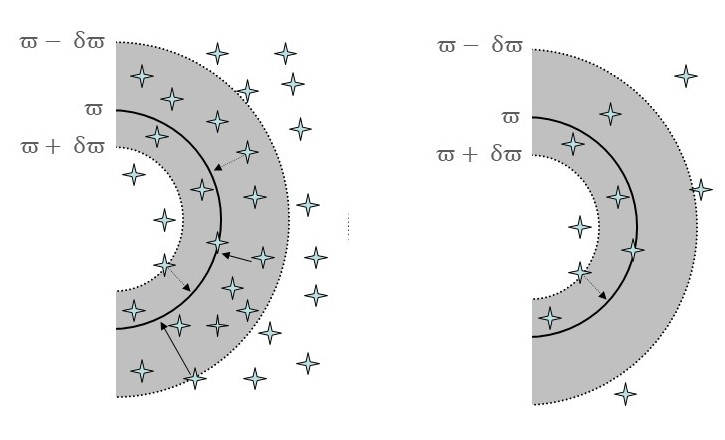}
    \caption{An illustration of Lutz--Kelker--Hanson population bias. Assume each parallax measurement can be in error by up to $\pm \delta\varpi$. In the figure on the left, there is a constant spatial density of stars. Then, the region $(\varpi, \varpi-\delta\varpi)$ has a greater number of stars that can scatter to the observed parallax $\varpi$ than the region $(\varpi, \varpi+\delta\varpi)$ and parallaxes are biased too low. Conversely in the figure on the right, the stellar density drops sharply beyond $\varpi$ due to either the edge of the population or magnitude limitations. More stars are available to scatter out than in, and parallaxes are biased too high.}
    \label{fig:LKbias}
\end{figure}
 
Conversely, suppose we were observing close to our magnitude limitations. Now the opposite bias would occur: we cannot see the further stars, so they can't crowd in. But the same number of closer stars are available to crowd out, so our observed parallaxes will now be biased too high. This is the well-known Malmquist bias \citep{Malmquist1922}, and is a major issue for surveys as naturally we will try to see as far as we can!
 
To deal with bias then involves \textit{modelling} of the instrumental error, the selection function, the population scatter and so forth. In modern surveys, this is normally done by constructing simulated catalogs with known physical parameters and some assumptions, and putting those catalogs through the same analysis pipeline as the real data to see what biases emerge. For example, \citet{Riess2018a} compute distance modulus biases of between 0.03 and 0.12 mag for MW Cepheids using a model for galactic stellar distributions. If working in a Bayesian framework, a posterior distribution for the distance may be derived using the method of \citet{Schonrich2019}. An alternative is to work directly with the parallaxes, instead converting Cepheid magnitudes to predicted parallaxes as done by \citet{Riess2018b}. As the magnitudes are measured considerably more accurately than the parallaxes, bias corrections to the magnitude to parallax conversion are not necessary. To check the predictions of LKH bias, \citet{Oudmaijer1998} compared ground-based to Hipparcos parallaxes, finding a bias towards brighter magnitudes up to relative error $\approx 30\%$, and dimmer magnitudes for larger error when Malmquist bias is dominant, as one would expect from the discussion above. 
 
We end our digression on biases here and move on to discuss other geometrical distances.

\subsection{Detached eclipsing binaries}
Imagine we had a star for which we knew the surface radiant flux density $J$, and the physical size $R$. The stellar luminosity would straightforwardly follow as $L = 4\pi R^2 J$, and we would have a standard candle. 

Cool, stable, helium-burning giants (that is, red clump) are sufficiently bright to be seen outside the Milky Way. For these \say{late-type} stars, just such an empirical relationship can be established for the surface brightness $S_V$ defined as 
\begin{equation}
    S_V \equiv V + 5 \log (\phi) \;\;\mbox{mag} \;, 
\end{equation}
where $\phi$ is the stellar angular diameter, and $V$ the visual band magnitude. This relationship has been calibrated by angular diameters obtained from optical interferometry of nearby stars, and is given by 
\begin{equation}
    \label{Eq:DEBdistance}
    S_V = (1.330 \pm 0.017) \times ((V-K) - 2.405) +(5.869 \pm 0.003) \;\;\mbox{mag} \;,
\end{equation}  
where $V-K$ is the colour difference between magnitudes in the $V$ and $2.2 \;\mu$m near-infrared $K$ band \citep{Pietrzynski2019}. The scatter is just 0.018 mag. Rearranging the definition of surface brightness and with $\phi = 2R / d$ it then follows that 
\begin{equation}
    d = 9.2984 \times \left(\frac{R}{R_{\odot}}\right) (10^{0.2(V - S_V)} \;\mbox{mas}) \;\; \mbox{parsecs}, 
\end{equation}
where $R$ and $\phi$ have been converted to solar radii and milli--arcseconds respectively. The pre-factor is purely geometric.

But how can we know the radius of distant stars? Eclipsing binaries allow just that. If the stars are well separated enough to spectroscopically resolve each one\footnote{We should correctly refer to these systems as \say{double-lined}, as \say{detached} means physically separate rather than resolvable. However the literature on $H_0$ almost always uses \say{detached} so we stick with it here.}, but close enough to eclipse each other, we can obtain their individual surface brightnesses, colours, radial velocities, eclipse depths and shapes, and the orbital period. With this data, the radius (and other physical parameters such as mass, eccentricity and inclination of the orbital plane) of each star can be solved for. Such an alignment of the stars is of course rare, but sufficient numbers do exist! By painstakingly observing 20 systems in the LMC over more than 20 years (covering many eclipses) \citet{Pietrzynski2019} determine the stellar radii to 0.8\% accuracy. Crowding can be easily spotted in the light curve and removed. They derive $\mu_{\rm{LMC Centre}} = 18.477\pm 0.004\; \mbox{(stat)} \pm 0.026\; \mbox{(sys)}$, where the main contribution to the systemic error budget is the $S_V$ relation above. 

This is the most accurate measurement of the distance to the LMC to date. As we shall see shortly when we discuss standard candles, this result has become key to many recent $H_0$ results: every standard candle can be found in the LMC, and the low error budget allows for a very accurate calibration.

Looking forwards, although DEBs have been found in M31 they are \say{early-type} stars with hot atmospheres, for which a reliable surface brightness to colour relation has not been established. The hope is future 30m-class telescopes will have sufficient spectroscopic resolution to extend this to late-types in M31 and other local group galaxies \citep{Beaton2019}.

\subsection{Masers}
Maser emission occurs when thermal collisions in warm gas in an accretion disk around the central black hole of a galaxy drive a population inversion of molecular energy levels.  Such systems are rare: the disk must be \say{just right}, not too hot, and not too cold, and have suitable local molecular abundance. The Type 2 Seyfert galaxy NGC 4258 at a distance of 7.5 Mpc is just such a system. Isolated bright spots of 22.235 GHz maser emission (from a  hyperfine transition of $\mathrm{H_{2}O}$) can be seen in three regions on the sky stretching 20 mas ($\sim 0.1$ ly) long, with the overall shape of a warped line. After subtracting the overall system redshift, the spots on one side are blue-shifted by $\sim\SI{1000}{\km\per\second}$, the other side is red-shifted by the same amount, and the spots in the middle are low velocity \citep{Argon2008}. If the spots are observed for long enough, their accelerations and proper motion can be obtained from the steady drift of their Doppler velocities and positions. The central spots show the lowest l.o.s.\ velocity and highest acceleration, and the outer spots having the highest velocity and lowest l.o.s.\ acceleration. 

The key assumption these observations support is that all the spots form an orbital system with shared parameters, as it is anticipated that disk viscosity will have reduced the orbits to close to circular. The disk shape is then modelled (including such parameters as inclination, warp, residual eccentricity and so forth), fitted and the orbital parameters of the spots derived. The outer dot velocities $v$ show a Keplerian behaviour with radius: $v^2 \propto GM/r$, and as the acceleration is then $\dot{v} \propto GM/r^2$, the physical radius of the system can be determined. The angular diameter distance is then
\begin{equation}
d_A = \frac{v^2}{\dot{v} \times \theta} \;,
\end{equation}
where $\theta$ is the angular impact parameter. \citet{Humphreys2013} have observed NGC 4258 for 10 years at 6 monthly intervals using VLBI interferometry, which resolves the relative position of the maser spots to $<3\;\mu$as accuracy (even the tectonic drift of the ground radio telescopes must be corrected for). The Doppler shifts are measured to within an accuracy of $1 \;\SI{}{\kilo\meter\per\second}$, and the central spots accelerate by $\sim 10\;\SI{}{\kilo\meter\per\second} \rm{yr}^{-1}$. The proper motions $\dot{\theta} \;\mu\mbox{as yr}^{-1}$ provide additional information as $d_A = v / \dot{\theta} \; \mbox{Mpc}$. This pattern fits an accretion disk orbiting a $10^7 \,M_{\odot}$ black hole.
\begin{figure}[h!]
\begin{subfigure}{1.0\textwidth}
\includegraphics[width=1.0\linewidth, height=3.5cm]{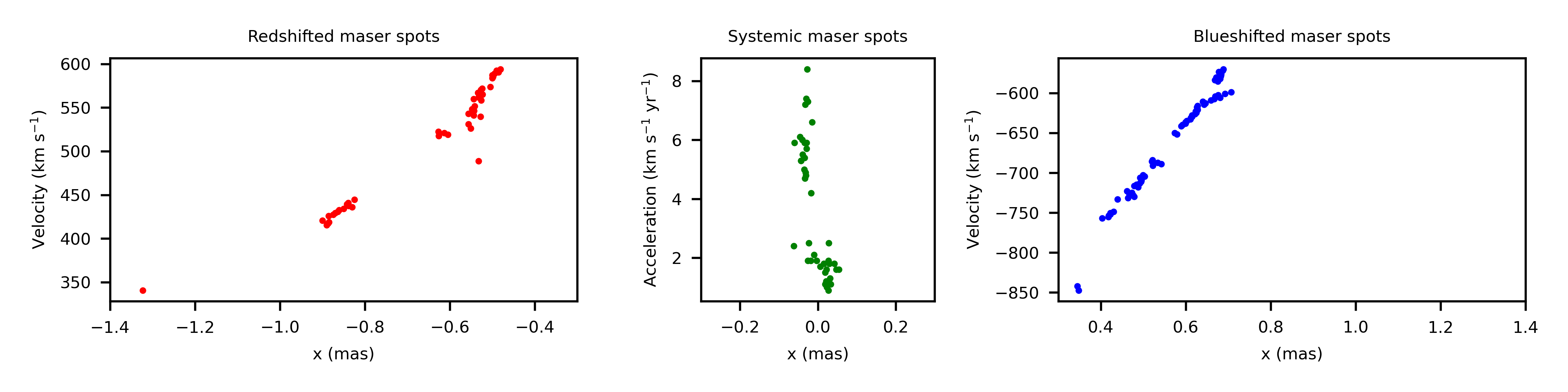} 
\caption{Maser positions, velocities and acceleration}
\label{fig:NGC3789spots}
\end{subfigure}
\begin{subfigure}{1.0\textwidth}
\includegraphics[width=1.1\linewidth, height=5cm]{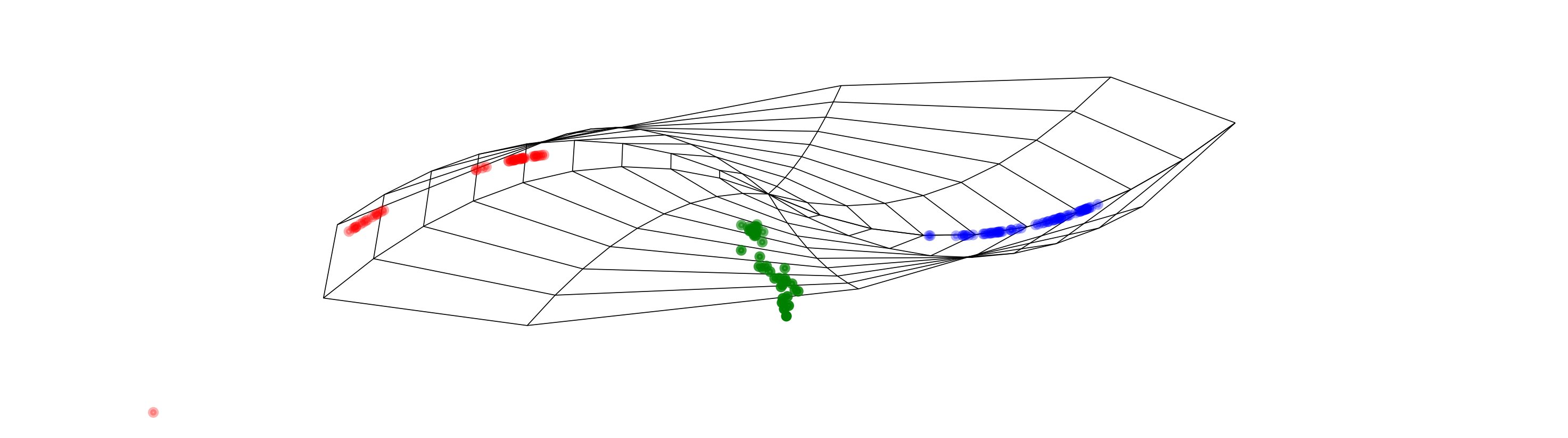}
\caption{The accretion disk of UGC 3789 (maser spot positions are schematic, not actual).}
\label{fig:NGC3789disk}
\end{subfigure}
\caption{The UGC 3789 maser system, showing the warped disk, velocities and acceleration of maser spots. The top left and top right figures show the LOS velocities in $\SI{}{\kilo\meter\per\second}$ relative to the central blackhole, offset by position in milliarcseconds; the $GM/r$ curve shape is apparent. The top centre panel shows the LOS acceleration in $\SI{}{\kilo\meter\per\second}$ per year, calculated from the velocity drifts over 3 years of observations. In the bottom panel, we schematically show the warped disk and projected spots. UGC 3789 is at a distance of $46$ Mpc so 1 mas $\simeq 0.24$ pc. The disk and maser spot data is available in \citet{Reid2013}. }
\label{fig:NGC4258}
\end{figure}

\citet{Reid2019} derive the distance as $d=7.576 \pm 0.082 \mbox{(stat)} \pm 0.076 \mbox{(sys)}$ Mpc, an accuracy of 1.4\%, competitive with the LMC distance error. The main statistical contribution to the error budget is the positional error of the $\sim 300$ spots.
 
Only 8 megamaser galaxies have been detected, with the furthest being NGC 6264 at 141 Mpc, and it is unlikely more will now be found at usable distances. We illustrate the data for UGC 3789 from \citet{Reid2013} in Figs.~\ref{fig:NGC3789spots} and \ref{fig:NGC3789disk}. The Megamaser Cosmology project \citep{Pesce2020} have used six of these galaxies to find $H_0 = 73.9 \pm 3.0 \;\SI{}{\kilo\meter\per\second\per\mega\pc}$, independently of distance ladders.

Although it is too close to determine $H_0$ directly (as its peculiar motion could be a large fraction of its redshift), NGC 4258 has become a key calibrator of the distance ladder, owing to its low error budget. Its particular usefulness is that, unlike the LMC or SMC, it is a fairly typical barred spiral galaxy, similar in morphology and environmental conditions (metallicity, star-formation rate and so forth) to the ones in which Cepheids and Type Ia supernovae are seen at greater distances. Using Cepheid and SN Ia data from \citet{Riess2016}, \citet{Reid2019} find $H_0 = 72.0 \pm 1.9 \;\SI{}{\kilo\meter\per\second\per\mega\pc}$ using solely NGC 4258 as the geometric calibrator.

\subsection{Cepheids}

Having discussed calibrators, we can now talk about the \say{engine room} of distance ladders. Cepheids form two classes, but it is the younger, population I, classical Cepheids which are of interest. These are yellow bright giants and supergiants with masses 4-20 $M_{\odot}$ and their brightness cycles over a regular period, between days and months, by around 1 magnitude. They are bright, up to 100,000 $L_{\odot}$, and they can be seen out to 30 Mpc with the HST. Although Milky Way Cepheids had been observed and catalogued since the 18th century, it was first discovered by Henrietta Swan Levitt in 1908 that there was linear relation between the logarithm of 
their oscillation period and absolute magnitudes, the \textit{Leavitt period-luminosity law} (we use the term $P$-$L$ \textit{law} for Miras). She had been observing Cepheids in the SMC and LMC, using the Harvard College Observatory telescope, and decided to rank them in order of magnitude. As stars in the LMC will have roughly the same distance from the Earth, the Leavitt law was immediately clear from their apparent magnitudes. In fact, one could say modern extragalactic astrometry began with her groundbreaking discovery. 
 
That stars can pulsate is not so surprising; after all, a star is in local equilibrium, and would be expected to oscillate about the equilbrium given any perturbation. However, something must \textit{drive} the oscillation otherwise it would dissipate. For Cepheids, the driver is heat-trapping by an opaque layer of doubly-ionised He surrounding a He-burning core. The trapped heat increases pressure, which expands the star, cooling the ionised He to the point where it can recombine and therefore becomes transparent. As the radiation escapes, the core cools and re-contracts, which in turn releases gravitational energy into the He layer. The He heats, re-ionises and the cycle repeats, with period proportional to the energy released. The Cepheid population lies in an instability strip  in the horizontal branch of the Hertzsprung--Russell diagram; the cool (red) edge of the population is thought to be due to the onset of convection in the He layer, and the hot (blue) edge by the He ionisation layer being too far into the atmosphere for pulsations to occur. They therefore form a well-defined population. 

A straightforward understanding of the origin of the Leavitt law can be found in thermodynamic and dynamic arguments. The luminosity of a Cepheid will depend on its surface area via the Stefan-Boltzmann law $L = 4\pi R^2 \sigma T^4 $, which expressed in bolometric magnitudes is 
\begin{equation}
\label{eq:stefanboltz}
    M = -5 \log_{10} R - 10 \log_{10} T + \mbox{const.} \;\; .
\end{equation}
The stellar radius can be mapped to the period by writing an equation of motion for the He layer as
\begin{equation}
    m \ddot{r} = 4 \pi r^2 p - \frac{G M m}{r^2} \;\; ,
\end{equation}
where $r$ denotes the radial position of the layer, $p$ is the pressure and $m$ the mass of the layer. For adiabatic expansion, it is then straightforward to solve for the period $P$ and we find $P \sqrt{\bar{\rho}} = \mbox{const.}$ where $\bar{\rho}$ is the mean density (for further details see \citealt{Cox1960}). With $\bar{\rho} \propto \; R^{-3}$ and temperature mapped to colour $B-V$, we obtain the Leavitt law as
\begin{equation}
\label{eq:leavitt}
    M = \alpha \log_{10} P + \beta (B-V) + \gamma \;,
\end{equation}
where $P$ is the period in days, $\alpha$, $\beta$ are empirically calibrated from the Cepheid data, and for the zero-point $\gamma$ we need a distance measurement. 

It is preferable to use so-called Wesenheit magnitudes, which are constructed to be reddening-free, and have a reduced colour dependence (intrinsically redder Cepheids are fainter). Let's see how this works in practice. \citet{Riess2019} define an NIR Wesenheit magnitude $m_H^W \equiv m_{\rm F160W} - 0.386(m_{\rm F555W} - m_{\rm F814W})$ using HST filter magnitudes, where the numerical constant is derived from a reddening law. They observed 70 LMC Cepheids with long periods, and after setting $\beta = 0$, they derive $\alpha = -3.26$ with intrinsic scatter 0.075 mag. By comparison, the scatter using solely optical F555W magnitudes is 0.312 mag. After subtracting the DEB distance modulus \citep{Pietrzynski2019}, we obtain $\gamma = -2.579$. The formal error in the Cepheid sample mean is 0.0092 mag, equivalent to 0.42\% in distance.
\begin{figure}[htbp]
    \centering
    \includegraphics[width=0.8\textwidth]{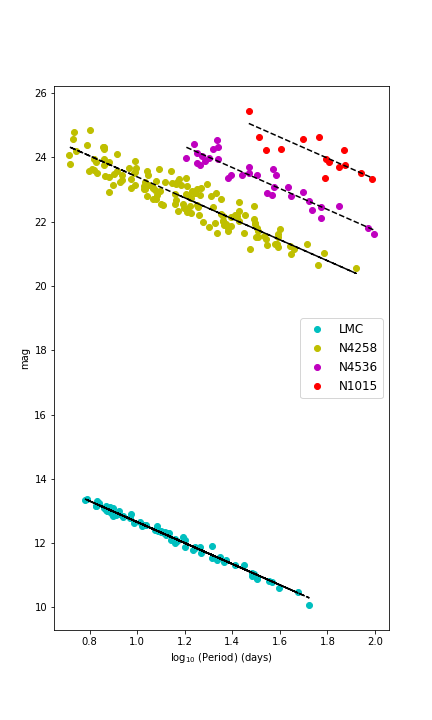}
    \caption{Illustration the Leavitt law in four galaxies used for the local distance ladder. The LMC and N4258 are the two main calibrators of Cepheid distances. For the two example SN Ia hosts N4536 and N1015, it is harder to observe the fainter, shorter-period Cepheids. The slope is fixed at -3.26, corresponding to the best estimate global fit in \citet{Riess2016}, and magnitudes are $m_H^W$. We have inverted the normal decreasing magnitude axis used in the literature for presentation purposes.}
    \label{fig:Cepheid PL}
\end{figure}

\begin{figure}[h!]
    \centering
    \includegraphics[width=\textwidth]{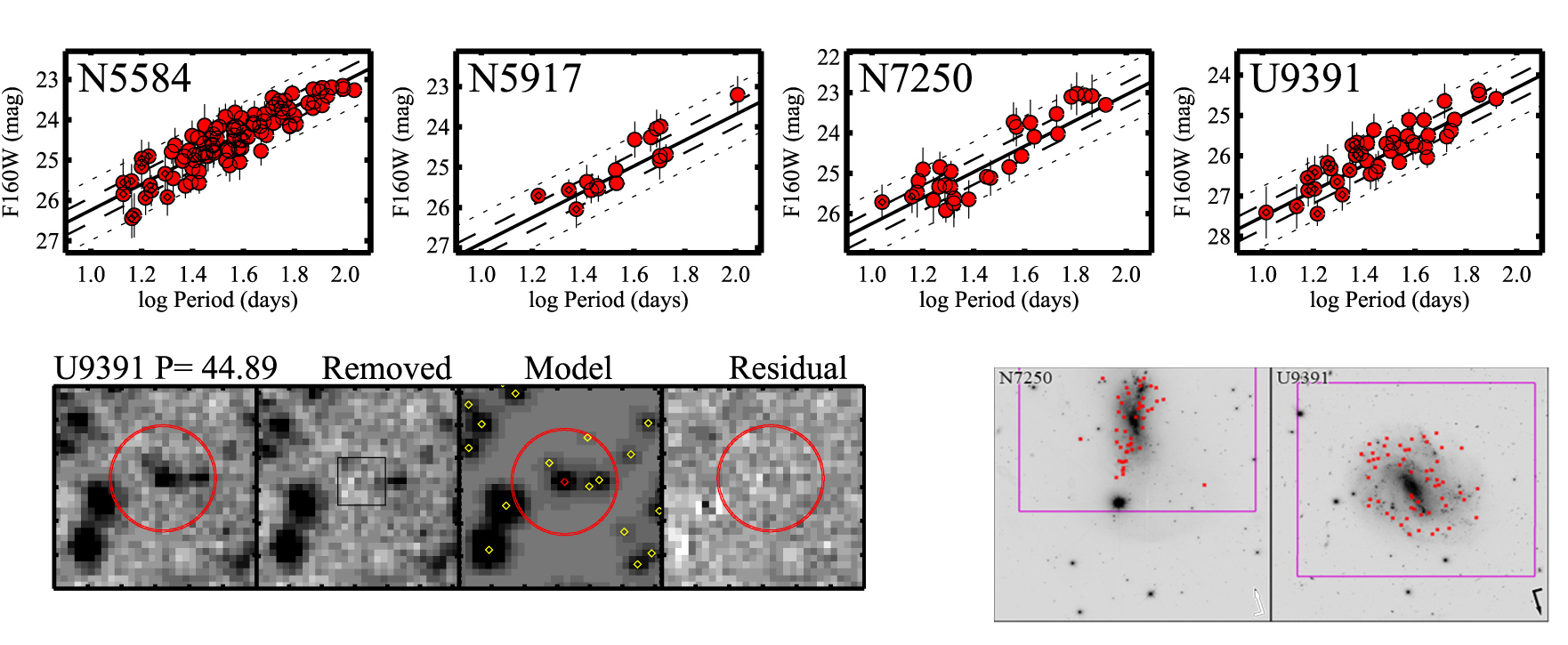}
    \caption{Illustration of Cepheid photometry from SN Ia host galaxies from Riess et al \citep{Riess2016}. U9391 is the most distant with a distance modulus of $32.92 \pm 0.06$. The association of Cepheids with spiral arms is clearly visible, and in the bottom left is an example crowding correction for point sources and background flux. The scatter around the Leavitt law is $\sim 0.7$ mag, thought to be due to residual crowding effects.}
    \label{fig:Cepheid montage}
\end{figure}

We would now like to feel that we can deduce the distance to any galaxy we can find Cepheids in, by applying this law to convert periods to absolute magnitudes, and comparing to the apparent magnitudes we observe. As is usual, though, things are not that simple! There are three principle objections:
\begin{itemize}
    \item \textbf{Are Cepheids in the LMC the same as the ones in distant galaxies?} The LMC has a lower metallicity than a typical spiral galaxy, so can we expect the Cepheids found there to have the same brightness? Unfortunately, it is hard to determine the metallicity of a Cepheid from its colour alone, and studies on the effects of metallicity are variable \citep{Ripepi2020}. One might try extending the Leavitt law to add a metallicity term $\kappa$ [Fe/H]. \citet{Riess2019} have estimated Cepheid metallicity in the LMC based on optical spectra of nearby HII regions, and find $\kappa = -0.17 \pm 0.06$, which is consistent with an earlier estimate from \citet{Freedman2010} that LMC Cepheids are 0.05 mag dimmer than galactic Cepheids, and later results \citep{Gieren2018, Breuval2021}.
    \item \textbf{Is the Leavitt law linear?} This matters for $H_0$, because the Cepheids seen in distant galaxies have longer periods than those used to calibrate the Leavitt law : they are brighter and more easily observed at distance. So, $\log{P} \sim 1.5$ for Cepheids in SN Ia host galaxies, whereas the average for the LMC is $\log{P} \sim 0.8$. If the Leavitt law is curved, a bias would be introduced. This can be dealt with by either introducing a separate calibration for short and long period Cepheids, or selecting only long period nearby Cepheids for calibration. 
    \item \textbf{Can we obtain clean astrometry of very distant Cepheids?} We want to observe Cepheids as far away as we can, to maximise the overlap with the SN Ia observations. But pushing the limits of resolution brings the risk of crowding, whereby the Cepheid photometry is blended with nearby dimmer, redder and cooler stars. Indeed, this seems to be the main cause of the increased scatter in residuals for distant galaxies. There are some ways to deal with crowding: \citet{Riess2016} (hereafter R16) add random Cepheids to images of the same galaxy and put them through the same analysis pipeline, to see how well input parameters are recovered. Figure~\ref{fig:Cepheid montage} shows an example of this. Cepheids which are outliers in colour, indicative of high blending, may be discarded (removing the colour cut lowers $H_0$ by $1.1 \;\SI{}{\kilo\meter\per\second\per\mega\pc}$). Another way to test for a crowding bias is to look for compression of the relative flux difference from peak to trough (a more blended Cepheid will be more compressed) as is done in \citet{Riess2020a}. Using an optical Wesenheit magnitude, in which stars may be less crowded than the NIR one, reduces $H_0$ by $1.7 \SI{}{\kilo\meter\per\second\per\mega\pc}$ (R16), although the argument can be made this is due to higher metallicity effects in the optical \citep{Riess2019}. As with metallicity, the matter of crowding continues to debated. 
\end{itemize}

Systematic error analysis is provided in Section~4 and Table~8 of R16, where the effect of some different analytical choices such as breaks in the Leavitt law, different assumed values for the deceleration parameter $q_0$, and methods of outlier rejection are shown to be $\pm 0.7\%$. This may not cover all potential systematics. In a recent talk, \citet{Efstathiou2020} re-examined the Leavitt laws of galaxies presented in R16. Calibrating each galaxy individually, the slopes of SN Ia host galaxies are generally shallower than M31 or the LMC, which should not be the case if Cepheids are a single population. A change in the slope will alter the zero-point, as would a change in the distance calibration. It is then noted that forcing the slope of the Leavitt law to $-3.3$ (the M31 value), in combination with using only NGC4258 as the anchor, lowers the R16 $H_0$ value to $70.3 \pm 1.8 \;\SI{}{\kilo\meter\per\second\per\mega\pc}$ (Equation 4.2b). Additionally, there appears to be tension between the relative magnitudes of Cepheids in the LMC and NGC4258, and the distance modulus implied by Masers and DEBs; calibrating with solely the LMC gives a higher $H_0$ value by $4.4 \;\SI{}{\kilo\meter\per\second\per\mega\pc}$. It is a feature of $\chi^2$ fits (as used in R16) that the joint solution will be drawn to the data with the lowest error, which in this case is the LMC value. But if two subsets of the data are in tension, it is uncertain that the one with the lowest $\chi^2$ would have the least systematics.
 
Such analyses do not show one value is preferable over another, nor can they show where any discrepancy may lie -- it might be metallicity effects, crowded Cepheid photometry, the NGC4258 distance, or some other systematic. Re-analyses of the results of R16 by various authors \citep{Feeney2018, Cardona2017, Zhang2017, Follin2018, Dhawan2017} use the same Cepheid photometric reduction data so are not independent as such.

Research has accelerated to close down these potential issues. We have already mentioned replacing LMC and NGC4258 Cepheids with Milky Way Cepheids in the section on parallax. In a recent paper, \citet{Riess2020a} show crowding effects can be detected as a light curve amplitude compression, and that their correction method has been robust. Finally, \citet{Javanmardi2021} fully re-derive the Cepheid periods and luminosities from the original HST imaging for NGC 5584 (which is face-on to the line of sight), intentionally using different analytical choices, finding no systematic difference in the light curve parameters. 

All this said, it would be clearly preferable if we had some other candles to check against Cepheids. We now turn to two possibilities, Miras and the Tip of the Red Giant Branch.

\subsection{Miras}

Miras are variable stars that have reached the tip of the Asymptotic Giant Branch (AGB), comprising an inert C-O core, and a He-burning shell inside a H-burning shell. Their mass is 0.8--8\,$M_{\odot}$, although they are typically at the lower end of this range. Their large size of $\sim 1$ a.u. means they are actually brighter by 2--3 mag than Cepheids in the NIR. That they follow a $P$-$L$ law was first established in 1928 \citep{Gerasimovic1928}, but being tricky to categorise and observe, were not extensively studied until the demand for cross-checks on Cepheids has brought renewed interest in them.

Miras form two distinct populations with O-rich and C-rich spectra, with the O-rich ones exhibiting somewhat less scatter than the C-rich \citep{Feast2004}, but more than Cepheids. Miras have very long periods $90<P<3000$ days and their light curves have many peaks of variable amplitude superimposed (see for example Fig.~3 in \citealt{Yuan2017}), so investment in observation time is needed to reliably determine $P$. The $P$-$L$ law curves upwards at around $P \sim 400$ days, where it seems likely the extra luminosity has been due to episodes of Hot-Bottom-Burning, so an extra quadratic term is required. Miras have high mass loss rates, and the surrounding dust cloud means Wesenheit magnitudes will be less reliable at subtracting reddening compared to Cepheids (because a standard reddening law is assumed in constructing them). Lastly, a size of 1 a.u. would mean their angular diameter is comparable to their parallax, and in addition the photocentre moves around the star, making parallax measurements challenging.

So why bother with them? Their advantages versus Cepheids is that they are (a) more numerous, and therefore easier to find in halos where there is less crowding (b) older, so can be found in all types of galaxies including SN Ia hosts with no Cepheids (c) $2-3$ magnitudes brighter than Cepheids in the near infrared. Given imaginative observation strategies and careful population analysis, the issues above can be addressed, and modern studies now exist for Miras in the LMC \citep{Yuan2017}, NGC 4258 \citep{Huang2018}, and the SN Ia host NGC 1559 \citep{Huang2019}.

In the first of the above, \citet{Yuan2017} establish a $P$-$L$ curve using 600 Miras in the LMC. They have sparse $JHK$-band observations from the LMC Near Infrared Synoptic Survey (LMCNISS, \citet{Macri2015}), which on their own wouldn't be enough to establish the period, but using much denser Optical Gravitational Lens Experiment (OGLE) I-band observations \citep{Szymanski2011}, they are able to establish a regression rule for the relationship between passbands, and use OGLE to \say{fill in} the light curves. The classification of Miras into O-rich or C-rich is also obtained from OGLE. The reference magnitude is defined as the median of maxima and minima of the fitted light curves. The period is obtained by fitting to a sum of sine functions, progressively adding harmonics if supported by a Bayesian Information Criterion\footnote{
Whilst there are well known issues with the Bayesian Information Criteria, mainly an unnecessary penalty to fits with a large a number of data points, \citet{Yuan2017} uses this statistic, and it is beyond the scope of this review to repeat their analysis using a different statistic.
}. 
They fit a law\footnote{It is good practice to center around the mean $P$ to reduce correlation between coefficients.}
\begin{equation}
\label{Eq:MiraPL}
M = a_0 + a_1 (\log_{10} P - 2.3) + a_2 (\log_{10} P - 2.3)^2 \;.
\end{equation}
The K-band for O-rich Miras shows the lowest scatter of 0.12 mag, with $a_0 =-6.90 \pm 0.01$, $a_1 = -3.77 \pm 0.08$, $a_2 = -2.23 \pm 0.2$. The zero-point is obtained using the DEB distance to the LMC given by \cite{Pietrzynski2013}. By comparison, the $m_H^W$ scatter for LMC Cepheids obtained by SH0ES was 0.075 mag \citep{Riess2019}.
\begin{figure}[h!]
    \centering
    \includegraphics[width=\textwidth]{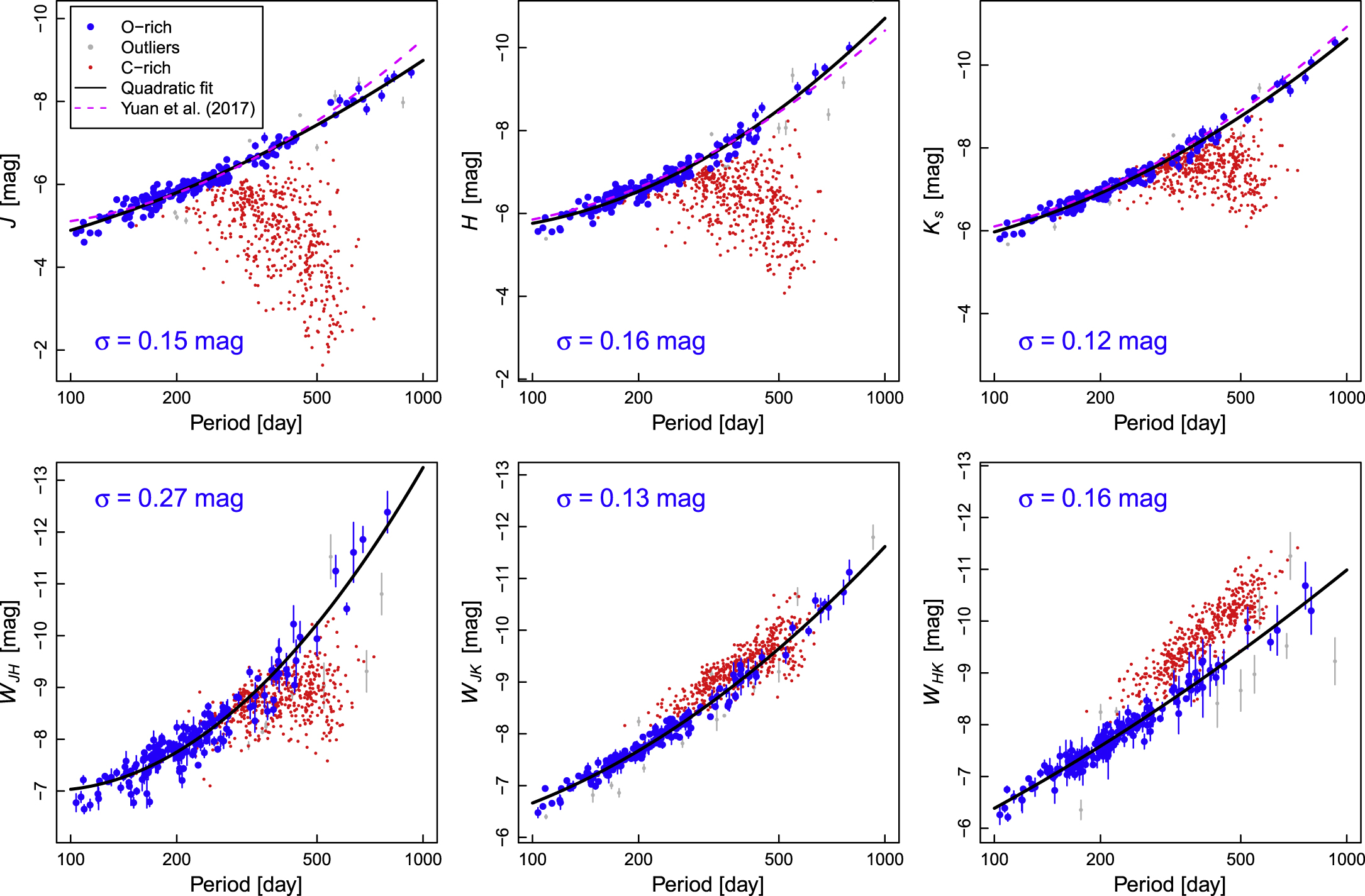}
    \caption{A PL calibration for Miras in the LMC from \citet{Yuan2017}. While non-linearity can straightforwardly be fitted, contamination must be kept under control. As seen in the bottom three panels, Wesenheit magnitudes have more scatter compared to standard magnitudes (the opposite of applying them to Cepheids), and they also hinder separating O-rich from C-rich stars.}
    \label{fig:LMCMiras}
\end{figure}

\citet{Huang2018} observed NGC4258 with the HST WFC in 12 epochs over 10 months. Mira candidates were identified by their $V$ and $I$ amplitude variation, and it was possible to use LMC data to show that using this method contamination between O-rich and C-rich could be kept to a minimum. Their ``Gold'' sample size was 161 Miras, and fitting Eq.~(\ref{Eq:MiraPL}) for apparent magnitudes gave $a_0 = 23.24 \pm 0.01$ for F160W with scatter 0.12. Adjusting the ground-based photometry of LMCNISS to F160W equivalent, the authors calculate the relative distance modulus between the LMC and NGC4258 of $\Delta\mu = 10.95 \pm 0.01 \mbox{(stat)} \pm 0.06 \mbox{(sys)}$. This is consistent with the Cepheid value of $\Delta\mu = 10.92 \pm 0.02$ \citep{Riess2016} and the Maser-DEB value of $\Delta\mu = 10.92 \pm 0.041$ \citep{Reid2019, Pietrzynski2019}. Similar methods are used by \citet{Huang2019} for observations of NGC 1559, with the addition of a low period cut to deal with potential incompleteness bias of fainter Miras. Using the LMC DEB and NGC 4258 distances as calibrators, they obtain $H_0 = 73.3 \pm 4.0 \;\SI{}{\kilo\meter\per\second\per\mega\pc}$. So Miras are consistent with Cepheids, but due to their larger error budgets, they are also consistent with Planck results. 

\subsection{Tip of the Red Giant Branch}

Red giant branch stars are stars of mass $\sim 0.5 - 2.0 M_{\odot}$ that are at a late stage in their evolution, having moved off the main sequence towards lower temperatures and higher luminosities. The core of the star is degenerate He, and as He ash rains down from the H-burning shell surrounding it, it contracts and heats up. Since the temperature range for He fusion ignition is rather narrow, and the degenerate matter means there is a strong link between core mass and temperature, the core in effect forms a standard candle inside the star, just prior to ignition. As soon as the helium flash ignites, the star moves over the course of 1 My or so (almost instantaneously in stellar evolution terms) to the red clump at a higher temperature and somewhat lower luminosity. Therefore, there is a well-defined edge in a colour-magnitude diagram that can be used to identify the tip of the red giant population just before the flash (TRGB). The TRGB is then not one single star, but a statistical average for a suitably large population. 

The TRGB offers many advantages compared to other standard candles. Although red giants are fainter than Cepheids in the optical, in the K-band they are $\sim 1.6$ magnitudes brighter. Stars at or near the TRGB are relatively common, and can be observed readily in uncrowded and dust-free galactic halos. They are also abundant in the solar neighbourhood, meaning great numbers are available for calibration by parallax (by contrast, Cepheids with good parallaxes in Gaia DR3 will number in the hundreds). Sufficient numbers to resolve the tip can be seen in globular clusters such as $\omega$ Centuari, where well-formed colour-magnitude diagrams can resolve metallicity and extinction effects, and an absolute magnitude calibration can be made either from Gaia DR3 parallaxes \citep{Soltis2020} or DEB distances \citep{Cerny2020}. 

Like Miras, they are found in all galactic types. Observation is efficient: one does not need to revisit fields in order to determine periods. The James Webb Space Telescope is capable of observing red giants in the near IR at distances of $\sim 50 \SI{}{\mega\pc}$, which is comparable to Cepheid distances obtainable from the HST. 

It is also beneficial that the modelling of late-stage stellar evolution is quite well understood. Empirical calibrations of absolute magnitude and residual dependence on metallicity, mass and age can be checked against the results of stellar codes. For example, the dependency on metallicity is somewhat complex: the presence of metals dims optical passbands by absorption in stellar atmospheres, and brightens the NIR.  \citet{Serenelli2017} express the TRGB as a curve with linear and quadratic terms in colour. The authors find an $I$-band median $M_I^{\rm TRGB} = -4.07$ with a variation with colour of $\pm 0.08$, and a downward slope in the colour-magnitude diagram as metallicity increases. This median value is consistent with empirical calibrations. 

Fitting the TRGB is equivalent to finding the mode of the gradient of the (noisy) stellar distribution in the colour-magnitude diagram, and techniques are borrowed from image processing to do this. The tip has a background of AGB stars, and the edge is sensitive to the distribution of them close to it. If the contamination is large or variable, there is a risk the mode can shift. It is therefore important to have a dense population of stars, and enough fields to test the robustness of the fitting process. For example, \citet{Hoyt2021} bin data by separation from the galactic centre and fit the tip separately for each bin.

The Carnegie-Chicago Hubble program (CCHP) have determined TRGB distances for 10 SN Ia host galaxies that also have Cepheid distances, using HST $I$-band imaging \citep{Freedman2019}. They calibrate the zero point from the TRGB of the LMC and SMC using the DEB distance, finding $M_I = -4.049 \pm 0.022 \;\mbox{(stat)} \pm 0.039 \;\mbox{(sys)}$, which is consistent with the theoretical value above.

\begin{figure}[h!]
    \centering
    \includegraphics[width=\textwidth]{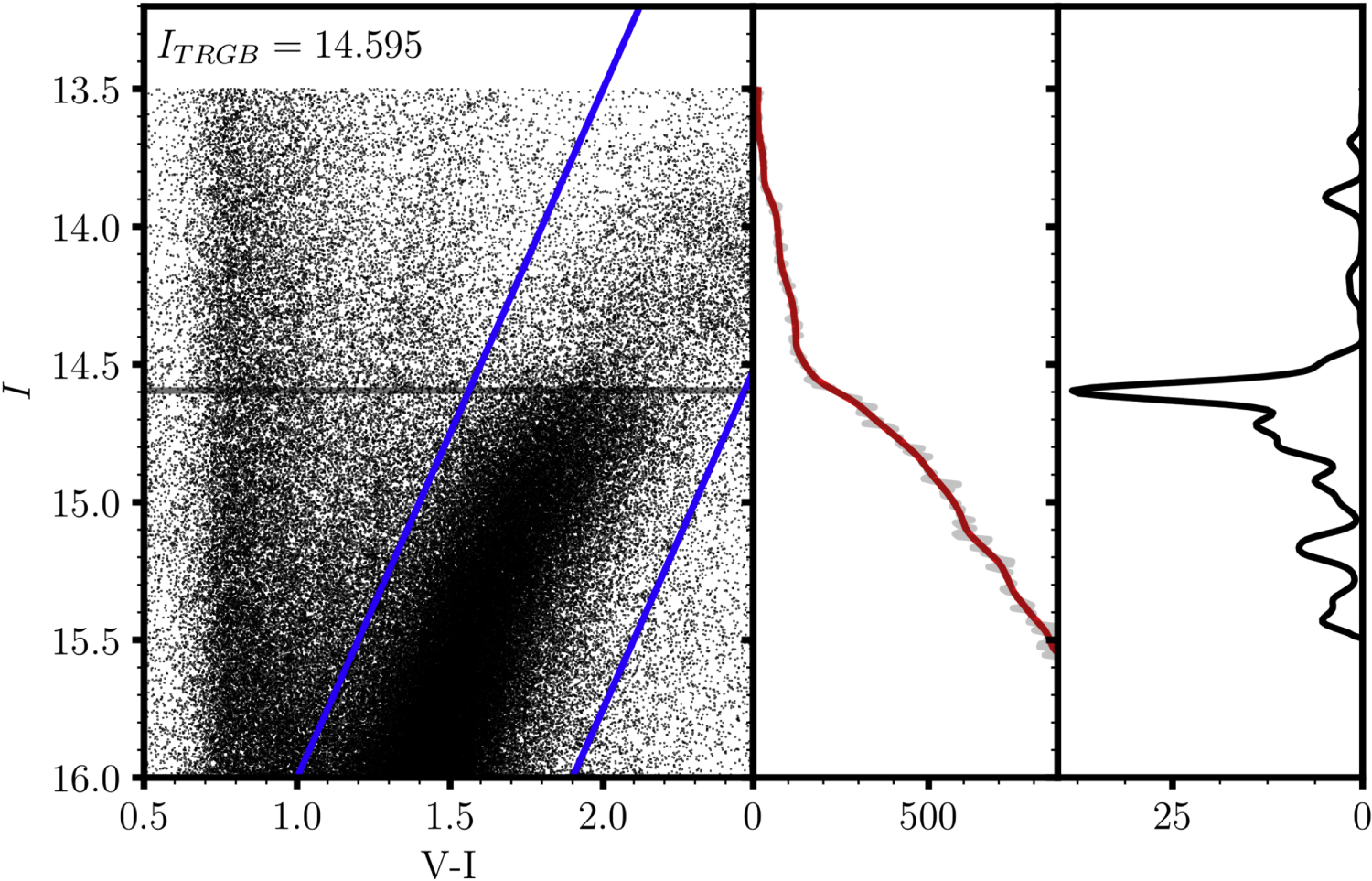}
    \caption{Fitting the TRGB edge from the colour-magnitude diagram of \citet{Freedman2019}. The middle panel shows the horizontally-summed (between the blue lines) and smoothed number count, from which the edge is determined by the maximum (vertical) gradient, which is the peak in the right hand panel. The background AGB population is visible in the stars above the tip.}
    \label{fig:TRGB fitting}
\end{figure}
\begin{figure}[h!]
    \centering
    \includegraphics[width=0.9\textwidth]{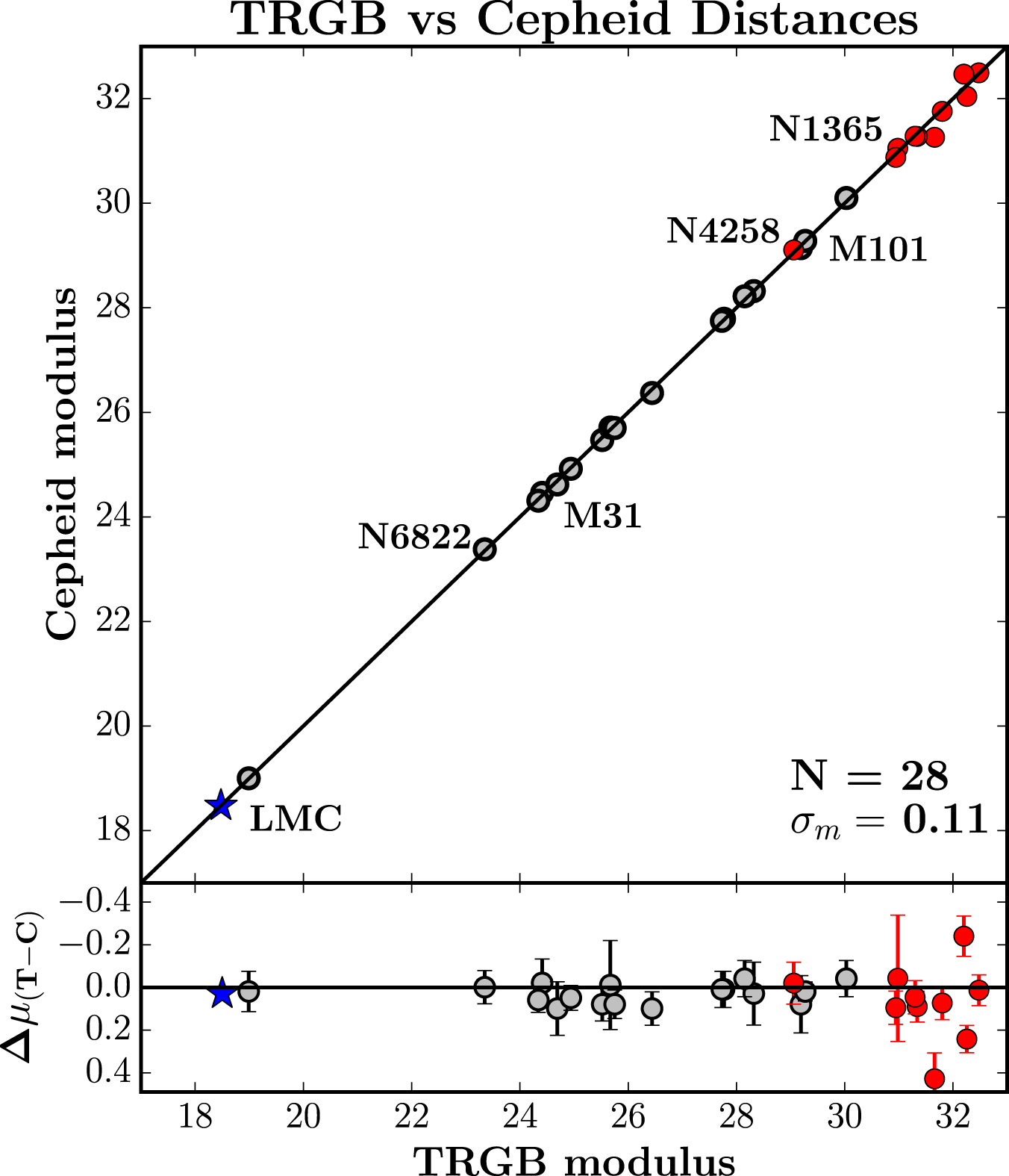}
    \caption{Comparison of TRGB and Cepheid distance moduli, from which it can be seen TRGB distances are consistently larger across a range of galaxies.  The red dots are SN Ia host galaxies, the grey dots other galaxies, and the blue star is the LMC TRGB calibrator. Figure is from \citet{Freedman2019}.}
    \label{fig:TRGB vs Cepheid}
\end{figure}

We show in Fig.~\ref{fig:TRGB vs Cepheid} their comparison of TRGB and Cepheid distances, covering a range of distances from 7 Mpc to almost 20 Mpc. By expanding their sample to non-SN Ia host galaxies, the authors show the TRGB has a lower scatter versus their Hubble diagram than do Cepheids, by a factor of 1.4. Calibrating the Pantheon SN Ia sample on TRGB distances alone gives $H_0 = 70.4 \pm 1.4 \SI{}{\kilo\meter\per\second\per\mega\pc}$, with the difference to SH0ES results due to the TRGB distances to the SN Ia host galaxies being further than Cepheid distances.

Currently, this result remains the subject of debate. \citet{Yuan2019} find $M_{I} = -3.99$, which gives $H_0 = 72.4 \pm 2.0 \SI{}{\kilo\meter\per\second\per\mega\pc}$. They attribute their result to a different methodology to determine the LMC extinction (\citet{Freedman2019} determine the extinction applicable to the TRGB by comparison of LMC and SMC photometry in $VIJHK$ bands, and using the lower SMC extinction as a reference point. \citet{Yuan2019} adopt the standard reddening law of \citet{Fitzpatrick1999}). They also revise the blending corrections of the older, lower resolution photometry of the SMC. Conversely, a calibration of the TRGB in the outskirts of NGC 4258 to the maser distance by \citet{Jang2020} gives a value fully consistent with \citet{Freedman2019}. In a followup paper, \citet{Freedman2020} present an upgraded methodology in which stellar photometry is itself used to separate metallicity and extinction effects (by exploiting their different sensitivities in $J$,~$H$ and $K$-band magnitudes), confirming their earlier results. Other results (\citealt{Soltis2020, Reid2019, Capozzi2020, Cerny2020} -- see for example Table~3 in \citealt{Blakeslee2021}) cluster evenly around these two values. We also note the difference between SN Ia zero points (Fig.~6 of \citealt{Freedman2019}) of the TRGB vs Cepheids appears to increase with distance, and is particularily large for one galaxy, N4308. 

More recently, an updated value of $H_0 = 69.8 \pm 0.6 \;\mbox{(stat)} \pm 1.6 \;\mbox{(sys)} \SI{}{\kilo\meter\per\second\per\mega\pc}$ was published by \citet{Freedman2021}. New and updated values for $M_I$ based on the anchors of the LMC, SMC, NGC 4258 and galactic globular clusters are derived. Additionally, potential zero-point systematic problems with the direct application of Gaia EDR3 data to globular clusters (as done by \citealt{Soltis2020}) are discussed. Each anchor is consistent with each other (although the closeness of the agreement suggests the errors may be over-estimated), and the paper brings the TRGB method on par with Cepheids in terms of number of anchors used.

The TRGB continues to attract attention because the CCHP result is very interesting: it is a late universe distance ladder that is in reduced $2\sigma$ tension with the Planck value of $H_0 = 67.4 \pm 0.5 \SI{}{\kilo\meter\per\second\per\mega\pc}$. While it cannot by itself fully resolve the Hubble tension, in combination with some other systematic -- perhaps in the calibration of SN Ia (see next section) -- it may offer a resolution of it that does not involve new physics.

What is also clear is that quality of a distance ladder is less a matter of \textit{quantity} of objects, but rather the \textit{accuracy} of calibration and \textit{control} of systematics. The TRGB seems very promising however. Once the calibration is agreed among the community, they offer the tantalising prospect of either confirming the $H_0$ tension through two independent data sets, or reducing it to a lower statistical significance. 

\subsection{Type Ia supernovae}

It is hard to overstate the impact of Type Ia supernovae in cosmology. At $M \sim -19$, they are both bright enough to be seen well into the Hubble flow (the furthest to date is SN Wilson at $z = 1.914$), and have a standardisable luminosity. The mainstream view is that they originate from the accretion of material from a binary companion onto a carbon-oxygen white dwarf. When the white dwarf reaches the maximum mass that can be sustained by degeneracy pressure, the Chandrasekhar mass $M_{\rm ch} = 1.4 M_{\odot}$, a runaway fusion detonation occurs destroying the white dwarf. Approximately $\sim 0.6 M_{\odot}$ fuses into heavier elements during the first few seconds of the explosion, and the observed light curve, which peaks at $\sim 20$ days and lasts for $\sim 60$ days is powered by radioactive decay of $^{56}$Ni to $^{56}$Co and then to $^{56}$Fe. Hence, a SN Ia is a \say{standard bomb}, primed to explode when it reaches critical mass. 

However, there is surprisingly little observational evidence to confirm the accretion theory. Both the lack of H lines and computer modelling suggest the donor star should survive the explosion, but a search of the site of the nearby SN2011fe both pre- and post- explosion have revealed no evidence of a companion. This may support an alternative \say{double degenerate} theory that some, or all, SN Ia originate from the merger of two white dwarves (for a review, see \citealt{Maoz2014}). But if that is the case, why are their luminosities so uniform? Whether SN Ia have one or two types of progenitors has important implications for the Hubble constant, which we return to shortly.

Individual SN Ia luminosities can vary by up to a factor of $2$, but they are empirically standardisable by the Tripp estimator \citep{Tripp1999}: 
\begin{equation}
    \mu = m_b - M_{\rm fid} + \alpha x - \beta c + \Delta_M + \Delta_B \;,
\end{equation}
where $\mu$ is the distance modulus, $m_b$ ia the apparent AB magnitude, and $M_{\rm fid}$ is the absolute magnitude of a typical SN Ia. The parameter $x$ is the \say{stretch} of the light curve, which is a dimensionless measure of how long the bright peak of the light curve lasts: longer duration SN Ia are brighter. $c$ is a measure of colour: redder SN Ia are dimmer. $\Delta_M$ is a correction for the environment of the SN Ia, and $\Delta_B$ a statistical bias correction for selection effects, analogous to the LKH bias of parallax we described earlier. $M_{\rm fid}$ can be calibrated by finding SN Ia in galaxies for which there are Cepheid or TRGB luminosity distances, or in the inverse distance ladder approach by combination with angular diameter distances from baryon acoustic oscillations, or the CMB. 
 
SN Ia are rare events: approximately 1 per galaxy per century, but by scanning a reasonably sized patch of sky, supernova surveys can detect hundreds per year\footnote{Supernovae that are photometric candidates to be SN Ia outnumber those that are confirmed by a factor of 10; the limitation is the availability of spectroscopy.}. Hence, supernovae catalogs are not uniformly distributed on the sky. However, SN Ia close enough to calibrate their magnitudes are few: there are just 19 SN Ia in galaxies with Cepheid distances from HST observations, although more are expected soon from new cycles (see for example HST proposal 16198).

To maximise statistics and sky coverage, it is common to use SN Ia from multiple surveys in cosmological analyses. \citet{Scolnic2018} have compiled the Pantheon sample of 1,048 SN Ia from CfA, CSP, PS1, SDSS, SNLS and HST surveys in a uniform light curve calibration, representing almost 40 years of observational data. The sample divides into 180 mostly older Low-$z$ ($z < 0.1$) SN Ia which are uniformly distributed on the sky, and 868 recent High-$z$ SN Ia concentrated in the survey fields, notably the thin \say{Stripe 82} along the celestial equator. The sets have a small overlap at $z \sim 0.1$, and cross-checks on their spectral and other parameters show they are likely to be from the same underlying population (that is, High-$z$ SN Ia are not intrinsically different to Low-$z$). However, the mean values of light curve parameters like stretch and colour do drift with $z$, possibly due to selection effects or changes in the host galaxy properties at higher redshift. Large numbers of Type Ia supernova have also been found, or are in the process of being surveyed, by the Dark Energy Survey \citep{Abbott2019}, All-Sky Automated Survey for Supernovae \citep{Holoien2018}, Foundation Survey \citep{Jones2019}, and Zwicky Transient Facility \citep{Yao2019}. 

\begin{figure}[h!]
    \centering
    \includegraphics[width=0.8\textwidth]{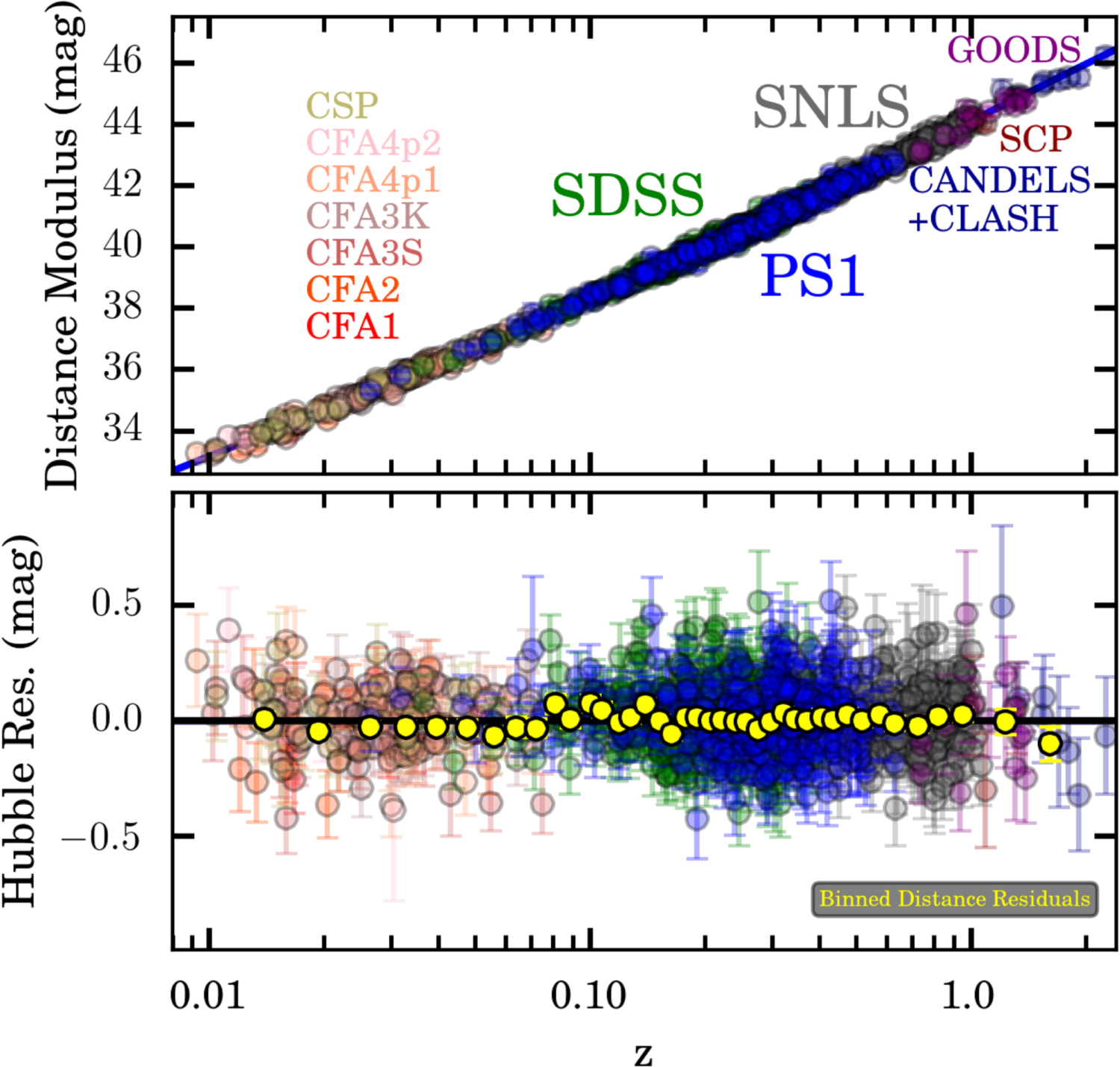}
    \caption{Pantheon SN Ia distance moduli and residuals versus a best fit $\mathrm{\Lambda CDM}$ cosmology. Figure is from \citet{Scolnic2018}.}
    \label{fig:Pantheon residuals}
\end{figure}

The process of fitting a SN Ia lightcurve is straightforward. \textit{ugriz} photometry is corrected for the background sky obtained from imaging the site after the nova has faded (a bonus of SN Ia being transient), and the light curve is fitted to a template, outputting the parameters $x, c$ and $m_b$. However, the bias correction $\Delta_B$ is challenging and subtle: it depends on the astrophysical modelling of sources of scatter and contamination by mis-classified supernovae, and the selection function of survey. For example, more distant SN Ia may be preferentially targeted for spectroscopic confirmation when their host galaxy is faint, so if their luminosities do depend on the galactic environment, this could bias the High-$z$ population relative the to Low-$z$. The selection function of early Low-$z$ surveys are not easy to model, and their biases can be up to 0.06 magnitudes. This could account for a somewhat elevated scatter of Pantheon residuals around $z \sim 0.1$ in the Hubble diagram where the two sets join (Fig.~\ref{fig:Pantheon residuals}). After fitting $\alpha$ and $\beta$, the residual intrinsic scatter of the SN Ia population is $\sigma_{\rm int} \sim 0.1$ magnitudes, which compares favourably with other standard candles. 

We return now to the question of whether SN Ia have additional environmental correlations that affect their luminosity (that is, what the nature of the $\Delta_M$ term in Tripp estimator is), and why this may matter for $H_0$. It is generally accepted that SN Ia in galaxies with stellar mass $M_{*} > 10^{10} M_{\odot}$ are intrinsically brighter by $\Delta_{M,*} \sim 0.04 - 0.06$ than those in lower mass galaxies, and that the transition is sharp, rather than a gradual evolution (see \citealt{Smith2020} and references therein). It has therefore become common to use a step-function for $\Delta_{M,*}$ in fits for $H_0$. \citet{Riess2016} calculate the effect is to lower $H_0$ by 0.7\%, as the mass of the calibrator galaxies is slightly lower than the mass of the Hubble flow set. The key point is then this: given that Cepheids are young, bright stars formed in active galaxies, is $\Delta_{M,*}$ sufficiently discriminating to ensure the 19 supernovae in galaxies with Cepheid distances are representative of the thousands in the Hubble flow? 

Host galaxy mass itself can't matter to an individual SN Ia, so it must be a proxy for some other environmental condition. Metallicity does vary with galactic mass, but the step-like feature suggests a connection with star formation: galaxies with mass below the threshold tend to be active, whereas those above are mostly passive. One hypothesis is that there is a prompt/bright SN Ia type that is continuously renewed by star formation, and an delayed/dimmer type originating from stars formed when the galaxy was young \citep{Rigault2015, Rigault2013, Maoz2014}. But any link between this and SN Ia formed by accretion onto or mergers of white dwarfs is speculative. Also, with estimates of \say{global} properties being effectively light-weighted, they are biased to galactic centres rather than the outskirts where most supernovae are seen.

If some SN Ia are associated with sites of star-formation, we can expect a better proxy to be local specific star formation rates (that is, star formation normalised by local stellar mass or LsSFR for short), on the grounds that the younger SN Ia population will not have had enough time to disperse from their birth region. Because the pixel resolution of typical SN Ia surveys can resolve a $2$ kpc aperture only out to $z \sim 0.1$, studies concentrate on the Low-$z$ sample. Star formation can be estimated from photometry, but is best tracked by H$_{\alpha}$ lines from ionised gas or UV imaging, when available. 

\citet{Rigault2018} have examined 141 SN Ia from the Nearby Supernova Factory sample (which has high resolution spectroscopy) and find a significant correlation between LsSFR and $\alpha$, the stretch slope, and no correlation for $\beta$. This suggests there is indeed an intrinsic difference between SN Ia originating in active areas versus passive ones. The size of the LsSFR step is $\Delta_{M,LsSFR} = 0.163 \pm 0.029$ mag, and including it in the Tripp luminosity estimator eliminates the need for a further global host mass step. The size of the step is interesting as it is significantly larger than $\Delta_{M,*}$, and because most SN Ia in the Cepheid calibrators are in regions of high LsSFR, whereas in the rest of the Low-$z$ sample the fraction is about a half. Hence, a potential bias to $H_0$ could be as much as 3\%. In contrary results, \citet{Jones2018} find a local stellar mass step the same size as $\Delta_{M,*}$ and little effect on $H_0$, and \citet{Riess2016} found just $0.1 \SI{}{\kilo\meter\per\second\per\mega\pc}$ difference in their $H_0$ calculation after restricting to SN Ia found in spiral galaxies. In a recent paper, \citet{Brout2020} argue the true factor is host galaxy extinction (correlated to star formation and hence mass), and present evidence of a distribution separated by colour into blue \say{clean} SN Ia with lower scatter than their red \say{dusty} counterparts. The host mass step is absent in the blue sample, and they suggest to use solely these in cosmological analyses to avoid bias (another possibility is to use near-infrared photometry \citep{Dhawan2017}).

In summary, while there is strong evidence SN Ia depend on local specific star formation rates, there is not yet a consensus on the cause, whether they constitute one or two populations, and the level of bias (if any) to $H_0$. They are numerous and therefore high \textit{relative} precision beyond $z>0.1$, and have been dubbed the \say{guard rails} of the Hubble diagram. However, their \textit{absolute} accuracy in distance ladders is limited by the small numbers available to calibrate their luminosities, and some uncertainty about the underlying astrophysics. Work is underway to address both of these issues: for example, the Foundation and Zwicky Transient Facility surveys \citep{Jones2019, Yao2019} are targeting a new high-quality Low-$z$ set with local spectroscopy to test environmental effects or extensions to the Tripp estimator. New HST observation cycles aim to extend the number of Cepheid calibrators up from 19, and the JWST will access a greater volume, perhaps increasing the calibration set beyond a hundred for both TRGBs and Cepheids. This would make a 1\% calibration from at least two independent sources achievable. 

\subsection{Time delay of lensed quasars}

Quasars are bright and can be variable on short timescales. Some are seen to be gravitationally lensed by a foreground galaxy, and each image will have a time delay caused by the extra path length and the gravitational time dilation of general relativity: 
\begin{equation}
\label{Eq:Lens_dt}
    \Delta t_i = D_{\Delta t}\left[ \frac{1}{2} (\Vec{\theta_i} - \Vec{\beta})^2 - \psi(\Vec{\theta_i}) \right] ,
\end{equation}
where $\Vec{\theta_i}$ is the sky position of the image, $\Vec{\beta}$ the (unknown) source position, and $\psi(\Vec{\theta_i})$ the potential of the lensing galaxy integrated along the line of sight. The relevant quantity for $H_0$ in the above equation is 
\begin{equation}
    D_{\Delta t} \equiv (1+z_l) \frac{D_l D_s}{D_{ls}} \propto H_0 ^{-1} \;,
\end{equation}
where $D_l, D_s, D_{ls}$ is the angular diameter distance to the lens, source, and between the lens and the source respectively as specified by Equations (\ref{Eq:LumDist}, \ref{eq:Etherington}) -- in the case of $D_{ls}$ the integration would be done between the redshifts of the lens and source. Then, if we have a system with multiple images, the relative time delay $\Delta t_{ij} = \Delta t_i - \Delta t_j$ of the quasar variability between the images scales with $H_0^{-1}$. The proportionality follows from Eqs.~(\ref{Eq:LumDist}, \ref{eq:Etherington}) with a weak dependence on other cosmological parameters. Acquiring the time delays $\Delta t_{i}$ requires extensive observations: relative delays range from weeks to a few months. Errors are typically $\sim 6\%$, although these are reducing over time. A typical lensing system may have $z_s \sim 1.5$ and $z_l \sim 0.5$, so we are probing intermediate cosmological distances.

Each single lens system may provide an independent estimate of $H_0$. But to do so, we need an idea of the lensing potential $\psi(\Vec{\theta_i})$ above. There is also a tricky degeneracy to deal with, known as the \textit{mass sheet transformation}. For our purposes, it is sufficient to understand that the same observed time delays may be produced by simultaneously scaling the line-of-sight surface density $\Sigma (\Vec{\theta})$ of the lensing system by a constant factor $\Sigma (\Vec{\theta})\rightarrow \epsilon \Sigma (\Vec{\theta})$ (the \say{mass-sheet}) and $H_0 \rightarrow \epsilon H_0$. For a clear account of the details, see \citet{Falco1985}.

The ingredients for modelling are then this: (a) a range of plausible mass profiles for the lensing galaxy with associated nuisance parameters to be marginalised over (b) a way to deal with the mass sheet degeneracy, through independent constraints on the mass distribution of the lensing galaxy and along the line-of-sight. 

This is a challenging, time-consuming and model dependent process for each system! Some other choices must be made, such as which of the galaxies along the line of sight will be modelled and which will be absorbed into the line of sight mass average \citep{Chen2019}. Galactic stellar velocity dispersions provide information on the lensing potential, but as a function of which galactic mass profile is assumed, and with an anistropy nuisance parameter. Also, the accuracy of velocity dispersions at such distances are $\sim 10\%$. If there are visible background galaxies, they may also be lensed and the shape distortions provide extra information on the mass profile. The line of sight mass may be estimated from the density of foreground galaxies, by searching in computer simulations of large scale structure for similar lines of sight, and calculating the line of sight density from them (lenses are associated with over-densities and therefore focusing is more likely). But a major advantage is that this modelling can be done blind to final value of $H_0$, eliminating the risk of confirmation bias on the part of the experimenter.

The H0LiCOW team have analysed six such systems so far \citep{Wong2019}, and we show their results below. The main contributions to the errors are uncertainty in the time delay, and velocity dispersion of the lens. H0LiCOW use flat priors of $\Omega_{M} \in (0.05, 0.5)$ and $H_0 \in (0,150) \;\SI{}{\kilo\meter\per\second\per\mega\pc}$, to avoid any dependence on other cosmological datasets (although they do assume a flat universe). The results are : 

\begin{center}
\renewcommand{\arraystretch}{1.5}
\begin{tabular}{ c c c c } 
 \hline
 Lens name & $D_{\Delta t}$ (Mpc) & $D_l$ (Mpc) & $H_0\;(\SI{}{\kilo\meter\per\second\per\mega\pc})$ \\ 
 \hline
 PG1115+080 & $1470^{+137}_{-127}$ & $697^{+186}_{-144}$ & $81.1^{+8.0}_{-7.1}$  \\
 RXJ1131-1231 & $2096^{+98}_{-83}$ & $804^{+141}_{-112}$ &  $78.2^{+3.4}_{-3.4}$ \\ 
 HE0435-1223 & $2707^{+183}_{-168}$ & - & $71.7^{+4.8}_{-4.5}$ \\
 WFI2033-4723 & $4784^{+399}_{-248}$ & - & $71.6^{+3.8}_{-4.9}$ \\
 B1608+656 & $5156^{+296}_{-236}$ & $1228^{+177}_{-151}$ &  $71.0^{+2.9}_{-3.3}$ \\ 
 SDSS1206+4332 & $5769^{+589}_{-471}$ & $1805^{+555}_{-398}$ & $68.9^{+5.4}_{-5.1}$ \\
 \hline
\end{tabular}
\end{center}

Their combined result is $H_0 = 73.3 \pm 1.8 \;\SI{}{\kilo\meter\per\second\per\mega\pc}$, a final error budget of 2.4\%. However, a key assumption in the combination is independence, and there are many modelling features in common between the lenses, in particular the specific mass profiles used. There is also an apparent drift in $H_0$ values with lensing distance, although the statistical significance is not strong enough to say this is not just random chance. 

A re-analysis has been undertaken by the TD Cosmo team \citep{Birrer2020}. By treating the mass sheet degeneracy as a fully unknown parameter at the population level, their analysis reduces reliance on specific mass profiles. For same set of systems, they find $H_0 = 74.5^{+5.6}_{-6.1} \;\SI{}{\kilo\meter\per\second\per\mega\pc}$. The central value is consistent with H0LiCOW, but the wider confidence intervals reflect the greater freedom allowed on lens mass distribution, combined with the small number of time-delay lenses. To re-narrow the uncertainty range, the paper investigates what effect introducing a set of non-time delay lensing systems from the SDSS survey may have. The purpose of this larger set is to use their imaging to constrain the mass-profile of lensings galaxy, on the assumption they are drawn from the same parent population as the H0LiCOW sample. TD Cosmo then find $H_0 = 67.4^{+4.1}_{-3.2} \;\SI{}{\kilo\meter\per\second\per\mega\pc}$. However, the H0LiCOW sample is selected as those systems capable of producing a clear time-delay signal, which may not have the same characteristics as those selected as having clean shear images. So the assumption they share the same mass profile as each other may not be valid.

It is therefore premature to claim lensed quasars are consistent with either SH0ES or Planck.  More data is needed: upcoming surveys by the Vera Rubin Observatory, Euclid and Nancy Grace Roman telescope will result in several hundred new time delay systems being discovered (so improvements in analysis efficiency will certainly be needed!). Adaptive optics can provide spatially resolved velocity dispersions to improve the mass models. TD Cosmo estimate 200 extra systems will be needed to constrain $H_0$ within $1.2\%$, which may be within reach in the next five years.

\subsection{Gravitational waves}
The gravitational wave signal emitted by the merger of two compact objects can be used as a self-calibrating standard candle. There are now operational detectors at LIGO Hanford and LIGO Livingston in the USA, Virgo in Italy, and KAGRA in Japan, with a further planned LIGO India \citep{Abbott2020}. The detectors measure the strain amplitude of a gravitational wave by using laser interferometry to detect the miniscule changes in the length of perpendicular beams as a wave passes by. The purpose of the two sites in the USA is to filter out local seismic vibrations. The wave amplitude is related to the \textit{chirp mass} $\mathcal{M}$ which is in turn derivable from the waveform calculated for a merger. A simplified form of the relevant equations are
\begin{align}
    \mathcal{M} &= \frac{(m_1 m_2)^{3/5}}{(m_1 + m_2)^{1/5}} \\
    &= \frac{1}{G} \left[ \frac{5}{96} \pi^{-8/3}f^{-11/3}\dot{f} \right]^{3/5} \\
    \mathcal{M}_z &= (1+z_{\rm obs}) \mathcal{M}  \\
    h(t) &= \frac{\mathcal{M}_{z}^{5/3} f(t)^{2/3}}{d_L} F(\theta, i) \cos{\Phi(t)} \label{eq:gravwave} \;,
\end{align}
where $f$ is the frequency, $m_1$ and $m_2$ the merging masses, $\Phi(t)$ the phase, and $h(t)$ the measured dimensionless strain of the strongest harmonic \citep{Abbott2016}. The rest-frame chirp mass is redshifted by $z_{\rm obs}$, and $F$ is a function of the angle between the sky position of the source and detector arms, and the inclination $i$ between the binary orbital plane and line of sight \citep{Arun2009}. Note that as we are measuring amplitude rather than energy flux, $h(t) \propto d_L^{-1}$. 

In a loose sense, every wave cycle is a measurement of the chirp mass $\mathcal{M}$ as it sets $f, \dot{f}$, although in practice the full waveform is fitted. The relative precision is floored by sensitivity of the strain measurement, currently around $5\%$. If the redshift and angles were known, $d_L$ and hence $H_0$ would be determined to the same precision. 

When a binary neutron star (BNS) system merges, there is an accompanying burst of light from matter outside the combined event horizon. For this reason, it is known as a \say{bright siren}. If the flash can be observed, the host galaxy is identified and one can use its redshift in Eq.~\eqref{eq:gravwave}. The event GW170817 was just such a BNS merger \citep{Abbott2017}. The gravitational wave was measured in Hanford and Livingston, which was enough to locate the sky position to 28 $\mbox{deg}^2$ (see next paragraph for how). Given the search region, an optical counterpart was found in NGC 4993 at a distance of $\sim 40$ Mpc\footnote{As the host galaxy was relatively close, the peculiar velocity adjustment was significant at $\sim 10\%$ of the Hubble flow velocity and the adjustment needs some care \citep{Mukherjee2019, Howlett2019,Nicolaou2020}.}. Around 3000 cycles of the wave resolved the chirp mass in the detector frame as $\mathcal{M} = 1.197M_{\odot}$ to accuracy of 1 part in $10^3$, consistent with a BNS merger. The main remaining uncertainty is then the inclination angle $i$. Using a flat prior for $\cos{i}$, \citet{Abbott2017a} obtain $H_0 = 70.0^{+12}_{-8} \SI{}{\kilo\meter\per\second\per\mega\pc}$. \citet{Hotokezaka2019} improved this to $H_0 = 70.3^{+5.3}_{-5.0} \SI{}{\kilo\meter\per\second\per\mega\pc}$ by constraining $i$ with observations of the interaction of the ejecta with the ISM.   

For black hole mergers, no optical counterpart is generated, and these are called \say{dark sirens}. However, it is still possible to constrain $H_0$ using them if a probable redshift can be estimated. To do this, the relative amplitudes and time delay between detectors located around the globe are used to approximately determine the sky position \citep{SoaresSantos2019}. For example, if 3 detectors observe the wave with perfect accuracy, the two independent time delays and three measured amplitudes will in turn determine the two sky position angles, two polarisation amplitudes, and one phase lag between polarisations. Hence, simultaneous detections of an event are essential in narrowing the size of the region on the sky where the source is located. A reasonable prior for $H_0$ will constrain the redshift range, and hence determine a localisation volume. Given a suitably complete galaxy catalog with sufficient sky coverage (the best available currently being SDSS and DES), galaxies within this volume can be averaged over with a suitable weighting to determine a value for $H_0$ \citep{Schutz1986}. Although each individual event is not very accurate, there are many more black hole mergers than neutron star mergers, so the errors are competitive in aggregate. It is also straightforward to mix dark and bright sirens to produce a combined result \citep{Palmese2020}.

\begin{figure}[h!]
    \centering
    \includegraphics[width=\textwidth]{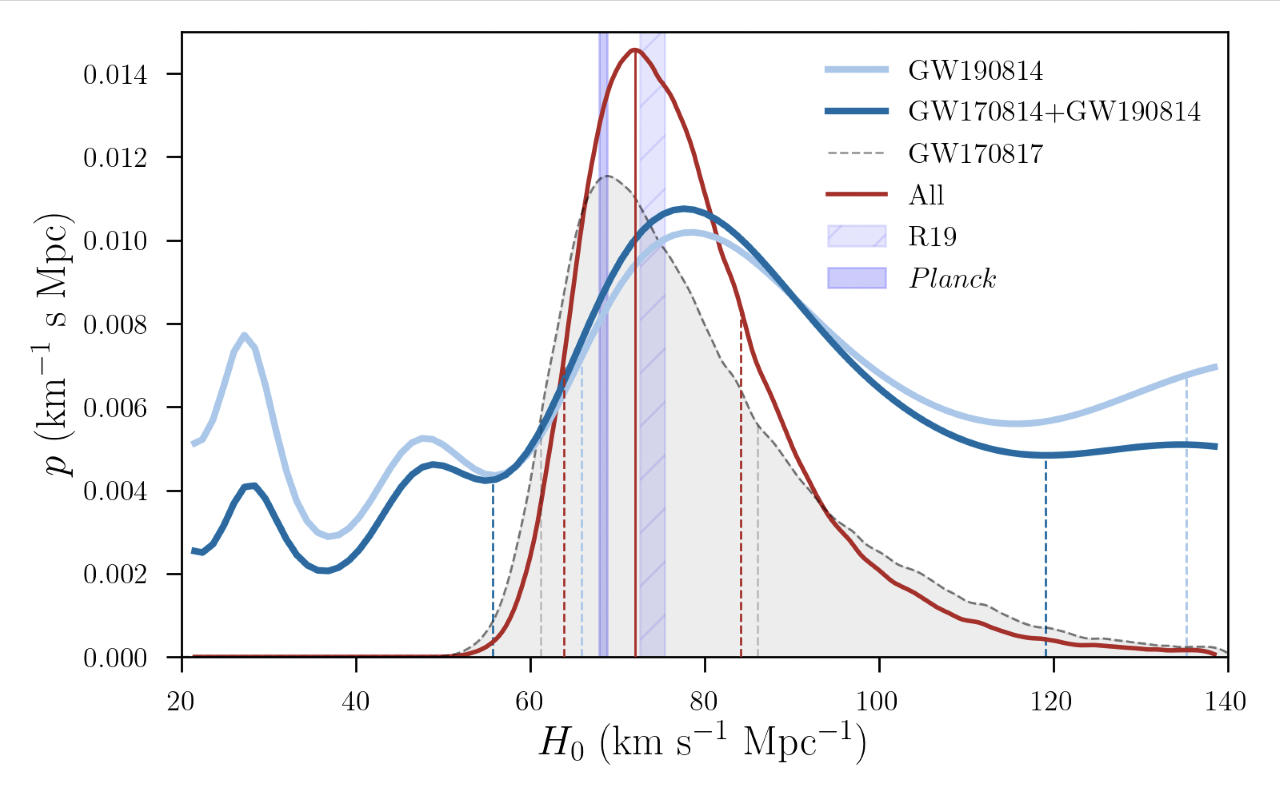}
    \caption{$H_0$ posterior distribution for the events GW190814 and GW170814, which was localised to a region in the DES survey footprint containing $\sim 1,800$ possible host galaxies. The light and dark blue lines represent the dark approach for each siren, which can be compared to the gray dashed line for GW170814 which was localised to one galaxy by its electromagnetic counterpart. The estimate for this bright standard siren can be improved by the addition of the dark sirens, as shown in dark red line. The posteriors of Planck \citep{PlanckCollaboration2018} and SH0ES \citep{Riess2019} are shown in purple boxes as a guide. Figure from \citet{Palmese2020}.}
    \label{fig:gravwavesh0}
\end{figure}

As each event's inclination angle is independent, averaging over $N$ events will improve the errors relative to a single event by a factor of $N^{-1/2}$. The main limiting factor is then detector calibration and sensitivity, which determines the event rates. Work is underway at the LIGO sites to improve this, using quantum engineering techniques such as squeezed light\footnote{Squeezed light is a state of light where the distribution of the Heisenberg uncertainty between observables is controlled for a specific application. In this case, the aim is to reduce photon shot noise at the expected frequency of the gravitational waves, and it is a nice crossover from techniques developed for quantum engineering to fundamental physics.}. \citet{Chen2018} estimate a 5yr observing run by the upgraded LIGO, Virgo, KAGRA and LIGO India detectors will be enough to measure $H_0$  to $1\%$ precision by 2030. 

In summary, the appeal of gravitation waves is that if one believes General Relativity correct, they are self-calibrating systems independent of the distance ladder and with minimal astrophysics input. Event rates and improvements in detector calibration are sufficient to converge to a $\sim 2\%$ precision within the next decade.

\subsection{Baryon acoustic oscillations}
Baryon acoustic oscillations (BAO) are the relic of density fluctuations propagating in the pre-recombination universe. The baryons in an initial localised seed of overdensity (such as would originate from inflation) are subject to radiation pressure and expand outwards at the speed of sound $c_s$, leaving the dark matter behind. At recombination, temperatures have dropped enough so that protons and electrons can combine to neutral hydrogen and the Thomson scattering rate quickly drops below the Hubble expansion rate. At the drag redshift $z_{\rm drag}$ shortly after, baryons are released from radiation pressure and the baryon overdensity shell is \say{frozen in} at a characteristic distance relative to the central dark matter overdensity. This distance
\begin{equation}
\label{Eq:soundhorizon}
    r_d  = \int_{z_{\rm drag}}^{\infty}\frac{c_s(z)}{H(z)} dz \;,
\end{equation}
in comoving coordinates is what is referred to as the sound horizon. As the universe continues to evolve, both the baryon shell and the central dark matter overdensity attract further gravitational infall of matter, forming galaxies. Thus, the sound horizon becomes visible as a characteristic (statistical) physical separation between galaxies. The sound speed $c_s = 1/ \sqrt{3(1+R)}$ depends on the baryon-to-photon ratio with $R = 3 \rho_b/4 \rho_\gamma$. $z_{\rm drag}$ has weak cosmological dependence, and Planck data gives $z_{\rm drag} = 1059 \pm 0.3$. The key point is that given a cosmological model, $r_d$ is calculable and its scale is large enough that the evolution of structure from $z_{\rm drag}$ to $z \sim 1$ is nearly linear. Thus we can compare observations of galaxy clustering in the late universe to $r_d$. Armed with data for $r_d$ we can use Eq.~(\ref{Eq:soundhorizon}) above to solve for $H_0$ in our model. 

The theory of measuring BAO was largely worked out by the mid 1990s \citep{Feldman1994}. One assumes galaxies are a Poisson sample of the relative matter density field $1+\delta(\Vec{r})$ and so
\begin{equation}
    P(\mbox{Volume element $\delta V$ contains a galaxy}) = \delta V \bar{n} (1+\delta(\Vec{r})) ,
\end{equation}
where $\bar{n}$ is the expected mean space density of galaxies. It is possible to work either with the two-point correlation function $\xi(\Vec{r}) \equiv \langle \delta(\Vec{r}^{'}) \delta(\Vec{r}^{'}+\Vec{r}) \rangle$ or equivalently the power spectrum
\begin{equation}
    P(\Vec{k}) \equiv \int d^3 r \xi(\Vec{r}) \exp{(i \Vec{k}\cdot \Vec{r})} .
\end{equation}
It might seem preferable to use the power spectrum, as we are looking for a characteristic wavelength, and the modes of the power spectrum are independent if the density field is Gaussian. Also, the galaxy survey footprint in real space becomes an easily understood point spread function in Fourier space. In practice though, because of issues with binning such as cut-offs in Fourier space, most research papers calculate both the correlation function and power spectrum. Adjusting $\bar{n}$ above can correct for bias introduced by edge effects of angular and redshift cuts of the survey.

Expanding the power spectrum in spherical harmonics, the monopole determines $d_A^2 H^{-1} (z)$, and the quadropole determines $d_A H (z)$ \citep{Padmanabhan2008} in a given redshift bin centred at $z$. With $\Delta \theta$ and $\Delta z$, the density peak separation transverse to and along the line of sight, we can solve for $d_A$ and $H$. We start with a mock galaxy catalog constructed in a fiducial cosmology seeded with density perturbations. This is matched to our real data with stretches $\alpha_{\perp}, \alpha_{\parallel}$:
\begin{align}
\label{eq:alphaperp}
    \alpha_{\perp} &= \frac{d_M(z) r_{d,\rm fid}}{d_{M,\rm fid}(z) r_d} & \alpha_{\parallel} = \frac{H_{\rm fid}(z)r_{d,\rm fid}}{H(z) r_d} \\
    \Delta \theta &= r_d / d_{M}(z) \\
    \Delta z &= H(z) r_d ,
\end{align}
where $d_M = (1+z) d_A$ is the comoving angular distance. Here $\alpha = \alpha_{\perp}^{2/3} \alpha_{\parallel}^{1/3}$ is the monopole term and $\epsilon = \alpha_{\perp}/\alpha_{\parallel}$ is the quadropole. 

There is the suggestion of circularity in this: a random catalog constructed in $\mathrm{\Lambda CDM}$ (with small scale power suppressed) is then \say{tweaked} to compare to the real sky to justify $\mathrm{\Lambda CDM}$. However, this method is essentially perturbative. If the real cosmology is far from the fiducial one, a bias might be introduced, but if it is close it should work well. 

A technique called \textit{reconstruction} is used to sharpen the BAO signal \citep{Eisenstein2007, Anderson2012}, and we describe it briefly here as we will mention it in Sect.~\ref{sec:four}, when we discuss modified gravity. The observed positions of galaxies will be somewhat moved from their original positions \say{on} the sound horizon, due to their subsequent infall towards overdensities. This has the effect of blurring the peak of the correlation function, and reduces precision. Reconstruction reverses the infall in the following way: (a) take a volume and smooth out small scale structure $< 20$ Mpc (b) embed it into a larger scale random structure to remove edge effects (c) estimate the displacement $\Vec{q}$ of each galaxy by the continuity equation $\nabla \cdot \Vec{q} = -\delta_{\rm gal}/b_{\rm gal}$ where $\delta_{\rm gal}$ is the fractional galaxy overdensity and $b_{\rm gal}$ the galaxy bias (d) shift the galaxies by $-\Vec{q}$ plus an additional shift for redshift space distortions (e) do the same to the mock catalog. 

Historically, the first galaxy surveys to report a BAO detection were the Two Degree Field survey \citep{Cole2005}, and the Sloan Digital Sky Survey \citep{Eisenstein2005}, followed by WiggleZ \citep{Blake2011,Blake2012} and others. We discuss here results from Baryon Oscillation Spectroscopic Survey (BOSS) \citep{Anderson2012}. BOSS has spectroscopically surveyed 1.5 million luminous galaxies over 10,000 deg$^2$ between $0.2 < z <0.7$, with angular separations greater than 62\hbox{$^{\prime\prime}$} (which is 0.5 Mpc at $z \sim 0.7$). \citet{Alam2017} analyse the BAO signal in Data Release 12. The galaxies are divided into three bins each of redshift and $\cos{\theta}$. From a fiducial cosmology with $h = 0.676$, they find $\alpha = 1.042 \pm 0.016$ after reconstruction. Their result is summarised as a set of three values for $d_m (z_i)$ and $H(z_i) \times r_d/ r_{d,\rm fid}$ for $z_i =0.38, 0.51, 0.61$, with precision $\sim 2.5\%$.

\begin{figure}[ht!]
    \centering
    \includegraphics[width=\textwidth]{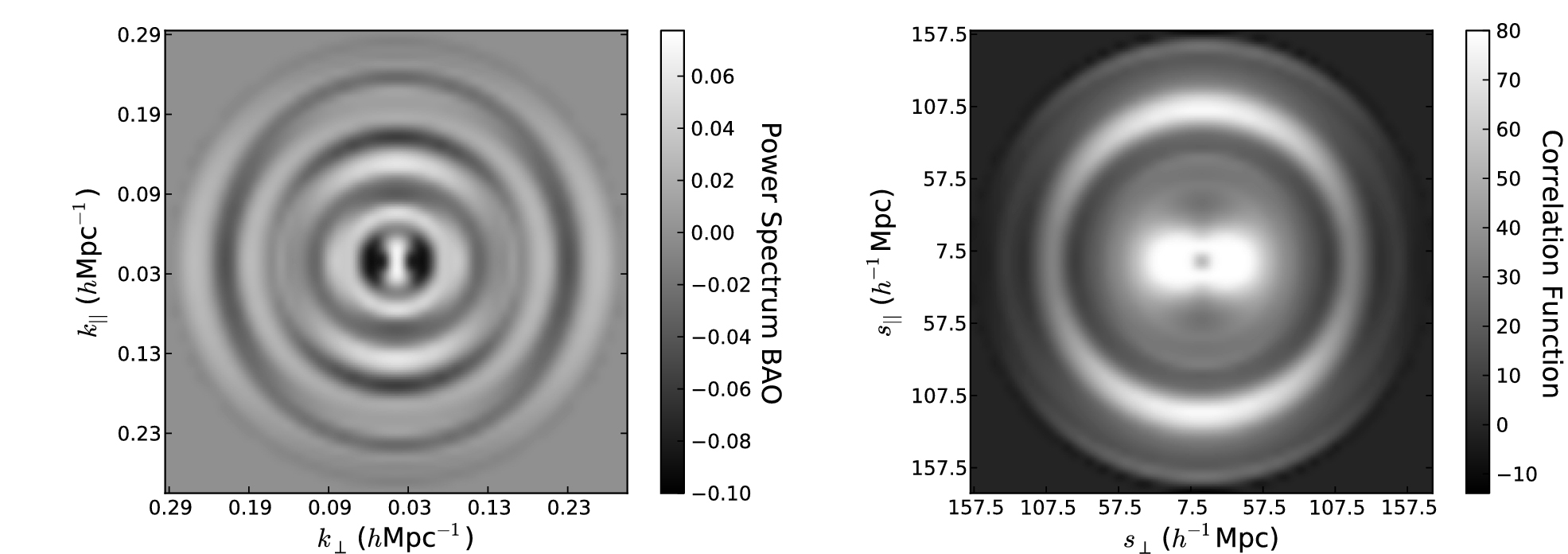}
    \caption{A beautiful illustration of the BAO peak in the redshift bin $0.4 < z< 0.6$ of BOSS DR12. The left hand panel shows the power spectrum along and transverse to the line of sight, and the right hand is the two-point correlation function. The anisotropy visible in the graphs breaks the $d_A^2 H^{-1}$ degeneracy. Figure is from \citet{Alam2017}.}
    \label{fig:BAO ripples}
\end{figure}

The construction of mock galaxy catalogs is a key part of this analysis. An important parameter is $b_{\rm gal}$ which is the (assumed linear) ratio of the galaxy overdensity to the matter overdensity. Constructing a mock galaxy catalog by brute force of N-body simulations is prohibitively expensive. Instead, construction starts with a fast, approximate gravity solver for the matter density field seeded with initial fluctuations, and resolution down to halo size. Galaxies are inserted into the matter halos using the bias ratio, while preserving the two- and three- point correlation functions. BOSS assumed $b_{\rm gal}$ is constant, but more sophisticated models exist. For example, bias tends to be higher for red galaxies (see for example \citealt{Zehavi2011}). Attention also needs to be paid to what types of galaxies are being counted: massive luminous galaxies of the type measured by BOSS reside in high density nodes of the cosmic web, whereas emission line galaxies are strung out along the filaments. For further details, see \citet{Kitaura2016}. BOSS DR12 uses two independent mock catalogs to mitigate against biases introduced by simulation effects.

BAO in isolation do not constrain $H_0$, but rather the combination of $H(z)$ and the sound horizon $r_d$ described above. To obtain $H_0$, we must either fit a cosmological model with constraints on its density parameters supplied by other datasets (for example SN Ia, CMB, or even the full shape of the matter-power spectrum), or we must supply a prior for $r_d$ (which is often obtained from the CMB, but see also below). We describe three independent results below. 

\citet{Alam2017} used $\mathrm{\Lambda CDM}$ calibrated to a combination of the Planck power spectrum and the JLA sample of SN Ia. They find $H_0 = 67.6 \pm 0.5 \;\SI{}{\kilo\meter\per\second\per\mega\pc}$ . This value is consistent with the Planck result \citep{PlanckCollaboration2018}, but is not independent of it. 
 
\citet{Lemos2019} use a parametric formula for $H(z)$, in conjuction with a WMAP-derived prior for $r_d$ and the Pantheon SN Ia sample, to derive $H_0 = 67.9 \pm 1.0 \;\SI{}{\kilo\meter\per\second\per\mega\pc}$. This result is now independent of Planck, and their parametric $H(z)$ allows for deviation from $\mathrm{\Lambda CDM}$ by separating early and late universe expansion histories. 

\citet{Addison2018} use data (almost) independent of CMB in $\mathrm{\Lambda CDM}$. They determine $\Omega_m = 0.292 \pm 0.02$ by combining galaxy and Ly$\alpha$ BAO\footnote{In the analysis of \citet{Addison2018} the two data sets are in mild tension, however recent data has a greater level of compatibility - see for example Figure 5 of \citet{Alam2021}.}. $\Omega_r$ is derived from the CMB temperature as measured by COBE/FIRAS and numerous balloon experiments \citep{Fixsen2009}. Setting $\Omega_k = 0$ determines $\Omega_{\Lambda} = 1 - \Omega_m - \Omega_r$. To get the sound horizon, we now just need the baryon density $\Omega_b$ (see Eq.~\eqref{Eq:soundhorizon}). Without using the CMB power spectrum, the best available way to get it is from primordial deuterium abundances. In Big Bang nucleosynthesis (BBN), deuterium is burned to create $^{4}$He, with some uncertainty due to the reaction rate of $d(p,\gamma)^3$He. The reaction rate increases with physical baryon density so [D/H] decreases with $\Omega_b h^2$. The authors use the primordial [D/H] abundance measured from metal-poor damped Ly-$\alpha$ systems, which is $2.547 \pm 0.033 \times 10^{-5}$. Putting this all together, they derive $H_0 = 66.98 \pm 1.18 \; \SI{}{\kilo\meter\per\second\per\mega\pc}$. The only CMB data that has been used is the temperature, and the result corroborates the Planck value\footnote{Standard $\mathrm{\Lambda CDM}$ is assumed, and in particular that the expansion rate around the time of BBN as constrained by He abundances (parametrised by the effective number of relativistic species in the early universe $N_{\rm eff}$) is consistent with the CMB.}. Repeating the analysis with galaxy data from the Dark Energy Survey (DES) Year 1 produces a consistent result \citep{Abbott2018}.

These results are consistent with more recent ones. DES Y3 survey data has been used to derive a set of three correlation measures (referred to as \say{3x2pt}). These measure the angular correlation of galaxy positions with each other, with the tangential shear of their shapes caused by weak lensing, and the cross-correlation in the two components of their ellipticities also due to weak lensing. Combining this with Planck, SN Ia data, BAO and redshift space distortions, the \citet{Abbott2021} find $H_0 = 68.0^{+0.4}_{-0.3} \;\SI{}{\kilo\meter\per\second\per\mega\pc}$ when fitting to the $\mathrm{\Lambda CDM}$ model. The eBOSS survey, the culmination of more than 20 years of survey work at the Apache Point Observatory, find $H_0 = 68.19 \pm 0.37 \;\SI{}{\kilo\meter\per\second\per\mega\pc}$ when fitting to a $\mathrm{\Lambda CDM}$ model using their data plus the above additional probes to also fit to $\mathrm{\Lambda CDM}$ \citep{Alam2021}. These seem to be the tightest constraints on $H_0$ yet published.

We also note that the logic explained above to derive $H_0$ can be reversed: BAO can be combined with a $H_0$ prior from nearby universe results, instead treating the sound horizon $r_d$ as a free parameter \citep{Bernal2016}. The $r_d$ values so inferred are in tension with the CMB, and we return to this point in Sect.~4.2. 

In summary, BAO serve as a standard ruler in the late universe which are almost independent of astrophysical assumptions. They may be used in extended models beyond $\mathrm{\Lambda CDM}$, albeit with the caveat they have been derived perturbatively against a $\mathrm{\Lambda CDM}$ background simulation. The two main data sets of BOSS and DES are soon to be joined by others. The Dark Energy Spectroscopic Instrument is operational, the ESA survey satellite Euclid is scheduled for launch in 2022 which is also when science operations on the Vera Rubin Observatory LSST in Chile commence. All of these surveys are complementary and will increase the precision and range of BAO measurements. Combined with deuterium abundances, BAO are an important corroboration of the Planck and WMAP results for $H_0$.

\subsection{Cosmic microwave background}
The CMB power spectrum carries the imprint of the same acoustic oscillations we described for BAO, sampled on the surface of last photon scattering. The Planck 2018 value of $H_0 = 67.4 \pm 0.5 \; \SI{}{\kilo\meter\per\second\per\mega\pc}$ \citep{PlanckCollaboration2018} is notable for its precision, and is consistent with previous results such as  Boomerang $64 \pm 10 \; \SI{}{\kilo\meter\per\second\per\mega\pc}$ \citep{Percival2002} and WMAP $70.0 \pm 2.2 \;\SI{}{\kilo\meter\per\second\per\mega\pc}$ \citep{Hinshaw2013}. More recently, the Atacama Cosmology Telescope (ACT) finds $67.9 \pm 1.5 \;\SI{}{\kilo\meter\per\second\per\mega\pc}$  \citep{Aiola2020}. There is, however, a moderate tension between Planck and the South Pole Telescope (SPT) result of $71.6 \pm 2.0 \;\SI{}{\kilo\meter\per\second\per\mega\pc}$, and the origin of the difference is not yet clear \citep{Henning2018, Handley2020}. 

Two questions are immediately raised: how is $H_0$ derived from a power spectrum sampled on the two-dimensional sky of the early universe, and why is the Planck value so precise?

\begin{figure}[ht!]
    \centering
    \includegraphics[width=0.9\textwidth]{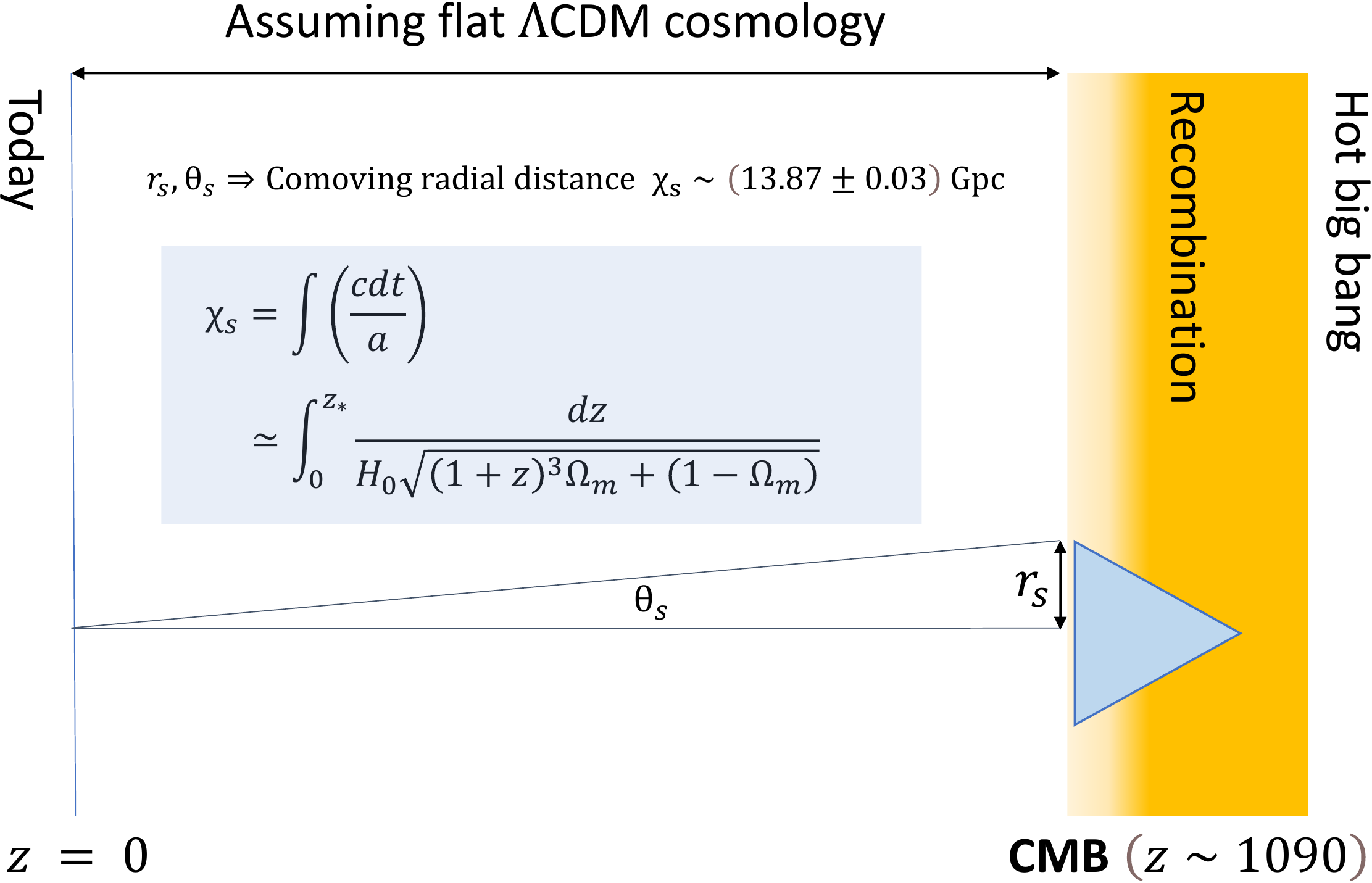}
    \caption{A schematic of how the sound horizon determines $H_0$. We observe the angular peak spacing $\theta_s$, and calculate $r_s$ from the pre-combination universe given a model for the sound speed and expansion history there. The comoving distance $\chi_s$ is then used to solve for $H_0$ given how $\chi_s$ evolves in the post-recombination universe. The box gives the flat $\mathrm{\Lambda CDM}$ formula for this. This figure is our version of Antony Lewis' original at https://cosmologist.info}
    \label{fig:H0 CMB}
\end{figure}

\begin{figure}[ht!]
    \centering
    \includegraphics[width=\textwidth]{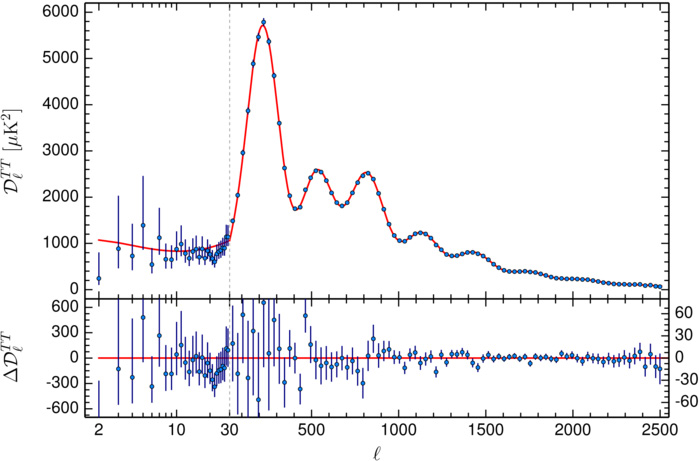}
    \caption{The binned Planck power spectrum for TT and residuals to the best fit $\mathrm{\Lambda CDM}$ model. The vertical line delineates the different methodology used to resolve the power spectrum at $l<30$. Note that the figure  shows the power spectrum of the temperature T, not of the anisotropies  $\Theta$. Figure from \citet{PlanckCollaboration2018}.}
    \label{fig:Planck Spectrum}
\end{figure}

To answer the former, we need to explain how the CMB power spectrum determines cosmological parameters. The spectrum is created by primordial fluctuations that evolve as a function of their scale and the sound speed, until recombination. Whilst this radiation is highly isotropic, we can obtain cosmological results from measurements of small anisotropies in both the temperature and the polarization of CMB photons. The former of these provides the largest amount of cosmological information, and can in fact measure the Hubble constant at high enough accuracy to be in tension with direct measurements without relying on polarization. These temperature anisotropies ($\Delta T \equiv T(\vec{n}) - T_0$) can be expanded in a basis of spherical harmonics: 
\begin{equation}
    \Delta T (\vec{n}) = \sum_{\ell=0}^{\infty} \sum_{m = - \ell}^{\ell} a_{\ell m} Y^{m}_{\ell} (\vec{n}) \;. 
\end{equation}
The power spectrum of the coefficients is commonly expressed as $\mathcal{D_{\ell}} = \ell(\ell+1) C_{\ell}/ 2\pi$, where
\begin{equation}
    C_{\ell} \equiv \frac{1}{2\ell+1} \sum_{m} a_{\ell m} a_{\ell\, -m}
\end{equation}
is the estimator of the power spectrum (we have only one CMB, so averaging over $m$ is a way to estimate the true \say{universe average} we want). The spectrum is shown in Fig.~\ref{fig:Planck Spectrum} and contains all the statistical information of the temperature anisotropies, as long as these are Gaussian (no evidence supporting non-Gaussianity in the CMB has been detected at the time of writing). This power spectrum has three main features: 

\begin{itemize}
    \item \textbf{Large-scale plateau} at scales ($\ell < 100$) that are not affected by post-recombination physics, and therefore reflect the primordial power spectrum. As noted above, we have just one \say{sampling} of the random seeds, so the amount of information we can extract from low multipoles is limited by cosmic variance $\propto (2\ell+1)^{-1/2}$.  
    \item \textbf{Acoustic oscillations} which depend on the sound horizon size
    \begin{equation}
    \label{rs}
        r_s(z_\star) = \frac{1}{H_0 \Omega_m^{1/2}} \int_{0}^{a_\star} \frac{c_s}{\sqrt{a+a_{\rm eq}}}da
    \end{equation}
    in flat ${\rm \Lambda CDM}$ where $a_{\rm eq} = \Omega_r / \Omega_m$,  $\Omega_b h^2$ determines $c_s$ and both determine $a_\star = a(z_\star)$, the scale factor at the surface of last scattering. These create the peaks in the spectrum. 
    \item \textbf{Silk damping} which is photon diffusion from hotter regions to colder ones, suppressing small scale power \citep{Silk1968}. The mean free path of a photon is $\lambda = 1/(\sigma_T n_e)$ where $n_e$ is the number density of charged particles and $\sigma_T$ the Thomson cross-section, so photons can diffuse a length $r_{\rm Silk} = \sqrt{\lambda/H}$. As the sound horizon $r_s \propto 1/H$, then $r_{\rm Silk}/r_s \propto \sqrt{H/n_e}$. Silk damping creates the downward slope in the spectrum, so this is sensitive to $n_e \propto \Omega_b$.
\end{itemize}
With all of this, how does the CMB constrain the Hubble parameter? Firstly, we can measure the acoustic scale length $r_s$, which serves as a standard ruler. As described in Eq. ~\ref{rs}, this can be calculated as a function of $\Omega_r$, $\Omega_m h^2$ and $\Omega_b h^2$: The radiation density $\Omega_r$ is determined by the temperature of the CMB. $\Omega_m h^2$ can be estimated from the effect of the integrated Sachs-Wolfe effect~\citep{1967ApJ...147...73S} -- the redshift of CMB photons through wells or hills of gravitational potential during eras of non-matter domination -- in the low multipoles of the power spectrum. Finally, the relative amplitudes of the acoustic peaks in the CMB serve to measure the ratio $\Omega_b/\Omega_m$. Once we have computed the acoustic scale, we need a measurement of its comoving angular diameter distance $\theta_s$, which is given by the distance between the acoustic peaks\footnote{This is to speak loosely - if we just knew the location of the peaks to $1\%$ accuracy then the accuracy of $H_0$ would be only $\sim 7\%$. Using the entire spectrum is the driver for the precision. For a very accessible account of the CMB power spectrum with semi-analytic equations we recommend \citet{Mukhanov2004}.}. In a flat $\mathrm{\Lambda CDM}$ model, this provides a measurement of the matter density $\Omega_m$, which combined with the measurement of $\Omega_m h^2$ gives an estimate on the Hubble parameter. This is illustrated graphically in Fig.~\ref{fig:H0 CMB}.

Planck measured full-sky temperature $T$ and polarisation $E$ fluctuations in 9 frequencies, with angular resolution up to 5 arcmins ($\ell \simeq 2000$) and sensitivity $2 \times 10^{-6}\mu$K. Before calculating the $TT, TE$ and $EE$ power spectra, foreground effects must be calculated and removed, and much of the work done by the Planck collaboration between successive data releases has been to develop the foreground model, and reduce the uncertainties it may introduce. Foreground effects include :
\begin{itemize}
    \item \textbf{Peculiar velocity of the earth-sun system} induces a dipole which is easily subtracted. The annual motion of the earth is actually helpful to calibrate the detector response.
    \item \textbf{Galactic synchrotron and free-free emission} caused by cosmic rays interacting with the magnetic field and ISM of the MW. This has a non-thermal frequency spectrum (in effect its colour is different to the CMB).
    \item \textbf{Galactic dust emission} whose contribution to the spectrum is also subtracted by estimating its mapped \say{colour} difference to the CMB. 
    \item \textbf{Point sources} of microwave emission are masked. Sky masking introduces some correlation between multipoles.
    \item \textbf{Sunyaev--Zel'dovich (SZ) effect} where CMB photons interact with hot intracluster gas and are Doppler-shifted (the kinetic SZ effect) by bulk motion or up-scattered (the thermal SZ effect) \citep{Sunyaev1972}. The former effect does not alter the black-body spectrum, but the latter does. The kinetic SZ effect is sensitive to epoch of re-ionisation at $z \sim 8$ \citep{Ade2014a}. The thermal SZ effect may be exploited to produce a cluster map from which information on inhomogeneity can be extracted \citep{Ade2014}. Modelling of intracluster gas predicts a spectral template (which peaks around $\ell \simeq 2000$) to subtract.
    \item \textbf{Weak lensing} of the CMB by matter. Weak lensing peaks between $1<z<2$, and may be used as a probe of the distribution of foreground matter. The effect increases as we move to smaller scales, particularily $\ell > 2000$ \citep{Zaldarriaga1998}. There are two signals of lensing: (a) its non-gaussianity, which can be measured by a four-point correlation function (b) a smoothing effect on the peaks and troughs of the spectrum, both of which can be calculated as a function of $\Omega_M$. However, there is more smoothing apparent in the spectrum than is predicted by the correlation function so a phenomenological parameter $\Omega_{m,\rm spectrum} = A_{\rm lens} \Omega_{m,\rm correlation}$ has been introduced to capture their relative amplitude. $A_{\rm lens} \simeq 1.2 \pm 0.1$ is seen in both Planck and South Polar Telescope (SPT) data \citep{Bianchini2020}, but appears to be absent in the Atacama Cosmology Telescope (ACT) data \citep{Aiola2020}.
\end{itemize}
The latter two are perhaps better characterised as secondary anisotropies, as they have been used to derive cosmological information in their own right. The secondary anisotropies then provide a consistency check of certain parameters derived from the primary anisotropies, if the modelling is correct. We examine $A_{\rm lens}$ in greater depth below.

Cross-checks on the effectiveness of foreground modelling are done by switching frequency bands, using alternative astrophysical models, and numerical simulations of parameter recovery from random CMB backgrounds \citep{Ade2016, PlanckCollaboration2018}. The spectrum is then compared to one computed in a standard code such as CMBFAST \citep{Seljak1996}, CAMB \citep{Lewis2000} or CLASS \citep{Blas2011} and a posterior for each cosmological parameter is computed. In each check, no serious discrepancy with the main results was found. Although each $C_{\ell}$ individually has $\simeq 7\%$ noise, there are 2,000 of them and the error is reduced further by the lensing signal. The binned spectrum and residuals is shown in Fig.~\ref{fig:Planck Spectrum}. We now consider two questions regarding the data.

Is Planck self-consistent? A feature which is (barely) visible by eye in Fig.~\ref{fig:Planck Spectrum} is an oscillating residual pattern lining up to the spectral peaks, indicating the Planck spectrum is slightly tilted versus theory (see Fig.~24 of \citet{PlanckCollaboration2018} for an enlarged view). As each multipole is nearly independent, we can split the spectrum and run the analysis on each half. \citet{Addison2016} find that $A_{\rm lens}$ is equivalent to a drift of cosmological parameters with different scales in the CMB. They fix $A_{\rm lens} =1$ and compare $H_0$ for $\ell<1000$ and $\ell>1000$ finding $H_0 = 69.7 \pm 1.7 \; \SI{}{\kilo\meter\per\second\per\mega\pc}$ and $H_0 = 64.1 \pm 1.7 \; \SI{}{\kilo\meter\per\second\per\mega\pc}$ respectively. Allowing $A_{\rm lens} = 1.3$, or a tilt parameter restores concordance. 

We caution that $A_{\rm lens}$ is not a physical parameter connected to lensing; it is a \say{fudge factor} that resolves internal inconsistency in CMB data. It does not resolve the $H_0$ tension, but does resolve matter power amplitude $\sigma_8$ tension with galaxy survey data such as the Dark Energy Survey and Kilo-Degree Survey. Could it be a foreground effect? Its value is relatively stable for various band channels and sky masks, which argues that it is not. The Planck Collaboration followed up on this curiosity \citep{PlanckCollaboration2018}. They note a dip in the power spectrum for $l<30$ pulls low-$\ell$ $H_0$ higher; this may account for the lower-resolution WMAP measurement being somewhat higher than Planck. The dip between $1420 \leq \ell \leq 1480$ mimics lensing but may be an unaccounted foreground effect. Potential explanations such as negative curvature density $\Omega_k <0$ and modified gravity are investigated but no convincing evidence is found in favour of these models when BAO data is added to the fit. A re-analysis of the Planck data using a different foreground subtraction methodology \citep{Efstathiou2019} found that $A_{\rm lens}$ decreased from $1.26$ to $1.06$ as the data was broadened from the temperature-only power spectrum to include polarisation and lensing data. Further, Planck data passes other consistency tests with the baseline $\mathrm{\Lambda CDM}$ model where $A$-type parameters may be introduced, such as the expected magnitude of the lensing correlation function and the Sachs-Wolfe effect. Hence, the preferred explanation of the Planck Collaboration is that $A_{\rm lens}$ is a statistical fluke. Nevertheless, while  $A_{\rm lens}$ does seem to vary over the sky, it is perhaps of concern that it seems to be larger for full-sky Planck data is and is a roughly $2\sigma$ effect.

Is Planck consistent with other current CMB measurements? The SPT covers a $6\%$ patch of the southern sky at a resolution 6$\times$ that of Planck, finding $H_0 = 71.3 \pm 2.1\; \SI{}{\kilo\meter\per\second\per\mega\pc}$ \citep{Henning2018}. SPT is limited to $\ell >650$ due its low sky coverage and atmospheric effects and has greater noise, but when the results are compared on a like-for-like basis (\say{Planck-in-patch} with SPT for $650 < \ell < 1800$), they are consistent with Planck. The analyses are independent and so argue against systematic errors (in this patch and multipole range) in either experiment. However, including higher SPT multipoles in the 150 GHz spectrum causes $H_0$ to drift higher: $H_0 \simeq 74 \pm 3\; \SI{}{\kilo\meter\per\second\per\mega\pc}$ for $650 < \ell < 3000$ \citep{Aylor2017}. While this is the opposite direction to Planck data, this may be due to the $l<30$ effect in Planck noted above. SPT confirm the Planck result of a greater lensing effect than predicted, and also note a spectral tilt versus the best-fit $\mathrm{\Lambda CDM}$. However, the trend is not apparent in 143 GHz data, so it is not at all clear if this is a physical effect, rather than merely chance or systematics. The ACT Data Release 4 derives cosmological parameters from a $6,000 \rm deg ^2$ patch of the southern sky up to $\ell \sim 4000$, finding $H_0 = 67.9 \pm 1.5\; \SI{}{\kilo\meter\per\second\per\mega\pc}$ \citep{Aiola2020}, however ACT appears to be in $2.6\sigma$ tension with Planck \cite{Handley2020}; this appears to be caused by differing physical baryon densities.

In summary, Planck 2018 results for $H_0$ are consistent with previous and current independent CMB data. The high precision for $H_0$ is a consequence of Planck's high resolution, lack of atmospheric interference, and full sky coverage. Checks have been performed on the codes, the methods used to remove foregrounds, and beam effects. Internal inconsistencies hint at new physics or foreground effects at small angular scales, manifesting either as a tilt or a smoothing of the spectrum, and non-physical parameters like $A_{\rm lens}$ have been introduced to capture them. However after their introduction the results for $H_0$ are not materially affected, and remain in tension with late universe results. Nevertheless, the persistence of these effects across differing sky patches, multipole ranges and experiments make them less likely to be a statistical fluke. It is hoped next generation CMB experiments with resolution up to $\ell=5000$ will shed further light on this. 

\subsection{Other methods}
Promising methods that we did not have space to discuss are Surface Brightness Fluctations \citep{Tonry1988, Blakeslee2021, Khetan2020}, Cosmic Chronometers \citep{Jimenez2002, Moresco2018}, the Tully--Fisher relation \citep{Tully1976, Kourkchi2020, Schombert2020}, Type II supernovae \citep{DeJaeger2020}, HII galaxies \citep{Terlevich1981, Arenas2018}, and Galaxy Parallax \citep{Croft2021}. Quasars \citep{Risaliti2019} and GRBs \citep{Schaefer2007} offer the prospect of extending the Hubble diagram up to $z \sim 5$, further testing $\mathrm{\Lambda CDM}$. However, despite improvements in characterising their intrinsic luminosity from observables (for example X-ray and UV flux in the case of quasars), it remains challenging at present to regard them as standard candles. We refer the reader to the above for the latest work in these fields. 

\section{Could new physics explain the tension?}
\label{sec:four}

We have now reviewed the main recent results on $H_0$, and while in the spirit of scientific scepticism we have identified potential sources of systematic error, we hope it is also clear how much effort has been made to root out potential biases. Therefore, a possibility that should be taken seriously is that the tension is real, and a sign of new physics. 

Some authors have interpreted the disagreement between CMB results and Cepheids as \say{early versus late} tension. In this view, $\mathrm{\Lambda CDM}$ with its fluids of radiation, standard model neutrinos, cold dark matter, baryons and dark energy is not the right model to derive $H(z=0)$ from the apparent $H(z \sim 1100)$. If late-time physics were changed, $H_0$ derived from the early universe might be reconciled to the local one. Alternatively, a change to pre-recombination physics would alter our calculation of the sound horizon $r_s$, and hence we would need to change distances to retrieve the same angular sizes of the CMB temperature fluctuations. More radically, some authors have argued our local $H_0$ is different from an average over randomly placed observers.

The essential problem for model builders is this: in other respects, $\mathrm{\Lambda CDM}$ fits the data very well across a huge span of the history of the universe. The CMB is not merely a \say{snapshot} of the universe, but carries the imprint of several epochs. The positions and heights of the peaks are sensitive to the content\footnote{In $\mathrm{\Lambda CDM}$, this is parametrised by $r_{\rm eq}$, the size of modes crossing the horizon at matter/radiation equality $z \sim 3300$.} between $z \sim 10^5$--$10^3$, the high-$\ell$ slope depends on the baryon-to-photon ratio close to recombination at $z \sim 1100$, and CMB lensing by matter peaks between $z \sim 1$--$2$. At earlier times, $\mathrm{\Lambda CDM}$ makes an accurate prediction of primordial element abundances, especially deuterium. At late times, galaxy surveys are consistent with the evolution of perturbations in a $\mathrm{\Lambda CDM}$ background\footnote{With the caveat of a moderate tension between the CMB and BAO values of the amplitude of fluctuations in the matter-power spectrum, parametrized by $\sigma_8$.}, and place tight limits on deviation from spatial flatness in the mid to late universe. The shape of $H(z)$ can be read off SN Ia luminosities between $0.01 < z < 1.4$, and is also consistent with $\mathrm{\Lambda CDM}$.

What we want to have is a model capable of modifying $H_0$ by the right amount, without disrupting $\mathrm{\Lambda CDM}$'s other successful predictions. The model must be testable, compatible with particle physics results, and ideally not fine-tuned. Two serious problems must be avoided : by dint of extra parameters, most models increase the range of allowed $H_0$ values. When a prior of $H_0$ from the local universe is convolved with the expanded likelihood, the posterior will -- by construction -- overlap with the prior. So to claim a resolution of the Hubble tension in this way is to use the same data twice: once to construct the posterior, and once to compare it. This can be avoided if the extra parameters have some preferred values -- for example, a range that is predicted by microphysics \citep{Vagnozzi2020}. A second problem with the use of $H_0$ priors has been discussed by \citet{Efstathiou2021}, who points out the difference between it and the actual data analysis performed by the SH0ES team, which is a calibration of the SN Ia fiducial absolute magnitude $M_B$, followed by a conversion from $M_B$ to $H_0$. 
To use an $H_0$ prior is then to transform and compress the data away from its true source; it would instead be better to use $M_B$ directly and priors for this are given in \citet{Camarena2021}. 
Another prevalent issue is the selective use of datasets -- in particular, BAO provide strong constraints on the late universe as we shall see shortly, and to omit them is again to risk a misleading analysis. 

A large number of creative proposals have been put forward, so much so that unfortunately we do not have space to introduce them all. Our preference here is to review selected models in broad classes that we hope will be informative, and readers are referred to extensive reviews by \citet{Mortsell2018}, \citet{Knox2019}, \citet{DiValentino2021}, and \citet{Vagnozzi2020} if they would like more detail.

\subsection{Late-universe physics}
In some respects, it is easier to propose changes to the late universe: one does not have to \say{negotiate} with the CMB power spectrum! It is straightforward to generalise the expansion history into any number of \say{cosmology-independent} forms, that do not rely on any specific form of matter-energy, or general relativity: all that is retained are the Copernican principles of homogeneity and isotropy, and the existence of a space-time metric (for example, see Eq.~\eqref{Eq:LumDist2}). The latter is important, as it implies the Etherington relation: $d_L = (1+z)^2 d_A$, which locks angular diameter distances against luminosity distances.

\subsubsection{Modified gravity}
Traditional models of large-scale modified gravity (normally referred to as MOND) would seem to have a low chance of solving the Hubble tension; the success of using reconstruction to sharpen the BAO peak argues that standard gravity works as we expect on these scales. \citet{Desmond2019} have proposed that small-scale modified gravity could distort the calibration of Cepheids in large galaxies like the Milky Way and NGC4258, however this appears to be disfavoured by the fact that using the LMC as a sole calibrator does not materially change $H_0$.

\subsubsection{Inhomogeneities and redshifts}
Clearly our universe is not completely uniform, so could we by chance live in an underdense region, measuring a local $H_0$ above the universal one? 

\citet{Shanks2019} have investigated the \say{local hole} idea, initially noted in galaxy survey data by \citet{Keenan2013}. They cite peculiar velocity outflows from $0 < z < 0.1$ in the 6dFGS galaxy survey as evidence for a local underdensity, finding the universal (ie CMB) $H_0$ to be lower than our local one by 1.8\%. However, we would expect to see the effect of a line of sight exiting our underdense region as a kink in the residuals of SN Ia luminosities fitted to $\mathrm{\Lambda CDM}$, or alternatively as an anistropy in $H_0$ depending on the footprint of the SN Ia survey used. While the sky distribution of low-$z$ SN Ia versus Hubble flow SN Ia are very different, and there is an uptick in magnitude residuals on the boundary between the two (see for example Fig.~11 of \citealt{Scolnic2018}), the lack of evidence for such a kink places limits on any local under-density \citep{Kenworthy2019}. Also, large-scale structure simulations find the probability for us to be in an underdensity of such magnitude is less than 1\% \citep{Odderskov2017, Macpherson2019} (although a different likelihood of a void may be obtained in a revision of $\mathrm{\Lambda CDM}$). That said, the observational tension between claims of local under-density based on galaxy surveys, and the lack evidence for it in SN Ia data, remains unexplained.

Density fluctuations also necessitate the adjustment of redshifts for peculiar velocities. Although heliocentric redshifts can be measured up to an accuracy of $10^{-7}$, care is needed in the conversion to cosmological ones. \citet{Davis2019} have analysed the effect of redshift biases on $H_0$. Changing $z$ by $+0.001$ for a set of standard candles between $0.01 < z < 0.15$ causes a bias in $H_0$ of $\sim 3\%$. Systematics might arise in peculiar velocity estimates or corrections for residual stellar motion (for example, if observing on one side of a spiral galaxy). \citet{Rameez2019} has pointed out differences between the redshifts for SN Ia in common between the JLA and Pantheon catalogs; there are 58 with differences $>0.0025$, concentrated in an arc opposed to the CMB dipole. Nevertheless, this does not appear to be biasing any $H_0$ estimates. 

\subsubsection{Modifying late-time $\mathrm{\Lambda CDM}$}
In $\mathrm{\Lambda CDM}$, dark energy is a property of the vacuum, with equation of state $\rho = wP$ where $w=-1$. This distinguishes it from standard inflation models, where $-1 < w < 0$ is a time-varying function of the potential and kinetic energy of a scalar field, but makes its microphysical origin rather obscure. This has caused some theorists to speculate with alternative models. 

In principle then, modifying dark energy in the late universe may be considered as a solution. The CMB has little to say about late-time dark energy: at early times, its physical density is much smaller than matter and radiation. The late time Integrated Sachs--Wolfe effect does imprint on the spectrum, but only as a large scale secondary anisotropy once the universe has entered the dark-energy dominated phase at around $z \sim 0.3$.

A simple modification of $\mathrm{\Lambda CDM}$ would be to allow $w$ to take on an arbitrary (constant) value. However as what is required is a boost to late-time expansion to match the local universe $H_0$, we would need $w<-1$ (known as phantom dark energy) to solve the tension. Phantom dark energy does not occur in standard single scalar field models, but is possible in models with more complex field configurations. If sustained, phantom dark energy leads to a \say{Big Rip} in a finite time, where all matter is pulled apart by accelerated expansion of the universe. 

Alternatively, dark energy may be considered a late-universe emergent phenomenon by allowing its equation of state to vary with time. This may be done writing $w(a) = w_0 + (1-a)w_a$, or by replacing $\Omega_{\Lambda,0}$ in Eq.~\eqref{eq:ez} with an phenomological $\Omega(z)$. This could be motivated by the action of scalar fields, dark matter decaying to dark energy, or some other novel microscopic theory of dark energy (see for example \citealt{DiValentino2021}). However, it can be shown that such models have a generic problem: the sound horizon. The sound horizon $r_d$ is normally introduced via its definition from early universe physics (Eq.~\ref{Eq:soundhorizon}). But it is also imprinted on the late-time universe, as the peak in the angular correlation function of galaxy number densities, which constrains the product $r_d H_0$. All that is needed is a late-time distance measurement to convert the angular size to a physical size at a given redshift, SN Ia to constrain the late-time expansion history, and $r_d$ may be calibrated \textit{independently} of the CMB.

\citet{Arendse2019} have used the H0LiCOW lens distances, BAO and Pantheon supernovae to obtain just such a late-time $r_d$. They show that two modified late-time dark energy models applied to Planck data can resolve its $H_0$ tension with the late-universe, but not its $r_d$ tension. \citet{Knox2019} investigated this in $\mathrm{\Lambda CDM}$ with the same result. 

Hence, if all data including BAO are considered -- as they should be -- it appears that late-time modifications to $\mathrm{\Lambda CDM}$ will not be satisfactory as a solution to the Hubble tension.

\begin{figure}[h!]
    \centering
    \includegraphics[width=\textwidth]{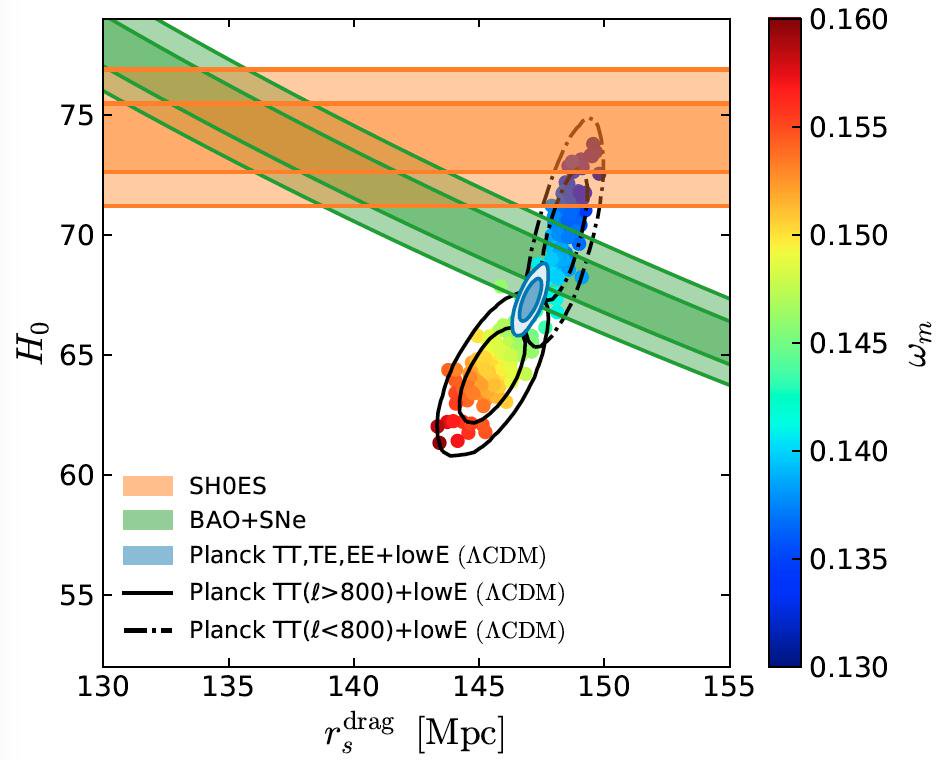}
    \caption{$\mathrm{\Lambda CDM}$ tensions in the $r_d - H_0$ plane. $\omega_m$ is the physical matter density $\Omega_m h^2$ and $r^{drag}_{s} \equiv r_d$ is the sound horizon from the end of the baryon drag epoch. There is no combination of values in standard, flat $\mathrm{\Lambda CDM}$ that can explain observations. Figure is from \citet{Knox2019}.}
    \label{fig:H0vsSound}
\end{figure}

\subsection{Early-universe proposals}
We have mentioned how the CMB is a set of rulers whose scale is fixed at different epochs. To recap, the observed angular scale of the sound horizon at a given redshift $z$ is
\begin{equation}
\label{eq:thetaz}
    \theta(z) = \frac{r_s}{d_A} = \frac{\int_{z_D}^{\infty} \frac{c_s (z') dz'}{H(z')}}{ \frac{1}{1+z} \int_{0}^{z} \frac{dz'}{H(z')}} \; ,
\end{equation}
where $c_s(\omega_b,z)$ is the sound speed which depends on the physical density of baryons, given by $\omega_b = \Omega_b h^2$. Decoupling happens when the mean free path of photons approaches the Hubble radius $c/H(z)$, hence $z_D$ is also a function of all of the densities $z_d \equiv z_d(\omega_{b}, \omega_{m}, \omega_{\gamma})$. Loosely, the goal is to reduce the sound horizon from (\ref{Eq:soundhorizon}) by $\simeq 7\%$, as shown in Fig.~\ref{fig:H0vsSound}.

We may isolate the free parameters of this formula, from those that are fixed by data as follows. As discussed, we may omit dark energy $\omega_{\Lambda}$ in the numerator. In the late universe the radiation density is not significant, and using the spatial flatness implied by BAO, we may write $\Omega_{\Lambda} = 1- \Omega_m$ .

$\Omega_m$ is determined by the heights of peaks in the CMB spectrum. Successive peaks represent modes that have crossed the horizon successively earlier in the universe when the ratio of matter to radiation was lower, and so will have had their growth suppressed by radiation pressure to a greater degree. $\Omega_m$ determined in this way from the CMB is consistent at a variety of redshifts with values from the late universe: Lyman$-\alpha$ absorbtion lines of quasars, BAO, galaxy lensing and the SN Ia Hubble diagram (see Fig.~\ref{fig:Cepheid and Sn1a Ladder}).
 
Further, $\Omega_b$ is determined by the Silk damping of high-$\ell$ modes, and also by the relative heights of odd and even peaks: these are respectively compression and rarefaction modes, sensitive to the pressure to density ratio. $\Omega_b$ from the CMB is consistent with the deuterium fraction predicted by BBN. Then, substituting for $H(z)$ using \eqref{eq:Friedmann2}, Eq.~\eqref{eq:thetaz} becomes
\begin{equation}
\label{eq:thetas}
    \theta(z) = \frac{\int_{z_D(\omega_b, \Omega_m h^2)}^{\infty} \frac{c_s (\omega_b, z') dz'}{[(1+z')^3 + \frac{\omega_r}{\Omega_m h^2}(1+z')^4]^{1/2}}}{\frac{1}{1+z}\int_{0}^{z} \frac{dz'}{[\frac{1-\Omega_m}{\Omega_m} + (1+z')^3]^{1/2}}} \; .
\end{equation}
Hence we see that for fixed $\omega_b$, $\omega_r$ and $\Omega_M$ -- all parameters whose CMB values are corroborated by non-CMB datasets -- $h$ only appears in the numerator. So to change $H_0$ we must either accept a departure from the Friedmann equations, change $\omega_r$, allow conversion between energy types, or add a new non-baryonic, non-dark matter component to the mix of the early universe. 

\subsubsection{Extra relativistic species}
Extra relativistic species, such as additional neutrinos beyond the Standard Model flavours, are well-motivated in extensions to the standard model in particle physics. There is a well-known constraint of 3 on the number of light neutral particles from the decay width of the Z-boson, but this bound may be avoided if the new particles do not couple to the Z (for example, if they are sterile neutrinos which only interact gravitationally). The number of species are parametrised by $N_{\rm eff}$ which is defined by 
\begin{equation}
    \omega_r = \{ 1+ N_{\rm eff} \frac{7}{8} (\frac{4}{11})^{4/3}) \} \omega_{\gamma} \; , 
\end{equation}
where the fractions are due to the fermionic nature of neutrinos, and in the standard model $N_{\rm eff} = 3.045$\footnote{The slight deviation from $3$ is due to non-thermal spectral distortions caused by electron-positron annihilation on the neutrino energy spectrum.}. Adding extra species would mean $H_0$ in Eq.~(\ref{eq:thetas}), would need to be increased to keep $\theta$ unchanged \citep{Bernal2016}.

But if we increase $\omega_r$, we have altered the redshift of matter-radiation equality $z_{\rm eq}$ which influences the relative heights of the peaks. If we restore the heights by also increasing $\Omega_m$ to keep $z_{\rm eq}$ unchanged, we conflict with $\Omega_m$ determined from SN Ia, BAO and lensing unless we accept dark matter decay between the CMB and then. Additionally, $N_{\rm eff}$ is constrained by He abundance to be $<4$ at the $3\sigma$ level during BBN \citep{Aver2015} (see also Fig.~39 in \citealt{PlanckCollaboration2018}). 

One way in which these constraints can be avoided is via self-interacting neutrinos. The new species couples to itself, but not the standard model neutrinos (although it mixes with them). Free streaming neutrinos have a small but measurable impact on the spectrum as they travel faster than the sound speed $c_s$ and drag some of photon-baryon fluid with them, resulting in a sound horizon that would be larger than a neutrinoless universe. Self-interactions slow the speed of the neutrinos, and by tuning the strength of the interaction it is possible to arrange for the CMB spectral peak heights to be unchanged (with some small changes to other parameters such as the primordial power spectrum tilt $n_s$ and $\Omega_{\Lambda}$). By adding an additional coupling to a new scalar particle, their mixing may be suppressed at BBN, in which case they evade the He abundance constraints. \citet{Kreisch2020} have analysed such a model and found $H_0 = 72.3 \pm 1.4\; \;\SI{}{\kilo\meter\per\second\per\mega\pc}$ for $N_{\rm eff} = 4.02 \pm 0.29$. 

The drawback is the self-interaction must be extremely strong. That is, the effective coupling $G_{\rm eff}$ required between neutrinos is $\sim 10^{10} G_{\rm Fermi}$ where the Fermi constant of weak interactions in the Standard Model is $G_{\rm Fermi} = 1.17 \times 10^{-5}$ GeV$^{-2}$. Such strength contradicts calculations of the decay time of muons and tau leptons, unless the self-interactions are restricted to the tau neutrino, or weakened to a level where they do not fully resolve the tension \citep{Das2020}. So this proposal does seem highly fine-tuned. However, it is accessible to testing by CMB S4 experiments, as it predicts a stronger decrease of the damping tail for modes $\ell > 2000$.

\subsubsection{Early dark energy}
Early dark energy (EDE) refers to a boost to the expansion of the pre-recombination universe caused by a scalar field potential $V(\phi) \sim \phi^n$ in a manner similar to (but milder than) inflation. This dark energy must be present at around $z \sim 10^4$, as that is the time from which most of the growth of the sound horizon occurs. BBN constraints mean it must be absent earlier. The dark energy must also decay by the time of recombination, in order not to disrupt the damping tail of the CMB spectrum. EDE then provides a short-term boost to the expansion speed, which decreases the sound horizon by modifying the numerator of Eqn.~(\ref{eq:thetas}).

\citet{Agrawal2019} have shown $H_0 \sim 72 
\;\;\SI{}{\kilo\meter\per\second\per\mega\pc}$ 
for a fraction $\Omega_{\rm EDE} \sim 5\%$ of the energy content and the onset of its decay at redshift $z_c \sim 5000$, for a range of potentials. It may be said that this feels like fine-tuning again (though it should be said that the same point may be made about late dark energy -- the \say{why now} problem)! The model shows a small residual oscillatory signal in the CMB spectrum versus a standard $\mathrm{\Lambda CDM}$ fit, so if $H_0$ is calculated in $\mathrm{\Lambda CDM}$ in an EDE universe, it is progressively biased lower for higher angular resolution \citep{Knox2019}. Although this trend is already seen in WMAP and Planck, it is opposite to the trend in SPT. 

EDE can then be tested by higher resolution in CMB S4 experiments. It may also be testable before then: it is necessarily present around the time of matter-radiation equality, and therefore will leave an imprint in the present-day matter power spectrum $P(k)$. The spectrum captures the strength of matter density fluctuations of wavenumber $k$ and decreases for $k>k_{\rm eq}$, the wavelength corresponding to matter-radiation equality, with a gradient weakly proportional to $k_{\rm eq}$. This spectrum is measured in galaxy surveys, and the greater precision of the Vera Rubin Observatory LSST and Euclid survey will constrain the presence of any additional components of the universe at $z_{\rm eq}$ when EDE needs to be close to its maximum effect. 

\subsubsection{$T_0$ tension}
In Eq.~(\ref{eq:thetas}), $\omega_r$ is determined by the temperature of today's CMB, measured by the COBE instrument FIRAS in 1996 to be $T_0 = 2.726 \pm 0.0013 K$ \citep{Fixsen2009} and the equation of state for radiation energy $T(t) = T_0 (1+z(t))^4$. Changes to $\omega_r$ in (\ref{eq:thetas}) can be reabsorbed into changes to $h$, so there is some degeneracy between the temperature of universe and $H_0$. A thought experiment can then be done: what if $T_0$ was allowed to vary outside FIRAS bounds, keeping other quantities fixed? \citet{Ivanov2020} have considered this, opting to keep fixed $\omega_i / T_0^3$, which corresponds to the absolute energy scale of components. On combining Planck data with SH0ES, they find the $H_0$ tension is resolved within in $\mathrm{\Lambda CDM}$, but with $T_0 = 2.582 \pm 0.033$.

While this is not new physics unless one proposes a new equation of state for radiation, it is a helpful recasting of the tension. Firstly, no balloon or other measurement of $T_0$ has produced such a low $T_0$ value \citep{Fixsen2009}. Secondly, the temperature of the universe may be estimated at earlier epochs via the Sunyaev--Zel'dovich effect, providing an independent probe of $T(z)$. \citet{Luzzi2015} use the Planck SZ cluster catalog to find that deviations from $T(z) = T_0 (1+z)$ are limited to $\sim 3\%$ at $2 \sigma$ confidence out to $z = 0.94$. 

\section{Conclusions and buyer's advice}
\label{sec:five}

\begin{figure}[htbp]
    \centering
    \includegraphics[width=1\textwidth]{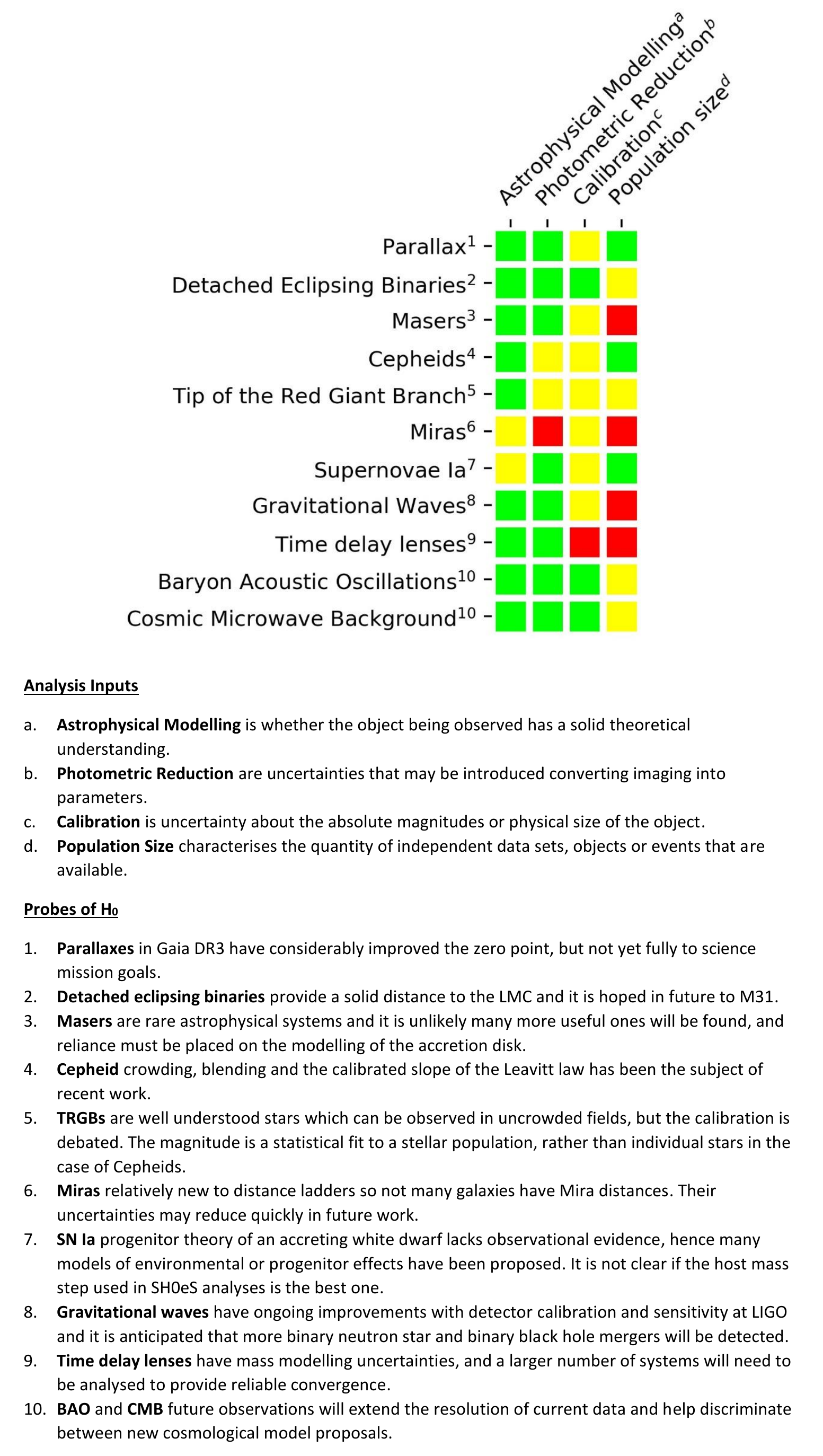}
    \caption{A `traffic light' colour coding corresponding to our view of the calculation of $H_0$ has uncertainties that may improve over the next few years. Green corresponds to a solid current position with low uncertainty, yellow is where some improvements can be expected, and red is where caution or cross-checks may be needed. For more detail the reader may consult the relevant subsection in Sect.~3. 
    This diagram was 
    inspired by a similar plot by Wendy Freedman at the ESO $H_0$ 2020 conference.}
    \label{fig:heatmap}
\end{figure}

Much effort has been expended by all the teams involved to reduce photometric biases, environmental effects, calibration error, lens mass modelling biases, CMB foreground effects and so on. Nevertheless, there remain some areas where a degree of healthy scientific scepticism might be focused, or improvements might be forthcoming. In the spirit of our Buyer's Guide, we offer our view on these areas in Fig.~\ref{fig:heatmap}.

The last 20 years have seen the error bars on $H_0$ shrink from 8\% to 1\%--2\% for the highest precision results\footnote{For a nice illustration of the evolution of Cepheid, TRGB and CMB distances, see Fig.~17 of \citet{Freedman2019}.}. For the late universe, the SH0ES team has calibrated the Cepheid luminosity zero point to 1.0\% precision, and the values are consistent whether parallaxes, the DEB distance to the LMC, or the maser distance to NGC4258 are used. The calibration of SN Ia luminosity adds another 1.3\% uncertainty, resulting in $H_0$ to 1.8\% accuracy \citep{Riess2021}. For the early universe, the high resolution and sky coverage of Planck results in 0.75\% accuracy \citep{PlanckCollaboration2018}. This has been characterised as \say{early versus late} tension, but this would be to ignore the results from the CCHP calibration of TRGBs which are a late universe result intermediate between Planck and SH0ES \citep{Freedman2019}. Additionally, the size of the BAO sound horizon in the late universe when calibrated with early universe BBN constraints is consistent with the CMB value \citep{Addison2018}.  

\subsection{Systematics or beyond $\mathrm{\Lambda CDM}$?}

Not too surprisingly there is at present no clear answer yet to the question \say{What is the true value of $H_0$?}. We do however think there is a compelling cosmological model, and that is $\mathrm{\Lambda CDM}$. It has been the best model around for the last 30 years, and no new model has yet presented a convincing case to replace it. 
Having said that, it is still disturbing that the two main ingredients in $\mathrm{\Lambda CDM}$, dark matter and dark energy, are not understood. The $H_0$ derived from the CMB is therefore part of a \say{package deal} which involves other cosmological parameters within the $\mathrm{\Lambda CDM}$ paradigm. Fortunately, the underlying physics of the CMB is well defined -- the CMB fluctuation spectrum is a solution of Boltzmann's equation. In contrast, the cosmic ladder measurements of $H_0$ probe directly this parameter and no others, and the astrophysics of the standard candles used is not fully understood.

Although many plausible extensions of $\mathrm{\Lambda CDM}$ can resolve the tension, none enjoy majority support. They either do not resolve other tensions related to $H_0$ such as the sound horizon, or seem somewhat contrived. Thus, at best they can be seen as effective theories of some unknown, and possibly more natural, microphysics. Given the lack of a compelling explanation from theory, and a greater understanding of how hard it is to \say{break} $\mathrm{\Lambda CDM}$, opinion in the community seems to be shifting. At a conference called \say{Beyond $\mathrm{\Lambda CDM}$} in Oslo in 2015, a poll suggested 69\% of participants believed new physics the most likely explanation (although we suspect some bias in the voter views given the title of the conference!). However, at a conference in 2021 entitled \say{Hubble Tension Headache} held virtually by Southampton University, UK, over 50\% of participants favoured the explanation that there were still systematics in the data.

\subsection{Open issues}

Here we comment on some frequently asked questions:
\begin{itemize}
\item \textbf{Is SH0ES $H_0$ supported?} Results from  Miras, Megamasers and HST Key Project have central values that are consistent with SH0ES, but due to their larger error bars are only in tension with Planck at the $1.5 \sim 3.0 \sigma$ level. Results from lensed quasars, TRGBs and surface brightness fluctuations vary between papers, and cannot be regarded as convincingly supportive of SH0ES.
\item \textbf{Is Planck $H_0$ supported?} The Planck value for $H_0$ is corroborated by the Atacama Cosmology Telescope to within 1\%, and is also consistent with the earlier WMAP satellite. However, it is in $2.1 \sigma$ tension with the South Pole Telescope. Planck's $H_0$ is also supported by BAO from the BOSS and DES galaxy surveys combined with an early universe prior for the baryon density.
\item \textbf{Are systematics still present?} Regarding Cepheids, considerable recent progress has been made in demonstrating crowding is under control and photometric reduction is free from bias. Nevertheless, potential weaknesses remain in the calibration and environmental corrections of SN Ia, and the calibration of TRGBs has been disputed. Re-analyses of lensed quasars show systematics are not yet under control. The nature of CMB analyses do not lend themselves easily to part-by-part decomposition or independent review, but a full re-analysis of Planck by \citet{Efstathiou2019} did not alter cosmological parameters. 
\item \textbf{Is characterisation of \say{early versus late} helpful?} The strongest evidence in favour of this are the Cepheid results and Planck, as while other results are broadly consistent with this view they are individually more uncertain. However, if the calibration of TRGB were resolved in favour of the CCHP value, it would weaken the case for this view. It is however helpful to consider new cosmological models in terms of those modifying the post- or pre-recombination universe, 
as at present those modifying the pre-recombination universe appear more viable. 

\item \textbf{Is there a convincing theoretical solution?} Simply put, $\mathrm{\Lambda CDM}$ is very hard to break. Proposals need to clear three main hurdles: fitting a wide range of cosmological data, consistency with other predictions of $\mathrm{\Lambda CDM}$, and providing a convincing Bayesian argument that the model is preferable to $\mathrm{\Lambda CDM}$. None yet do so. Nevertheless, there are strong hints what a solution might look like, and it is fair to point out that the innovation of $\Lambda$ in $\mathrm{\Lambda CDM}$ itself solved a previous tension between a low matter density and flat universe. 
\item \textbf{Are TRGBs and gravitational waves offering a way forwards?} It is likely that a consensus on TRGB calibration will be reached in the near-future, and TRGBs promise cleaner fields and better theoretical modelling than Cepheids. They are also ideal for JWST observations, which has four times the $J$ band resolution of the HST/WFC3 $H$ band. Miras are interesting, but are more expensive to observe and harder to calibrate. Gravitational waves offer an inherently transparent and accurate way to measure $H_0$, once enough bright and dark siren data has been collected by the middle of this decade. If these two were to corroborate SH0ES, it would be strong evidence in favour of the breakdown of $\mathrm{\Lambda CDM}$. Alternatively, if they agree with Planck, one would have to conclude SH0ES is the outlier.
\end{itemize}

\subsection{Buyer's advice}
We would suggest the following
in response to the question of \say{Which $H_0$ should I use?}, depending on the desired application.

\begin{itemize}
\item \textbf{For cosmological inference}: As the tension in the Hubble parameter is between local direct estimates, and the best-fitted ${\rm \Lambda CDM}$ model, 
the only two possible explanations are unknown systematic effects in the local direct estimate, or that our universe is described by a model different from ${\rm \Lambda CDM}$.
If it turns out that there are  systematics in the local estimate, it is clear that we should use 
Planck's value $H_0 = 67.4 \;\;\SI{}{\kilo\meter\per\second\per\mega\pc}$ \citep{PlanckCollaboration2018}. If the universe is not ${\rm \Lambda CDM}$, it remains the case that ${\rm \Lambda CDM}$ is a great fit to cosmological data with all other parameters consistent with their CMB-derived values. We therefore recommend the 
full package of Planck or Planck-like cosmological parameters should be 
used, so one can test the the ${\rm \Lambda CDM}$ model self consistently.
A specific example is in choosing parameters for N-body simulations.
It makes sense to select the Planck set of parameters 
($H_0$, $\Omega_m$ and $\sigma_8$, etc.) so one can test the growth of structure with cosmic time given ${\rm \Lambda CDM}$ fit to Planck. 
In contrast, the ${\rm \Lambda CDM}$ model with $H_0 = 73.2 \;\;\SI{}{\kilo\meter\per\second\per\mega\pc}$ is a poor fit to most existing cosmological observations (CMB and galaxy surveys). 

Another application of using $H_0$ and other parameters from Planck is when using them as priors for analyses of new data sets, in a Bayesian framework. Here the errors bars of parameters or the full Planck posterior of the probability distribution are important, as they propagate through the parameter marginalisation process.
We also maintain that it is not correct to inflate $H_0$ confidence intervals to \say{hedge bets}, unless one does that in a non-$\mathrm{\Lambda CDM}$ model and fully rederives the posteriors for all other parameters in a Bayesian fashion. 

\item \textbf{For the local universe:}. The SH0ES value of $H_0 = 73.2 \pm 1.3 \; \;\SI{}{\kilo\meter\per\second\per\mega\pc}$ \citep{Riess2021} using Gaia parallaxes and Cepheids is a good choice, as it is consistent with independent measurements such as Megamasers and Miras and re-analyses of the data (albeit generally starting with the photometric parameters, and not the original imaging) have produced mostly consistent results. Therefore, the SH0ES value should be used for calculations that require an expansion rate or distances alone, but are local enough to have minimal cosmological model dependence. 
Having said that, the CCHP TRGB result of $H_0 = 69.8 \pm 1.9 \; \;\SI{}{\kilo\meter\per\second\per\mega\pc}$ \citep{Freedman2019} is noteworthy for its consistency with Planck and the advantages TRGB observations may offer relative to Cepheids. It is very much a case of \say{watch this space} for developments.

\item \textbf{For pedagogical purposes:}
For the purpose of teaching cosmology, popular talks, or back-of-the-envelope calculations
it would make sense to use the intermediate round value of $H_0 = 70 \; \SI{}{\kilo\meter\per\second\per\mega\pc}$. Wherever possible we recommend indicating the dependence of your key results on $H_0$ either in formulae, or as a method to adjust the key results. Anticipating a future resolution of the tension, this will be of great assistance to future researchers.  

\end{itemize}

\subsection{Future prospects}
Looking to the future, the Zwicky Transient Facility and Foundation surveys of nearby SN Ia will be very helpful in reducing potential calibration issues. They will do this by resolving the underlying population characteristics, having cleaner selection functions, and providing more galaxies in which to calibrate the distant Hubble flow sample. The early signs of Gaia Early Data Release 3 is that it provides a considerable reduction, but not elimination, of the bias apparent in Data Release 2. But it is not the last word, and the next release is scheduled for 2022. The key point of Gaia is that it addresses lingering concerns about how the low metallicity environment of the LMC, or the maser disk modelling and crowded fields of NGC4258 may affect the calibration of standard candles. In particular, it will be useful to have accurate Gaia parallaxes to Milky Way globular clusters such as $\omega$Cen to provide additional calibrators of the TRGB. The JWST will greatly expand the range of TRGB observation, and provide continuity in the case of any further degradation of the ageing HST. The Extremely Large Telescope in Chile, scheduled for first light in 2025, can extend the range of DEB distances to M31 allowing it to contribute its Cepheid and TRGB populations to calibrations. 

In the field of gravitational waves, it is perhaps disappointing that there have been no more observations of \say{bright sirens} like the neutron star merger GW170817 so far. But the commencement of operations at the VIRGO detector in Italy, and recently the KAGRA detector in Japan offer hope that the more frequent event detections and better sky localization, combined with an improved instrumental calibration, will provide a 2\% measurement of $H_0$ within this decade. Many more time-delay lensing systems will be seen in future galaxy surveys, so there is a strong incentive to resolve remaining systematics in the modelling and speed up the analysis pipeline. 

The Simons Telescope in Chile aims to map the polarisation spectrum of the CMB to an order of magnitude higher than Planck, and will start taking data in the next two to three years. Close to 2030, two new CMB Stage 4 telescopes will be operational in Chile and the South Pole, which will further extend the spectral resolution. The depth of these surveys will be able to support or rule out many pre-combination modifications of $\mathrm{\Lambda CDM}$. Surveys conducted by the Dark Energy Spectroscopic Instrument (DESI), Vera Rubin Observatory (previously named LSST), Euclid satellite and Nancy Grace Roman Space Telescope (previously named WFIRST), combined with theoretical progress in the quasi-linear regime of structure formation, will enable the matter-power spectrum to be compared with $\mathrm{\Lambda CDM}$ predictions with much greater depth and resolution.

\begin{acknowledgements}

We thank George Efstathiou, Stephen Feeney, Wendy Freedman, John Peacock and Adam Riess for their helpful comments on the paper. We also thank the two anonymous referees for their thorough and helpful suggestions. We are also grateful to the BOSS Collaboration, Wendy Freedman, Lloyd Knox, John Peacock, the Planck Collaboration, Adam Riess, Dan Scolnic and Wenlong Yuan for the use of figures from their original manuscripts. OL and PL acknowledge support from an STFC Consolidated Grant ST/R000476/1, and PL acknowledges STFC Consolidated Grant ST/T000473/1. OL also acknowledges a Visiting Fellowship at All Souls College, Oxford. 

\end{acknowledgements}

\bibliographystyle{spbasic-FS-etal}      
\bibliography{H0_review_citations4}   

\end{document}